\documentclass[12pt,Bold,letterpaper]{mcgilletdclassmine}
\usepackage[dvips,final]{graphicx}
\usepackage{latexsym,amssymb, amsmath,amsfonts,epsfig, epic, eepic,subfigure}
\usepackage[dvips,letterpaper,left=3.8cm,top=3.2cm,bottom=2.5cm,right=2.5cm,includefoot]{geometry}
\usepackage[ps,dvips,matrix,arrow,frame,import,curve,color]{xy}
\usepackage{epsfig}
\usepackage{epstopdf}
\usepackage{epsf}
\usepackage{xspace}
\usepackage{multirow}
\usepackage{rotating}
\input{epsf.sty}
\usepackage{verbatim}

\def\blfootnote{\xdef\@thefnmark{}\@footnotetext}

\long\def\symbolfootnote[#1]#2{\begingroup%
\def\thefootnote{\fnsymbol{footnote}}\footnote[#1]{#2}\endgroup}

\newcommand{\be}{\begin{eqnarray}}
\newcommand{\ee}{\end{eqnarray}}
\newcommand{\ben}{\begin{eqnarray*}}
\newcommand{\een}{\end{eqnarray*}}

\newcommand{\bcent}{\begin{center}}
\newcommand{\ecent}{\end{center}}
\newcommand{\benum}{\begin{enumerate}}
\newcommand{\eenum}{\end{enumerate}}
\newcommand{\bdesc}{\begin{description}}
\newcommand{\edesc}{\end{description}}
\newcommand{\bitem}{\begin{itemize}}
\newcommand{\eitem}{\end{itemize}}
\newcommand{\bquote}{\begin{quote}}
\newcommand{\equote}{\end{quote}}
\newcommand{\bhalfp}{\begin{minipage}{0.45\textwidth}}
\newcommand{\ehalfp}{\end{minipage}}
\newcommand{\bhead}{\begin{center}\bf \Large}
\newcommand{\ehead}{\end{center}\bigskip}

 \newcommand{\del}{{\partial}}
 
\def\be{\begin{equation}}
\def\ee{\end{equation}}
\def\ba{\begin{eqnarray}}
\def\ea{\end{eqnarray}}

\newcommand{\roughly}[1]{\mathrel{\raise.3ex\hbox{$#1$\kern-0.85em
\lower1ex\hbox{$\sim$}}}}

\def\2pi{\left(2\pi\right)}

\def\beq{\begin{equation}}
\def\eeq{\end{equation}}
\def\bg{\begin{eqnarray}}
\def\nd{\end{eqnarray}}
\def\bea{\begin{eqnarray}}
\def\eea{\end{eqnarray}}

\def\D3{\overline{\mbox{D3}}}

\SetTitle{\Huge\bfseries{F-theory \& M-theory perspectives on $\mathcal N=2$ supersymmetric gauge theories in four dimensions. \\\vspace{1cm}}}%
\SetAuthor{\vspace{1.5cm}\Large\bfseries{Alisha Wissanji}}%
\SetDepartment{\Large{Physics Department}}
\SetUniversity{\Large{McGill University}}%
\SetUniversityAddr{\Large{Montreal, Quebec}}%
\SetThesisDate{\Large{July 2012}}%
\SetRequirements{\vspace{1cm}\Large{A thesis submitted in partial fulfilment \\ of 
the requirements for the degree of}}%
\SetDegreeType{\vspace{0.3cm}\Large{Doctor of Philosophy}}%
\SetCopyright{\Large{\copyright\ Alisha Wissanji, 2012. All rights reserved}}%

\begin{document}

\maketitle%

\newpage\thispagestyle{empty}
${}$ \clearpage
\newpage

\begin{romanPagenumber}{2}%
\SetAbstractEnName{Abstract}%
\SetAbstractEnText{Deformations of the original F-theory background are proposed.  These lead to multiple new dualities and physical phenomena.  We concentrate on one model where we let seven-branes wrap a multi-centered Taub-NUT space instead of $\mathbb R^4$.  This configuration provides a successful F-theory embedding of a class of recently proposed four-dimensional $\mathcal N=2$ superconformal (SCFT) \`a la Gaiotto.  Aspects of Argyres-Seiberg duality, of the new Gaiotto duality, as well as of the branes network of Benini-Benvenuti and Tachikawa are captured by our construction.  The supergravity theory for the conformal case is also briefly discussed.  Extending our construction to the non-conformal case, we find interesting cascading behavior in four-dimensional gauge theories with $\mathcal N=2$ supersymmetry.  Since the analysis of this unexpected phenomenon is quite difficult in the language of type IIB/F-theory, we turn to the type IIA/M-theory description where the origin of the $\mathcal N=2$ cascade is clarified.  Using the T-dual type IIA brane language, we first start by studying the $\mathcal N=1$ supersymmetric cascading gauge theory found in type IIB string theory on $p$ regular and $M$ fractional D3-branes at the tip of the conifold.  We reproduce the supersymmetric vacuum structure of this theory.  We also show that the IIA analog of the non-supersymmetric state found by Kachru, Pearson and Verlinde in the IIB description is metastable in string theory, but the barrier for tunneling to the supersymmetric vacuum goes to infinity in the field theory limit.  We then use the techniques we have developed to analyze the $\mathcal N = 2$ supersymmetric gauge theory corresponding to regular and fractional D3-branes on a near-singular K3, and clarify the origin of the cascade in this theory.}
\AbstractEn%
\SetAbstractFrName{R\'{e}sum\'{e}}%
\SetAbstractFrText{Diff\'erentes d\'eformations de la g\'eom\'etrie originale de la th\'eorie F sont propos\'ees.  Ces derni\`eres  g\'en\`erent une multitude de nouvelles dualit\'es ainsi que de nouveaux ph\'enom\`enes physiques.  Nous nous  concentrons sur un seul mod\`ele o\`u les membranes en sept dimensions spatiales s'enveloppent autour d'un espace Taub-NUT avec multi-centres au lieux de l'espace $\mathbb R^4$ original.  Cette configuration g\'en\`ere avec succ\`es la r\'ealisation, en th\'eorie F, d'une famille de th\'eories de jauges superconformes en quatres dimensions avec  $\mathcal N=2$ supersym\'etries nouvellement propos\'ees par Gaiotto.  Deplus, plusieurs aspects de la dualit\'e d'Argyres-Seiberg, de la nouvelle dualit\'e de Gaiotto ainsi que du r\'eseaux de membranes de Benini-Benvenuti et Tachikawa sont r\'ealis\'es par notre construction.  La th\'eorie de supergravit\'e pour le cas conforme est bri\`evement discut\'ee.  La g\'en\'eralisation de notre construction au cas non-conforme m\`ene \`a l'observation surprenante de cascade chez les th\'eories de jauges avec $\mathcal N=2$ supersym\'etries en quatres dimensions.  Puisque l'analyse de ce ph\'enom\`ene est difficile dans le language de type IIB/ th\'eorie F, nous nous tournons vers le type IIA/theorie M o\`u l'origine de ce ph\'enom\`ene est \'elucid\'ee.  En utilisant le langage des membranes en type IIA sous la dualit\'e-T, nous d\'ebutons par l'\'etude de cascade chez les th\'eories de jauges avec $\mathcal N=1$ supersym\'etrie tel que pr\'esent\'e en type IIB avec $p$ membranes D3 r\'eguli\`eres et $M$ membranes  D3 fractionnaires situ\'ees au bout d'un espace conifold.  Nous reproduisons avec succ\`es la structure du vide supersym\'etrique de cette th\'eorie.  Aussi, nous d\'emontrons que l'analogue en type IIA des \'etats non-supersymmetriques d\'ecouverts par Kachru, Pearson et Verlinde en type IIB sont m\'etastables en th\'eorie des cordes alors que la barri\`ere permettant de passer au vide supersymmetrique tant vers l'infinie dans la limite de la th\'eorie des champs.  Nous utilisons finalement les techniques que nous avons d\'evelopp\'ees afin d'analyser la th\'eorie de jauge supersymm\'etrique avec $\mathcal N=2$ correspondante \`a des membranes D3 r\'eguli\`eres et fractionnaires sur un espace K3 presque singulier et clarifions l'origine du m\'ecanisme de cascade dans cette th\'eorie. }%
\AbstractFr
%%%%%%%%%%%%%%%%%%%%%%%%%%%%%%%%%%%%%%%%%%%%%%%%%%%%%
%%         Acknowledgements                        %%
%%%%%%%%%%%%%%%%%%%%%%%%%%%%%%%%%%%%%%%%%%%%%%%%%%%%%

\SetDedicationName{\MakeUppercase{Dedication}}%
\SetDedicationText{\begin{flushright}
Bismillahir Rahmanir Rahim\\
Al Hamdu Lillahi Rabbil Alamin
\end{flushright}
\ \\\\
This thesis is dedicated to my family for their unwavering faith and support.
 }%
\Dedication%

\SetAcknowledgeName{\MakeUppercase{Acknowledgements}}%
\SetAcknowledgeText{I would like to thank my advisor Keshav Dasgupta and my co-advisor Alexander Maloney for their mentorship and guidance.  I would like to thank them for having opened me the doors of academia.  I also thank Keshav Dasgupta for carefully reading the draft of this thesis and answering my numerous questions.  I would like to thank my collaborators Paul Franche, Rhiannon Gwyn, Bret Underwood, Jihye Seo, Keshav Dasgupta and David Kutasov.  I am also grateful to the faculty members of the McGill Physics Department as well as to those of the Centre de Recherche Mathematiques (CRM) of Montreal University, especially Johannes Walcher, Robert Brandenberger and Veronique Hussin for their continuous guidance.  I would like to thank many postdocs for discussion and help:  Alexi Kurkela, Mohammed Mia, Anke Knauf, Andrew Frey, Josh Lapan, Shunji Matsuura, and especially Alejandra Castro and Simon Caron-Huot for their friendships and for having inspired me.  I am grateful to the faculty members met in different conferences and who have taken the time to sit down with me to discuss physics:  Nathan Seiberg, Amihay Hanany, Savdeep Sethi, Ashoke Sen, Sergei Gukov, and especially David Kutasov; I was deeply touched by the kindness of you all.  I am in debt to David Kutasov for the tremendous respect he conferred to me and for his patience.  Special thanks to Edward Witten for an advise that transcends physics.  I thank my friends for memorable times: Paul Franche, Johanna Karouby, Nima Lashkari, Aaron Vincent, Xue Wei, Fang Chen, Michael Pagano, Arnaud Lepage-Jutier, Olivier Trottier, Long Chen, St\'ephane Detournay, Geoffrey Compere, Balt van Rees and especially Nopaddol Mekareeya for his guidance.  I would also like to thank some people dear to me outside of High Energy Physics: Simon Langlois, Francoise Provencher and especially Nasreen Dhanji for her wisdom.  Most importantly, I thank my family for their continuous guidance, support and faith in me and without whom I would not have made it this far.
I would like to take the opportunity to thank all the administrative and professional  staff of the Physics Department of McGill University for their kind help.  I acknowledge financial support from the Government of Quebec,  the Government of Canada and the Physics Department of McGill University.   }%
\Acknowledge
%%%%%%%%%%%%%%%%%%%%%%%%%%%%%%%%%%%%%%%%%%%%%%%%%%%%%
%%         Preface                        %%
%%%%%%%%%%%%%%%%%%%%%%%%%%%%%%%%%%%%%%%%%%%%%%%%%%%%%
 \SetAcknowledgeName{Preface}%
\SetAcknowledgeText{
\noindent \textbf{\Large{Statement of Originality}}\\
\ \\
The results presented in this thesis constitute original work that was published in the following articles: \\
\begin{itemize}
\item \textbf{Chapter 3} K.Dasgupta, J. Seo and \textbf{A. Wissanji} (2012), ``F-theory, Seiberg-Witten curves and $\mathcal N=2$ Dualities," Journal of High Energy Physics \textbf{1202}, 146, 117pp.\\
\item \textbf{Chapter 4} D.Kutasov and \textbf{A. Wissanji} (2012), ``IIA Perspective on Cascading Gauge Theory" arXiv: 1206.0747[hep-th], 43pp.\\
\end{itemize}
Chapter 3 is based on what was referred in \cite{DasguptaSW} as model 2.  We present a deformation of Sen's original F-theory geometry which enabled us to embed in F-theory a class of Gaiotto new $\mathcal N=2$ SCFT as well as several aspects of Argyres-Seiberg duality, Gaiotto duality, and the Benini-Benvenuti-Tachikawa brane network.  We are therefore able to present a simple geometric brane picture which captures many intricacies of $\mathcal N=2$ supersymmetric gauge theories in four dimensions.  We also propose a type IIB/F-theory non-conformal  construction which seems to have all the right ingredients to lead to a cascade mechanism in four-dimensional $\mathcal N=2$ SYM theories.  Chapter 4 is based on \cite{KutasovKW} where we study the $\mathcal N=1$ cascade mechanism of Klebanov-Strassler and reproduce the supersymmetric vacuum structure of this theory using type IIA/M-theory brane constructions.  We show that the type IIA analog of the non-sypersymmetric state of Kachru-Pearson-Verlinde is metastable in string theory but the barrier for tunnelling to the supersymmetric vacuum goes to infinity in the field theory limit.  We finally analyze the $\mathcal N=2$ supersymmetric gauge theory using type IIA/M-theory and clarified the origin of the cascade in this theory. \\
\ \\
\noindent \textbf{\Large{Contribution of the author}}\\
\ \\
\cite{DasguptaSW} was work done in collaboration with Professor Keshav Dasgupta and Jihye Seo from McGill University.  I proposed the original idea that led to model 2 in \cite{DasguptaSW}.  This was the stepping stone which led to further studies and generalizations in the paper.  In this article, I participated in detailed discussions and calculations at every step of the analysis, leading to many results that were in included in the article.  Finally, I wrote various appendices in the paper.  \\
\ \\
\cite{KutasovKW} was work done in collaboration with Professor David Kutasov from the Enrico Fermi Institute at the University of Chicago.  In this paper, I proposed to study the $\mathcal N=2$ cascade, wrote the original draft of the associated  section in the article, and contributed to detailed discussion at every step of the analysis which led to many results that were included in the article.  
%In particular, we were able to solve the quantum moduli space of $\mathcal N=2$ SYM theory using M-theory after I worked out the analogy with $\mathcal N=2$ SQCD and pointed out that the solution was to look at the root of the baryonic branch.  
}%
\Acknowledge%

%\cleardoublepage${}$
\TOCHeading{\huge\bf{Table of Contents}\vspace{1cm}}%
\LOFHeading{List of Figures}%
\LOTHeading{List of Tables}%
\tableofcontents %
\listoffigures %
\listoftables %

\end{romanPagenumber}

\mainmatter
\setcounter{secnumdepth}{2}
%\title{F-Theory, Seiberg-Witten Curves and ${\cal N} = 2$ Dualities}

%%%%%%%%%%%%%%%%%%%%%%%%%%%%%%%%%%%%%%%%%%%%%%%%%%%%%%%%%%%%%%%%%%%%%%%%%%%%%%%%%%%%%%%%%%%%%%%%%% Introduction

\chapter{Introduction}

The primary goal of this thesis is to show that many new facets of non-abelian gauge theories with $\mathcal N=2$ supersymmetry (susy) in four dimensions can be revealed by using the language of branes in string theory.  
In their proper regime of validity, branes capture all the physics contained in supersymmetric field theories and provide insights on new physical phenomena by generating geometric pictures of the intricacies of these field theories.  Moreover, they are powerful tools for understanding certain dualities occurring in supersymmetric field theories.  In particular, branes in string theory shed new light on the strongly coupled regime of supersymmetric non-abelian gauge theories both in the conformal and non-conformal cases.  The long term goal of this research direction is that it might lead to a better understanding of some aspects of physical phenomena occurring at strong coupling in non-supersymmetric non-abelian gauge theories and about which little is currently known.  We will however not address this question in this thesis.   

Non-abelian, non-supersymmetric gauge theories such as Quantum Chromodynamics (QCD) are the foundation on which our understanding of the dynamics of elementary particles in the Standard Model lies.  Given the importance of such theories, it is surprising to realize that there is still much to learn about them.  For instance, these theories are often asymptotically free meaning that that they are free in the ultra-violet energy (short distance).  These theories are also strongly coupled at infra-red energy (long distance).  In the latter regime, all our perturbative (weakly coupled) field theory techniques fail.  Trying to study these strongly coupled non-abelian gauge theories is one of the biggest challenge of modern theoretical high energy physics.  

Seiberg and Witten provided in \cite{sw1,sw2} a much celebrated breakthrough in understanding the strongly coupled regime of a certain family of gauge theories namely Supersymmetric Yang-Mills (SYM) theories.  Their work was achieved through better understanding of the implications of supersymmetry and use of a known duality which allowed them to probe the strong coupling regime of supersymmetric field theories by providing a dual weakly coupled picture; inverting the electric matter for the magnetic one in the process.  The work of Seiberg and Witten led to exact results on the vacuum structure of non-abelian supersymmetric gauge theories, both with and without matter content.  These results became the stepping stone for many generalization to higher rank gauge groups, new superconformal field theories, and more complicated dualities.  In addition to the plethora of new mathematical applications they provided, these supersymmetric field theories became toy models for studying theories such as QCD since they capture some phenomena which also occurs in non-supersymmetric non-abelian gauge theories.   

In recent years, it was shown that branes - extended object in string theory \cite{PolchinskiDB}- are powerful objects that provide geometric and tractable descriptions of supersymmetric gauge theories.  From the point of view of theories living on branes, gauge theories appear as effective low energy descriptions which are valid in prescribed regions of the moduli space of vacua.  Different brane pictures have different descriptions depending on which region of the moduli space of vacua one is interested in studying; sometimes providing insights into regions which don't even have field theoretic descriptions.  In addition to shedding light on relations between such field theories, we will see that the simple nature of branes allows one to unveil new physics hidden in the language of field theory.  The brane description that we will mostly be concern with throughout this thesis is that of IIB and F-theory as proposed by Vafa in \cite{vafaF} as well as that of type IIA and M-theory put forward by Witten \cite{WittenMT}.  

In aiming to understand the interplay between the brane language and supersymmetric gauge theories, we will review in Chapter 2 what we believed to be the starting point of this research direction, namely the embedding of Seiberg-Witten theory in F-theory by Sen \cite{senF} and Banks, Douglas and Seiberg \cite{bds}.  We will then describe possible deformations of the original F-theory background which enabled us to not only describe recently proposed field theories and their associated dualities but also discover new physics using the language of branes. In Chapter 3, we will concentrate on one model where we let seven-branes wrap on a multi-centered Taub-NUT space instead of $\mathbb R^4$.  This configuration provides a successful embedding in F-theory of a class of recently proposed four-dimensional $\mathcal N=2$ SCFT \`a la Gaiotto \cite{gaiotto}.  Aspects of Argyres-Seiberg duality \cite{ArgyresSeiberg}, of the new Gaiotto duality \cite{gaiotto}, as well as of the brane network of Benini-Benvenuti and Tachikawa \cite{bbt} will be captured by our construction.  The supergravity theory for the conformal case will also be briefly discussed.  Extending our construction to the non-conformal case, we will find interesting cascading behavior in theories with $\mathcal N=2$ supersymmetry in four dimensions \cite{PolchinskiMX, BeniniIR}.  Since the analysis of this unexpected phenomenon will be limited by the difficulties of the type IIB/F-theory language, we will turn, in Chapter 4, to type IIA/M-theory where the origin of $\mathcal N=2$ cascade mechanism will become clear.  Using the T-dual type IIA brane language, we will first start by studying the $\mathcal N=1$ supersymmetric cascading gauge theory found in type IIB string theory on $p$ regular and $M$ fractional D3-branes at the tip of the conifold \cite{KlebanovHB}.  We will reproduce the supersymmetric vacuum structure of this theory \cite{DymarskyXT}.  We will then show that the IIA analog of the non-supersymmetric state found by Kachru, Pearson and Verlinde \cite{KachruGS} in the IIB description is metastable in string theory, but the barrier for tunneling to the supersymmetric vacuum goes to infinity in the field theory limit.  We will then use the techniques we will have developed to analyze the $\mathcal N = 2$ supersymmetric gauge theory corresponding to regular and fractional D3-branes on a near-singular K3, and clarify the origin of the cascade in this theory.  We will end with Chapter 5 where a discussion of possible extensions of this work will be presented.  An appendix to Chapter 4 is also included in this thesis. 

We start Chapter 2 with a review of some basic notions of string theory and branes that will be useful throughout the thesis.  The notation used in the thesis is as follows: $1+9$ spacetime dimensions in string theory are labeled by $(x^0,x^1,\cdots, x^{9})$.  The eleventh spatial dimension of M-theory is $x^{10}$.  The corresponding Dirac matrices are denoted by $\Gamma^\mu$, $\mu=0,1,\cdots, 9$ with algebra 
\bg
\{\Gamma^\mu,\Gamma^\nu\}=2\eta^{\mu\nu}
\nd
and the metric has the signature $(-+\cdots+)$.

\chapter{Aspects of String Theory and Branes}

In this section, we will review some useful properties of type IIA and type IIB String Theory as well as their embedding in M-theory and F-theory respectively.  This discussion will be accompanied by a description of the branes in each regime.    
%\subsubsection{supersymmetry?}
%strings, boundary conditions and Chan-paton factors.
%\subsection{strings in String theory}
%under T-duality, closed bosonic strings transforms to closed bosonic strings of the same type.  Open bosonic strings transform according to their boundary condition. 

\section{String theory parameters}

The protagonists of this story are strings.  These are one spatial dimensional objects with characteristic length scale denoted by $l_s$.  The string length scale $l_s$ is related to the string tension $T$ and to the open-string Regge slope parameter $\alpha'$ by the relations
\bg\label{alphaprime}
T={1\over {(2\pi\alpha')}}; \hskip 1cm \alpha'=\frac{1}{2}l_s^2.
\nd
Fundamental constants such as the speed of light $c$, Planck's constant $\hbar$ and Newton's gravitational constants $G$ form the Planck length $l_p$ and the Planck mass $m_p$:
\bg\label{plancklength}
l_p&=&\left(\frac{\hbar G}{c^3}\right)^{1/2}=1.6\times 10^{-33} \text{cm},\\
m_p&=&\left(\frac{\hbar c}{G}\right)^{1/2}=1.2\times 10^{19} \text{GeV}/c^2,
\nd
where $1$ GeV$\approx(1\times 10^9)\times(1.6\times 10^{-19})$ joules.  
The UV cutoff of the theory is given by $1/l_s$ and the relation between $l_s$ and $l_p$ is \cite{BBS}: 
\bg
l_p=g_s^{1/3}l_s.
\nd
At energies far below the Planck energy $E_p$ ($E_p=m_pc^2$), distances of the order of the Planck length can not be resolved and strings can be accurately approximated by point particles, like its the case in quantum field theory \cite{BBS}.

As it moves, the string sweeps out a two-dimensional surface in spacetimes called the string world sheet of the string.  We will see later that a one spatial dimensional string can be generalized to a $p$ spatial dimensional object denoted $p$-brane (the fundamental string having $p=1$).  The latter has tension $T_p$ and sweeps out a $p+1$ dimensional volume $V$  in spacetime.  The action of such $p$-brane is given by $S_p=-T_pV$ \cite{BBS}.

\section{Strings in String Theory}

The string sigma model action classically represents the world sheet action.  The former is given by:
\bg
S_\sigma={T\over 2}\int{\sqrt{-h}h^{\alpha\beta}\eta_{\mu\nu}\partial_\alpha X^\mu\partial_\beta X^\nu d\sigma d \tau},
\nd
where $\alpha,\beta$ are world sheet indices and $\mu,\nu$ are target space indices.  $h_{\alpha\beta}(\sigma,\tau)$ is the world sheet metric, $h=\det h_{\alpha\beta}$.  
Throughout the text, we will encounter the function $X^\mu(\sigma,\tau)$ which describes, in the string sigma model action, the spacetime embedding of the string world sheet.  The latter is parameterized by $\tau$ and $\sigma$ where $\tau$ is the world sheet time coordinate and $\sigma$ is the spatial coordinate, parametrizing the string at a given time \cite{polchinski1}.

\subsection{Open or closed strings}

Strings have different boundary conditions \cite{BBS, polchinski1} depending on whether they are opened or closed. A closed string is topologically a circle whereas an open string is topologically a line element.  We let $0\le \sigma \le \pi$.  The two types of boundary conditions which respect D-dimensional Poincar\'e invariance are stated below.  They stipulate that no momentum is flowing through the ends of the string for all values of $\mu$.

\begin{itemize}
\item The spatial coordinate $\sigma$ is periodic for closed strings, leading to the condition 
\bg
X^{\mu}(\sigma,\tau)=X^{\mu}(\sigma +\pi,\tau).
\nd
The endpoints are thus joined to form a loop and there is no boundary. 
\item Open string with Neumann boundary condition have a vanishing momentum with component normal to the boundary of the world sheet 
\bg
\partial^\sigma X^{\mu}(\tau,0)=\partial^\sigma X^{\mu}(\tau, \pi)=0.
\nd
The end of the open strings move freely in spacetime. 
\end{itemize}
Throughout the text, we will need to consider a different sort of boundary condition which breaks Poincar\'e invariance: 
\begin{itemize}
\item The Dirichlet boundary condition for open strings 
\bg
X^{\mu}|_{\sigma=0}&=&X^\mu_0,\\
X^{\mu}|_{\sigma=\pi}&=&X^\mu_\pi,
\nd
which means that the two ends of the open string are fixed i.e $\delta X^\mu=0$.
$X^\mu_0$ and $X^\mu_\pi$ are constant with $\mu=1,\cdots,D-p-1$ where $D$ is the dimension of spacetime and where Neumann boundary conditions are statisfied for the remaining $p+1$ coordinate.  
\end{itemize}
$X^\mu_0$ and $X^\mu_\pi$ represent the positions of $Dp$-branes e.g $p$ spatial dimensional Dirichlet branes.  The defining property of the latter is that fundamental strings can end on them: the coordinates of the attached string satisfy Dirichlet boundary condition in the direction normal to the brane and Neumann boundary condition in the direction parallel to the brane.  $Dp$-branes break Poincar\'e invariance unless $p=D-1$ \cite{BBS, polchinski1}.  We will elaborate more on Dirichlet branes in the next few sections. 
 
\subsection{Chan-Paton factor}

%As we have just seen, Dp-branes are remarkable objects which can introduce nonabelian gauge theories in string theory.  More precisely, nonabelian gauge fields seem to live on the worldvolume of multiple Dp-branes.  We review below the process responsible for this phenomenon, namely Chan-Paton factors.
Under T-duality ($R\to 1/R$), an open string with Neumann boundary conditions becomes a open string with Dirichlet boundary conditions whose end points ends on $Dp$-branes.  When one open string is in the presence of a stack of $N$ $Dp$-branes, the open string carries at its endpoints quantum numbers (initially though of as quarks and antiquarks) transforming respectively in the $N$-dimensional representations $\mathbf{R,\ \bar R}$ of a gauge group G.  These $N$-valued labels, referred to as Chan-Paton charges, associate $N$ degrees of freedom to each endpoints of the open string.    
Oriented open strings are characterized by complex representations $\mathbf R$ for which $\mathbf{ R\neq \bar R}$ and thus have distinguishable endpoints.    
Unoriented open strings, on the other hand, have real representations $\mathbf R$ which leads to $\mathbf{R=\bar R}$ and thus have indistinguishable endpoints \cite{GSW, BBS}.

Accordingly, for an oriented open string, one described the gauge group $U(N)$ by letting the charges located at $\sigma=0$ transform under the fundamental representation $\mathbf N$ while the degrees of freedom at $\sigma=\pi$ transform under the antifundamental representation $\mathbf{\bar N}$ of the gauge group.  Unoriented open strings lead to orthogonal or symplectic groups with real fundamental representations at both $\sigma=0$ and $\sigma=\pi$. This can be understood as follows.
For strings with Dirichlet boundary conditions $\partial_\sigma X^i=0$ at $\sigma=\{0,\pi\}$ for $\mu=i$,  the mode expansion for $X^i(\sigma, \tau)$ is given by :
\begin{equation}
X^i(\sigma,\tau)=x^i+p^i\tau+i\sum_{n\neq 0}{\frac{1}{n}\alpha_n^i e^{-in\tau}\cos {n\sigma}},
\end{equation} 
where $\sigma$ goes from 0 to $\pi$ along the string.
Under orientation reversal, one interchanges the two ends of the string and let the parametrization of $\sigma$ run in the opposite direction.  We obtain the following mode expansion: 
\begin{equation}
X^i(\pi-\sigma,\tau)=x^i+p^i\tau+i\sum_{n\neq 0}{\frac{1}{n}(-1)^n\alpha_n^i e^{-in\tau}\cos {n\sigma}}.
\end{equation} 
The coordinate transformation $\sigma\to\pi-\sigma$ and $\tau\to \tau$ used above is generated by the world sheet parity operator $\Omega$ whose properties are that $\Omega^2=1$ and its eigenvalues are given by $\Omega=\pm 1$.  As we see from the equations above, $\Omega$ exchanges $\alpha^i_n$ for $(-1)^n\alpha^i_n$ where $\alpha^i_n$ is a transverse oscillator

\bg
\Omega \alpha^i_n\Omega^{-1}=(-1)^n\alpha^i_n.
\nd
When the representations $\mathbf R$ and $\mathbf{\bar R}$ are the same at both ends of the string, it is sensible to think that the quantum wave function of the string $X^i(\tau,\sigma)$ is invariant under the above orientation reversal operation.  This leads to 
\begin{equation}\label{unorientedstring}
|\Lambda(\alpha_n^i);b,a\rangle=\epsilon|\Lambda((-1)^n\alpha_n^i);a,b\rangle,
\end{equation}
where $|\Lambda\rangle$ parametrizes the string state in the oscillator Hilbert space and $a,b$ are additional labels carried by the state: parametrizing the Chan-Paton charges at the endpoints of the strings (transforming in the same representation for the case of unoriented strings).  In the equation above, $\epsilon=\pm 1$.  Generically, $|\Lambda; a,\bar b\rangle$ describes massless vector particles where the quantum number $a\bar b$ transform in the adjoint representation of the gauge group since massless vectors in consistent interacting theories always transform in the adjoint representation of the gauge group.  Strings satisfying (\ref{unorientedstring}) are called unoriented open strings (with $b=\bar b$).  
%The states of the associated massless vector in the adjoint representation are either all symmetric or antisymmetric under orientation reversal.  
What differentiates the orthogonal from the symplectic group for unoriented strings is whether their adjoint representation forms symmetric or antisymmetric states.  In particular, for $SO(N)$ gauge group with $\mathbf{R\ \bar R}$ both equal to the N-dimensional real fundamental representation, the adjoint representation are given by antisymmetric matrices (antisymmetric part of the $\bf{R\times R}$ representation) and the associated massless vector satisfies (\ref{unorientedstring}) with $(-1)^N=+1$ and $\epsilon=-1$ (for superstring).  
For $Sp(N)$ gauge group where $\mathbf{R,\ \bar R}$ are the fundamental representation of the gauge group, the adjoint representation is the symmetric part of the $\mathbf{R\times R}$ representation.   
The massless vector are those of (\ref{unorientedstring}) with $\epsilon=+1$ (for superstring).
Recall that symplectic matrices are even-dimensional, leading to $Sp(N)$ with even $N$.  To make contact with what we have already seen: for the oriented string case $a$ and $\bar b$ run over the fundamental $\mathbf N$ and antifundamental $\mathbf{\bar N}$ representation of $U(N)$ respectively where the adjoint representation is given by $\mathbf{N\times \bar N}$  \cite{GSW}.
We focus on one oriented string in the presence of a stack of $Dp$-branes. Every state in the open-string spectrum has $N^2$ multiplicity, with $N^2$ massless vector states describing the $U(N)$ gauge fields.  
The basis of the open-string can be labelled as follows: 
\begin{equation}
|\phi,k,ij\rangle,
\end{equation}
where $\phi$ is the Fock space state, $k$ the momentum and $i,j=1,\cdots, N$ label the Chan-Paton factors.  As explained in \cite{BBS}, this state transform with charge $+1$ under $U(1)_i$ and charge $-1$ under $U(1)_j$.  Arbitrary string states are described by a linear combination
\begin{equation}
|\phi,k,\lambda\rangle=\sum_{i,j=1}^N{|\phi,k,ij\rangle\lambda_{ij}},
\end{equation}
where there are $N^2$ hermitian matrices $\lambda_{ij}$ called Chan-Paton matrices corresponding to the representation matrices of the $U(N)$ algebra.  The string states then become matrices transforming in the adjoint representation of $U(N)$.  
The $N^2$ degrees of freedom of the oriented open string are not visible unless one puts that string on a stack of coinciding parallel $Dp$-branes and let the Chan-Paton factor at the endpoints of the string be $i=1,\ j=N$ or vice versa.  
If the branes are separated, then there are $N$ different massless $U(1)$ vectors and the resulting gauge theory is $U(1)^N$ and the Chan-Paton indices running from $i,j=1,\cdots,N$ correspond to the endpoints of different open strings, ending respectively on the $i^{th}$ and $j^{th}$ branes.
This way of generating non-abelian gauge theories from the point of view of open oriented string with Chan-Paton factor ending on $Dp$-branes will be reviewed in the language of $Dp$-branes in a latter section. 
%[\textbf{read bound states of string and p-brane} arXiv:hep-th/9510135]
\newpage
\section{Type IIA and Type IIB}

Type IIA string theory is a non-chiral theory as it has $(1,1)$ spacetime supersymmetry where the spacetime supercharges generated by left and right moving degrees of freedom $Q_L,\ Q_R$ have opposite chirality \cite{GiveonSR}:
\begin{eqnarray}
\Gamma^0\cdots\Gamma^9Q_L=+Q_L,\\
\Gamma^0\cdots\Gamma^9Q_R=-Q_R.
\end{eqnarray}
Type IIB on the other hand is a chiral theory since it has $(2,0)$ spacetime supersymmetry, where the left and right moving supercharges have the same chirality \cite{GiveonSR}:
\begin{eqnarray}
\Gamma^0\cdots\Gamma^9Q_L=Q_L,\\
\Gamma^0\cdots\Gamma^9Q_R=Q_R.
\end{eqnarray}
%\textbf{Here explain notation:} $\Gamma,\ Q_L,\ Q_R,\ (n,m)$

Ten dimensional type IIA supergravity theories can be obtained by dimensional reduction of a unique eleven dimensional supergravity theory which arises as the low energy limit of M-theory.  We review here the field content of these theories. 

Eleven dimensional supergravity includes the following bosonic fields: a  metric $G_{MN}$ and an antisymmetric three-form potential $A_{MNP}\equiv A_3$ with field strength $F_4$.  Here $M,\ N,\ P=0,\cdots 10$.  The fermionic content is given by the gravitino $\psi^M_\alpha$ with $\alpha=1,\cdots, 32$.  The bosonic part of the action of eleven dimensional supergravity theory is given by \cite{polchinski2}: 
\begin{equation}
2\kappa_{11}^2S_{11}=\int{d^{11}x(-G)^{1/2}\left(R-\frac{1}{2}|F_4|^2-\frac{1}{6}\int{A_3\wedge F_4\wedge F_4}\right)}.
\end{equation}
Dimensionally reducing eleven dimensional supergravity along a circle, 
\begin{eqnarray}
ds^2&=&G_{MN}^{11}(x^\mu)dx^{M}dx^{N}\\
&=&G^{10}_{\mu\nu}(x^\mu)dx^\mu dx^\nu+\text{exp}(2\sigma(x^\mu))[dx^{10}+A_\nu(x^\mu)dx^\nu]^2,
\end{eqnarray}
one obtains type IIA supergravity.  By the above process, one obtains from $G_{MN}$ the following fields in type IIA: 
\begin{itemize}
\item metric $G_{\mu\nu}$
\item gauge field $A_\mu=G_{\mu,10}$
\item scalar $\Phi=G_{10,10}$ 
\end{itemize}
where $\mu,\nu,\lambda=0,\cdots 9$.  On the other hand, the antisymmetric tensor $A_{MNP}$ of eleventh dimensional supergravity gives rise to the following antisymmetric tensors in type IIA:
\begin{itemize}
\item $A_{\mu\nu\lambda}$
\item $B_{\mu\nu}=A_{\mu\nu,10}$
\end{itemize}
After appropriate reparametrizations, the type IIA metric is given by \cite{polchinski2}:
\begin{eqnarray}
S_{IIA}&=&S_{NS}+S_R+S_{CS}\\
S_{NS}&=&\frac{1}{2\kappa_{10}^2}\int{d^{10}x(-G)^{1/2}e^{-2\Phi}\left(R+4\partial_\mu\Phi\partial^\mu\Phi-\frac{1}{2}|H_3|^2\right)}\\
S_R&=&-\frac{1}{4\kappa_{10}^2}\int{d^{10}x(-G)^{1/2}\left(|F_2|^2+|\tilde F_4|^2\right)}\\
S_{CS}&=&-\frac{1}{4\kappa^2_{10}}\int{B_2\wedge F_4\wedge F_4},
\end{eqnarray}
where $\tilde F_4=dA_3-A_1\wedge F_3$.
Neveu-Schwarz (NS) sector fields are $G_{\mu\nu},\ B_{\mu\nu}$ and $\Phi$.  The field strength associated to the potential $B_2$ is denoted by $H_3$.  The Ramond-Ramond (RR) sector fields are the gauge fields $A_\mu$ and $A_{\mu\nu\lambda}$.  Their potentials and field strengths are respectively denoted by $C_p$ and $F_{p+1}$.  The vacuum expectation value of the exponential of the dilation $\Phi$ gives the coupling constant $g_s$ of string theory.  Consider the chain below where everything to the left of $(* F)$ electrically sources the $p$-brane denoted here by $D_p$ while $(* F)$ and what is on its right magnetically sources the $p$-brane.  $(*F)$ represents the Hodge dual of $F$, mapping a $k$-vector to an $(n-k)$-vector where $n=10$ here.  Note that although the notation used below refers to  RR sector fields, the relations between which fields sources which branes still holds for the NS sector. 
\begin{eqnarray}
D_p\to C_{p+1}\to F_{p+2}\to(* F)_{10-(p+2)}\to C_{10-(p+2)-1}\to D_{10-(p+2)-2}.
\end{eqnarray}
Here $C_{p+1}$ denotes the potential and $F_{p+2}$ is the field strength associated to the p-brane. 
We refer to branes  that couple to the NS sector gauge field as NS-branes.   On the other hand, branes charged under RR sectors fields are referred to as Ramond branes or D-branes.  As an example, the chain above makes it clear that $B_{\mu\nu}\equiv B_2$ electrically sources a $1$-brane (a fundamental string) and magnetically couples to a fivebrane ($NS_5$) through a six-form gauge field dual to $B_{\mu\nu}$.  Since $B_{\mu\nu}$ is an NS sector field, we refer to theses branes as NS-branes.  Similarly, $D0$-branes (point particles) are electrically charged under $A_\mu$ while the latter gauge field couples magnetically to $D6$-branes.  $D2$ and $D4$-branes are electrically and magnetically sourced by the antisymmetric tensor field $A_{\mu\nu\lambda}$.   

Type IIB supergravity is a ten-dimensional parity-violating theory.
% (\textbf{still leads to anomaly cancellation}). 
Its massless spectrum contains the same NS sector as in type IIA supergravity, namely $G_{\mu\nu}$, $B_{\mu\nu}$ and $\Phi$ which couple to the corresponding NS string and fivebranes.  The RR sector fields of type IIB is different than that of type IIA as it contains an additional scalar (0-form potential) called the axion $C_0$ which combines with the dilaton $\Phi$ to generate the complex coupling of type IIB: 
\begin{equation}
\tau=C_0+ie^{-\Phi}.
\end{equation}
The antisymmetric tensors in the RR sector of type IIB are $\tilde B_{\mu\nu}$ and $A_{\mu\nu\lambda\rho}$.  $\tilde B_{\mu\nu}$ couples electrically to D-string and magnetically to D5-branes.  $A_{\mu\nu\lambda\rho}$ sources D3-branes both electrically and magnetically since this four-form is self dual: $*d A=dA$.  This latter fact also implies the existence of a self-dual five-form field strength $*F_5=F_5$ leading to $|F_5|^2=0$.  
The action of type IIB is given by \cite{polchinski2}:  
\begin{eqnarray}
S_{IIB}&=&S_{NS}+S_R+S_{CS}\label{IIBmetric}\\
S_{NS}&=&\frac{1}{2\kappa_{10}^2}\int{d^{10}x(-G)^{1/2}e^{-2\Phi}\left(R+4\partial_\mu\Phi\partial^\mu\Phi-\frac{1}{2}|H_3|^2\right)}\\
S_R&=&-\frac{1}{4\kappa^2_{10}}\int{d^{10}x(-G)^{1/2}\left(|F_1|^2+|\tilde F_3|^2+\frac{1}{2}|\tilde F_5|^2\right)}\\
S_{CS}&=&-\frac{1}{4\kappa_{10}^2}\int{C_4\wedge H_3\wedge F_3},
\end{eqnarray}
where 
\begin{eqnarray}
\tilde F_3&=&F_3-C_0\wedge H_3,\\
\tilde F_5&=&F_5-\frac{1}{2}C_2\wedge H_3+\frac{1}{2}B_2\wedge F_3.
\end{eqnarray}
Although the condition $*\tilde F_5=\tilde F_5$ can not be imposed on the action (\ref{IIBmetric}) or else the wrong equations of motion result, the field equations of (\ref{IIBmetric}) are consistent with the aforementioned condition- even if they don't imply it.  The self duality of $\tilde F_5$
 can therefore be added by hand on the solutions of the equations of motion as an additional constraint  \cite{polchinski2}.
 
\subsection{S-duality}

An interesting fact about the low energy type IIB supergravity action (\ref{IIBmetric}) is that it can be written in an way that is invariant under $SL(2,\mathbb{R})$ symmetry \cite{polchinski2}.  To see this, consider the following coordinates: 
\begin{eqnarray}
G_{E\mu\nu}&=&e^{-\Phi/2}G_{\mu\nu}\label{Einstmetric}\\
\tau&=&C_0+ie^{-\Phi}\label{IIBtau}\\
M_{ij}&=&\frac{1}{\text{Im}\tau} \left[ \begin{array}{cc}
|\tau|^2 & -\text{Re}\ \tau  \\
-\text{Re}\tau & 1 
 \end{array} \right]\\
 F^i_3&=&\left[\begin{array}{c}
 H_3\\
 F_3
 \end{array}\right].
\end{eqnarray}
The metric (\ref{IIBmetric}) can then be rewritten in the following $SL(2,\mathbb R)$ invariant way: 
\begin{eqnarray}
S_{IIB}&=&\frac{1}{2\kappa_{10}^2}\int{d^{10}x(-G_E)^{1/2}\left(R_E-\frac{\partial_\mu \bar\tau\partial^\mu \tau}{2(\text{Im}\tau)^2}-\frac{M_{ij}}{2}F^i_3\cdot F^j_3-\frac{1}{2}|\tilde F_5|^2\right)}\nonumber\\
&-&\frac{\epsilon_{ij}}{4\kappa^2_{10}}\int{C_4\wedge F^i_3\wedge F^j_3},
\end{eqnarray}
where $G_E$ (\ref{Einstmetric}) is the Einstein metric and where the coupling and fields transform as:
\begin{eqnarray}
\tau'&=&\frac{a\tau+b}{c\tau+d}\\
F^{i'}_3&=&\Lambda^i_jF^j_3\hskip 1cm \Lambda^i_j=\left[\begin{array}{cc} 
d&c\\
b&a
\end{array}\right]\\
\tilde F'_5&=&\tilde F_5\hskip 1.2cm G'_{E\mu\nu}=G_{E\mu\nu},
\end{eqnarray}
with $a,\ b,\ c,\ d\in \mathbb R$ such that $ad-bc=1$. Although the low energy effective action of IIB supergravity has a global  $SL(2,\mathbb {R})$ symmetry, $\tau$ is invariant under an $SO(2,\mathbb {R})$ subgroup, so the moduli space is locally the coset space $SL(2,\mathbb {R})/SO(2,\mathbb {R})$.  It is in fact known that the full type IIB superstring theory (considering quantum effects) has the discrete subgroup $SL(2,\mathbb {Z})$ as an exact symmetry of the theory.  The latter transforms the fields in (\ref{IIBmetric}) as: 
\begin{eqnarray}
\Phi'&=&-\Phi\hskip 1cm G'_{\mu\nu}=e^{-\Phi}G_{\mu\nu}\\
B'_2&=&C_2\hskip 1.4cm C'_2=-B_2\\
C'_4&=&C_4,
\end{eqnarray}
where $G_{E\mu\nu}=e^{-\Phi/2}G_{\mu\nu}=e^{-\Phi'/2}G'_{\mu\nu}$.
The $SL(2,\mathbb Z)$ symmetry here makes sure that $(p,q)$ strings, sourced by $\left(\begin{array}{c}
B_2\\
C_2 \end{array}\right)$, a doublet of $SL(2,\mathbb R)$, carry integer charge under the two two-form gauge fields.  Recall that in this notation, an $F$-string has charge $(1,0)$ and a $D$-string has charge $(0,1)$.  
As mentioned previously, the string coupling $g_s$ is given by the expectation value of the exponential of $\Phi$ and the type IIB coupling is (\ref{IIBtau}).  The above field theory transformation taking the dilation $\Phi\to-\Phi$ is in fact a symmetry which takes $g_s\to 1/g_s$ and $\tau\to-1/\tau$ ($C_0=0$): taking a strongly coupled theory to a weakly coupled one.  This is called $S$-duality or strong-weak duality and it relates type IIB superstring theory to itself \cite{BBS}.  It is the first example we encounter of a duality that helps probing the strongly coupled regime of a theory.  

%This differs from low energy IIA supergravity where the theory is symmetric under SO(1,1,\textbf{R}). 
\section{D-branes, Orientifolds and NS5-branes}

Branes are extended $p$ spatial dimensional objects in String Theory.  Denoted as $p$-branes, these objects are classified into two categories depending on their tension $T_p$ (energy per unit p-volume) at weak fundamental string coupling $g_s$ \cite{GiveonSR}:  
\begin{enumerate}
\item Neveu-Schwarz (NS) or solitonic branes: if the tension behaves like $1/g_s^2$
\item Dirichlet or D-branes: if the tension behaves like $1/g_s$
\end{enumerate}
In the limit $g_s\to 0$, the above criteria indicate that  Dirichlet branes are lighter then NS-branes.  

\subsection{Dirichlet-branes}

Dirichlet p-branes \cite{PolchinskiDB}, denoted $Dp$-branes are objects stretched along the hyperplane parametrized by $(x^1,\cdots, x^p)$ and are point-like in the directions $(x^{p+1},\cdots,x^9)$.  Their defining property is that open strings with Neumann boundary conditions for $(x^0,\cdots, x^9)$ can have one of their ends ending on $Dp$-branes whereas open strings with Dirichlet boundary conditions in $(x^{p+1},\cdots,x^9)$ have both their ends starting and finishing on $Dp$-branes \cite{GiveonSR}.  We will see below that $Dp$-branes are sourced by Ramond-Ramond $(p+1)$-forms potentials in both type IIA and type IIB string theory.  As alluded to above, the tension of $Dp$-branes is given by
\begin{equation}\label{Dptension}
T_p=\frac{1}{g_sl_s^{p+1}},
\end{equation}
where $l_s$ is the fundamental string scale.  $D$-branes are BPS objects which preserve half of the thirty two supercharges of type II string theory.  In particular, they preserve supercharges of the form $\epsilon_L Q_L+\epsilon_R Q_R$ with
\begin{equation}\label{Dpsusy}
\epsilon_L=\Gamma^0\Gamma^1\cdots \Gamma^p\epsilon_R.
\end{equation}
The low energy worldvolume theory on a $Dp$-brane is a $p+1$ dimensional field theory, the action of which is given by the $p+1$ dimensional Born-Infeld action.  Expanding the square root in the latter and keeping only the bosonic part we obtain the following action \cite{GiveonSR} on the worldvolume of $Dp$-branes: 
\begin{equation}\label{Dpaction}
S=\frac{1}{g^2_{SYM}}\int{d^{p+1}x\left(\frac{1}{4}F_{\mu\nu}F^{\mu\nu}+\frac{1}{l^4_s}\partial_\mu X^I\partial^\mu X_I\right)},
\end{equation}
where the $U(1)$ gauge coupling $g_{SYM}$ on the brane is given by \cite{GiveonSR}:
\begin{equation}
g^2_{SYM}=g_sl_s^{p-3}.
\end{equation}
The low energy worldvolume of $Dp$-branes describes the dynamics of the ground states of open Dirichlet strings.  The massless spectrum includes a $p+1$-dimensional $U(1)$ gauge field $A_\mu(x^\nu)$, $9-p$ scalars $X^I(x^\mu)$ which parametrize the fluctuations transverse to the $Dp$-branes and some fermions.  Here $I=p+1,\cdots, 9$ and $\mu=0,\cdots, p$.
At high energy, one needs to decouple the massless gauge theory degrees of freedom from gravity and massive string modes if one wants to study SYM on the brane.  To do that, we send $l_s\to 0$ while holding $g_{SYM}$ fixed (this decouples gravity from the action, returning only the open string modes on the brane) \cite{GiveonSR}.  We obtain the following three cases: 
\begin{itemize}
\item $g_s\to0$ for $p<3$
\item $g_s\to\infty$ for $p>3$
\item $g_{SYM}$ independent of $l_s$ for $p=3$
\end{itemize}
The limit $l_s\to0$ in the latter case describes a $U(1)$ gauge theory in $\mathcal N=4$ SYM in $3+1$ dimensions.  More generally, the theory in the UV behaves as $p+1$-dimensional SYM for $p\leq 3$.
%It is obtained by dimensionally reducing ten-dimensions $\mathcal N=1$ SYM with gauge group $G=U(1)$ to $p+1$ $\mathcal N=4$ SYM
% for $p\leq 3$ the above limit leads to a consistent theory on the brane, whose UV behavior is that of $p+1$ dimensional SYM.  For $p>3$ SYM provides a good description in the IR but breaks down at high energy
%the low energy worldvolume describes a $p+1$ dimensional field theory invariant under sixteen supercharges.  It is obtained by dimensional reduction of $\mathcal N=1$ SYM with gauge group $G=U(1)$ from ten spacetime dimensions to $p+1$ dimensions.  
The generalization of (\ref{Dpaction}) to $N_c$ parallel $Dp$-branes is given by the following bosonic $p+1$ kinetic term and potential for the gauge field $A_\mu$ and the adjoint scalar $X^I$: 
\begin{eqnarray}\label{iib1}
&&\mathcal L_{kin}=\frac{1}{g^2_{SYM}}\text{Tr}\left(\frac{1}{4}F_{\mu\nu}F^{\mu\nu}+\frac{1}{l_s^4}\mathcal D_\mu X^I\mathcal D^\mu X_I\right),\\\label{iib2}
&&V\sim \frac{1}{l_s^8g^2_{SYM}}\sum_{I,J}{\text{Tr}[X^I,X^J]^2},
\end{eqnarray} 
where 
\begin{eqnarray}\label{covariantderiv}
&&\mathcal D_\mu X^I=\partial_\mu X^I-i[A_\mu,X^I],\\\label{covariantderiv2}
&&F_{\mu\nu}=\partial_{[\mu}A_{\nu]}-i[A_\mu,A_\nu],
\end{eqnarray}
%\textbf{check approx. sign+ check link with DBI action}\\
and where the Coulomb branch of the $U(N_c)$ $p+1$-dimensional SYM theory (4d $\mathcal N=4$ SYM if $p=3$) therein is parametrized by the flat directions of the above potential\footnote{Equations (\ref{iib1}) to (\ref{covariantderiv2}) can be obtained by taking the symmetric trace of the Born-Infeld action.}. 

\subsection{Nonabelian gauge groups from stacks of Dp-branes}
As we have just seen, $Dp$-branes are remarkable objects which can introduce nonabelian gauge theories in string theory.  The mechanism responsible of this is the Chan-Paton factor.  In the presence of a stack of $N_c$ parallel $Dp$-branes, the scalars $X^I$ (\ref{Dpaction}) turn into $N_c\times N_c$ matrices transforming in the adjoint representation of the gauge group $U(N_c)$.  The diagonal component of $X^I$ as well as the $N_c$ massless gauge fields in the Cartan subalgebra of $U(N_c)$ correspond to open strings with both ends ending on the same $Dp$-brane.  The off-diagonal components of $X^I$ and the charged gauge bosons corresponds to strings with endpoints lying on different branes.  
%(\textbf{check the link between this and Chan-Paton factor for one string with dof equal to $N^2$.  How does the latter end on Dp-branes and how to understand the $N^2$ dof of 1 strings instead of $N$- strings.})  
The $(i,j)$ and $(j,i)$ matrix element of $X^I$ and $A_\mu$ are associated with the two orientations of a fundamental string whose endpoints are connected to the $i^{th}$ and  $j^{th}$ $Dp$-branes \cite{GiveonSR}.
%More precisely, nonabelian gauge fields seem to live on the worldvolume of multiple Dp-branes. 
In summary, (four-dimensional $\mathcal N=4$ ) SYM theory with $U(N_c)$ gauge group- 16 supercharges- arise on the low energy worldvolume of a stack of $N_c$ parallel $Dp$-branes  (D3-branes) as a result of the ground states of open strings ending on $Dp$-branes.  We will see in a section below how ``webs" of branes can reduce the amount of supersymmetry, focusing on process leading to $\mathcal N=4\to \mathcal N=2\to \mathcal N=1$.\\
%\textbf{ non-abelian gauge theories on Dp-branes,  SYM theory on Dp-branes }\\

%anti-Dp-branes carries opposite RR charge and preserve the other half of the supercharges.
%The defining property of D-brane is that open string can end on them.  The fact that fundamental strings can end on D-branes implies that quantum field theories of the YM type -like SM- reside on the world volume of D-branes.  The YM fields arise as the massless modes of open strings attached to the D-branes. The NS-branes on 
%As we will soon discuss, both D-branes and NS-branes preserve half the supersymmetry in weakly coupled string theory.  

\subsection{Orbifolds and Orientifolds}

Orbifolds are a class of compactification objects in string theory for which the metric is explicitly known.    
Roughly speaking, orbifolds are singular spaces defined as quotient spaces of the form $X/G$ where $X$ is a smooth manifold and $G$ a discrete isometry group. A point $g\in X/G$ consists of an orbit of points on the manifold $X$.  Recall that an orbit of $x\in X$ is a point and all of its images under the action of the group $G$.  Singularities on $X/G$ arise when nontrivial group elements leave points in $X$ invariant. 
The orbifold $X/G$ is locally indistinguishable from the manifold $X$ at nonsingular points \cite{BBS}.

Example of such noncompact space obtained by identifying spacetime coordinates under the reflection of $k$ coordinates $X^k\to -X^k$ is given by $\mathbb R^k/\mathbb Z_2$ whereas the compact space version obtained under the same identification on a $k$-torus leads to $T^k/\mathbb Z_2 $.  The former orbifold has $k$ singularities whereas the latter has $2^k$ fixed points. 
%\textbf{(check this: value of k, fixed points, twist strings sectors, oriented strings)} \cite{polchinski1}.

While orbifolds preserve the orientation of strings, orientifolds denoted by\\ $O_p$-planes are generalization of $\mathbb Z_2$ orbifolds fixed plane to non-oriented strings. Namely, they act on spacetime coordinates and reverse the orientation of the string.  For example, a $\mathbb Z_2$ $O_p$-plane extending along $(x^1,\cdots, x^p)$ acts as $x^I(z,\bar z)\to-x^I(\bar z,z)$ for $I=p+1,\cdots, 9$ where $z,\ \bar z$ parametrizes the string worldsheet with $z=e^{\tau+i\sigma}$ .  As an example, consider the orientifold $T^2/\mathbb Z_2$ in type IIB.  The $\mathbb Z_2$ transformation is defined as $(-1)^{F_L}\cdot\Omega\cdot \mathcal I_2$  where $(-1)^{F_L}$ changes the sign of all the Ramond sector states on the left, $\Omega$ denotes the orientation reversal transformation (exchanges the left and right moving modes on the worldsheet) while $\mathcal I_2$ acts on the torus by inverting the sign of both the coordinates of the torus \cite{senF}.

Orientifolds break the same 16 supercharges of susy as a parallel $Dp$-brane would.  Their sole presence modifies the transverse space by replacing $R^{9-p}$ by $R^{9-p}/\mathbb Z_2$.  This has for consequence to generate mirror $\mathbb Z_2$ images of objects outside the orientifold plane if one still want to work in $R^{9-p}$ space\footnote{and implement appropriate (anti)-symmetrization of states}.  In particular, $D$-branes far from the $Op$-plane acquire mirror brane images. 
%(\textbf{check this:  mirror brane})

As shown in the previous section, $Dp$-branes are charged under RR $(p+1)$-form potential in type II.  Orientifolds carry charge under the same RR $(p+1)$-form gauge potential as $Dp$-branes.  Denoting the RR charge of $Op$-plane by $Q_{O_p}$, it is equal to the RR charge of $2^{p-4}$ $Dp$-branes or $2^{p-5}$ pairs of $Dp$-brane and its mirror.  The RR charge of $D_p$-branes is denoted by $Q_{D_p}$.  We summarize here some of the properties of the branes encountered so far in type II string theory\cite{GiveonSR}: 
\begin{itemize}
\item $Dp$-branes: charged under RR $(p+1)$-form potential in type II
\item Type IIA: $p$ even $\to$ $Dp$-branes with $p=0,2,4,6,8$
\item Type IIB: $p$ odd $\to$ $Dp$-branes with $p=-1,1,3,5,7,9$ 
\item $Dp$-brane RR charge $Q_{D_p}$ equals its tension (\ref{Dptension}) 
\item Orientifold: $Q_{O_p}=\pm 2\cdot 2^{p-5}Q_{D_p}$
\end{itemize}
The gauge theories living on a stack of $N_c$ $Dp$-branes parallel to an Op-plane are:   
\begin{eqnarray}
G=Sp(N_c/2)\ \text{with }N_c\ \text{even}&&\ Q_{O_p}=+2\cdot 2^{p-5}Q_{D_p} \\\label{opcharge}
G=SO(N_c)&&\ Q_{O_p}=-2\cdot 2^{p-5}Q_{D_p}
\end{eqnarray}
where the rank of the gauge groups are $[N_c/2]$ and they preserve 16 supercharges.

\subsection{NS5-branes}

Solitonic fivebranes are BPS object preserving half the supersymmetry of the theory. Their tension is given by 
\begin{equation}
T_{NS5}=\frac{1}{g_s^2l_s^6}.
\end{equation}
Type IIA NS-fivebranes stretched along $(x^1,\cdots, x^5)$ preserve supercharges of the form $\epsilon_LQ_L+\epsilon_RQ_R$ with
\begin{eqnarray}\label{NS51susy}
\epsilon_L&=&\Gamma^0\Gamma^1\Gamma^2\Gamma^3\Gamma^4\Gamma^5\epsilon_L,\\
\epsilon_L&=&\Gamma^0\Gamma^1\Gamma^2\Gamma^3\Gamma^4\Gamma^5\epsilon_R,
\end{eqnarray}
whereas type IIB fivebranes have
\begin{eqnarray}\label{NS52susy}
\epsilon_L&=&\Gamma^0\Gamma^1\Gamma^2\Gamma^3\Gamma^4\Gamma^5\epsilon_L,\\
\epsilon_L&=&-\Gamma^0\Gamma^1\Gamma^2\Gamma^3\Gamma^4\Gamma^5\epsilon_R.
\end{eqnarray}
Light fields living on the worldvolume of a type IIA NS5-brane form a tensor multiplet of six-dimensional  (2,0) susy.  The latter is made of a self-dual $B_{\mu\nu}$ field, five scalars and some fermions.  Four of the scalars parametrize the fluctuations of the NS-brane in the transverse directions.  The fifth scalar lives on a circle of radius $l_s$.  Close to the NS5-brane, $g_s$ is strongly coupled and the circle on which the fifth scalar lives  parametrizes the eleventh direction of M-theory.  On the other hand, there exists a vector multiplet living on a single type IIB NS5-brane.  It contains a six dimensional gauge field, four scalars and some fermions.  Here, all four scalars parametrize the transverse directions fluctuations of the fivebrane.  On type IIB fivebrane, the vector field's gauge coupling is given by \cite{GiveonSR}: 
\begin{equation}\label{ns5}
g^2_{SYM}= l_s^2.
\end{equation}
Now that we understand what kind of low energy theory lives on a stack of $N_c$ $Dp$-branes, one could ask the same question about a stack of type IIA NS-branes.   The answer is that the low energy theory describing $k$ parallel type IIA NS5-branes correspond to a non-trivial $5+1$ dimensional $(2,0)$ field theory.  There exists Dirichlet membranes stretched between the NS5-branes. These Dirichlet membranes are tensionless for coinciding NS5-branes.  These string-like low energy excitations are charged under the self-dual $B_{\mu\nu}$ fields.  The expectation value of the diagonal piece of the five scalars in the tensor multiplet parametrize the Coulomb branch, the origin of which is a non-trivial superconformal field theory.  $(2,0)$ susy in $d=6$ also appears on type IIB at A-D-E singularities and also appears on coincident M5-branes once we lift type IIA to M-theory.  On the other hand, the low energy worldvolume dynamics on a stack of $k$ parallel type IIB NS5-branes is a $5+1$-dimensional $(1,1)$ $U(k)$ SYM described by (\ref{iib1})-(\ref{iib2}) with gauge coupling (\ref{ns5}) with $p=5$.  It preserves 16 supercharges .  However, this is purely informational and we will not discuss about the gauge theory on a stack of $NS5$-branes in the present thesis.

%[\textbf{read: Witten and Hall on (2,0) theories}]

%Dualities
\subsection{Webs of branes}

We will see below how webs of branes can allow us to study SYM theories with lower supersymmetry than configurations preserving 16 supercharges.  Many branes systems exist which preserve 8 supercharges.  Some of them are given below \cite{GiveonSR}:
\begin{itemize}
\item $Dp-D(p+4)$ 
\item $Dp-D(p+4)+O_p, O_{p+4}$-planes
\item $Dp-D(p+2)$
\item $NS-Dp$
\end{itemize}
In all the above cases, the main idea is the same: the amount of susy preserved by a system of branes is found by imposing all the susy condition (\ref{Dpsusy}, \ref{NS51susy}, \ref{NS52susy}) present in the system on the spinors $\epsilon$.  
%Stacks of Dp-branes and NS-branes preserve respectively half the susy (16 supercharges) of the theory.  
We analyze the $Dp-D(p+4)$ system because it will be useful later when describing the F-theory embedding (using D3-$D7$-branes) of $\mathcal N=2$ SYM in $d=4$.  
Consider a stack of $N_c$ $Dp$-branes parallel to a stack of $N_f$ D$(p+4)$-branes.  As seen previously, each stack preserve respectively half the susy of the theory ie. 16 supercharges.  It thus makes sense to think that when present together, they preserve one fourth of the susy of the theory namely 8 supercharges. 
Let's now make this statement more precise by reviewing an example analyzed in detail in \cite{GiveonSR}.  Consider $N_c$ $Dp$-branes stretched along the hyperplane parametrized by $(x^1,\cdots,x^p)$ and $N_f$ D$(p+4)$-branes along $(x^1,\cdots,x^{p+4})$.
Combining both susy conditions we find the following constraint on $\epsilon_L$ and $\epsilon_R$: 
\begin{equation}
\epsilon_L=\Gamma^0\Gamma^1\cdots\Gamma^p\epsilon_R=\Gamma^0\Gamma^1\cdots\Gamma^{p+4}\epsilon_R.
\end{equation}
First, we notice that  the sole knowledge of $\epsilon_R$ fixes $\epsilon_L$. Second, we realize that the LHS of the above equation simplifies to
\begin{equation}
\epsilon_R=\Gamma^{p+1}\Gamma^{p+2}\Gamma^{p+3}\Gamma^{p+4}\epsilon_R.
\end{equation}
%First, we notice that  the sole knowledge of $\epsilon_R$ fixes $\epsilon_L$.
The above combination of gamma matrices which we denote by $\Gamma$ is traceless and squares to the identity matrix.  Thus the constraint on $\epsilon_R$ translates to $\epsilon_R=\Gamma\epsilon_R$ with $\Gamma=1$. The tracelessness condition constraints 8 of the 16 eigenvalues  of $\Gamma$  to equal $+1$ while the remain $8$ eigenvalues are equal to $-1$.  Thus, eight independent components of $\epsilon_R$ out of the 16 are preserved by the constraint.  In summary, having applied susy conditions of both types of branes onto $\epsilon$, we find that there are eight independent supercharges preserved by this given brane configuration, as expected.  

The light degrees of freedom on the $N_c$ $Dp$-branes comprise $N_c$ $(p+1)$-dimensional gauge fields $A_\mu$ generating a $p+1$ dimensional $U(N_c)$ gauge theory, $9-p$ scalars in the adjoint representation of $U(N_c)$ and some fermions.  Recall that $Dp$-branes have $9-p$ transverse directions whereas $D(p+4)$ branes have $5-p$ transverse directions.  Consequently, $5-p$ adjoint scalars of the $Dp$-brane naturally parametrize the fluctuations of the $Dp$-branes transverse to the $D(p+4)$-branes.  With the gauge field, these scalars form a vector multiplet while the remaining $4$ adjoint scalar parametrize an adjoint hypermultiplet.  The light degrees of freedom on $D(p+4)$ branes are the same as those just mentioned provided we replace $N_c$ by $N_f$ and $p$ by $p+4$.  
 
Strings stretched between the two stacks of branes are seen as $N_f$ flavors in the fundamental representation of $U(N_c)$ from the point of view of the $Dp$-branes.  They correspond to $N_c$ pointlike defects in the fundamental representation of $U(N_f)$ from the D$(p+4)$-branes perspective.  From the point of view of $Dp$-branes, the $U(N_f)$ symmetry is a global symmetry, the only dynamical fields generated from $D(p+4)$-branes being the $N_f$ flavors.  
%(\textbf{check equation (4.7) Giveon-Kutasov review}).  
The positions of the $D(p+4)$-branes in $(x^{p+5},\cdots, x^9)$ correspond to the masses of the $N_f$ fundamentals flavors labelled by $\vec{m}_i$ with $i=1,\cdots, N_f $.  These corresponds to couplings in the worldvolume theory of $Dp$-branes.  The locations $\vec x_a$ with $a=1,\cdots N_c$ of the $Dp$-branes in the transverse space $(x^{p+5},\cdots, x^9)$ are associated with the expectation values of the adjoint scalars $\vec X$ of $U(N_c)$.  Their expectation value parametrize the Coulomb branch of the $U(N_c)$ gauge theory with $\vec x_a=\langle \vec X_{aa}\rangle$.  The $\vec x_a$ correspond to moduli on the worldvolume theory of $Dp$-branes.  The expectation values of the adjoint hypermultiplet of $U(N_c)$ are parametrized by the position of $Dp$-branes parallel to $D(p+4)$-branes in $(x^{p+1},\cdots, x^{p+4})$.  
%(\textbf{check all of this and maybe add and image.})  
As a general rule: the location of the heavy $D(p+4)$-branes correspond to couplings on the worldvolume of $Dp$-branes whereas the location of the light $Dp$-branes are moduli on the same worldvolume \cite{GiveonSR}.  A $Dp$-branes inside a $D(p+4)$-branes can be though of as a small instanton.  A $Dp$-brane embedded in a stack of $N_f$ $D(p+4)$-branes is a small 4d $U(N_f)$ instanton which can reach finite size.  The full Higgs branch of the theory is parametrized by the moduli space of $N_c$ instantons in $U(N_f)$. 
%[\textbf{read Douglas 1995-1996}.]  

Since orientifolds $O_p$-plane preserve the same supercharges as $Dp$-branes, adding $O_p$-planes and/or $O_{p+4}$-planes to the $Dp-D_{p+4}$ branes system does not further break supersymmetry and thus still preserves 8 supercharges.  As seen previously, the presence of $O_p$-plane generate either $SO(N_c)$ or $Sp(N_c/2)$ gauge theory on the stack of $N_c$ $Dp$-branes.  In the presence of $O_p$-plane, the global symmetry of a theory preserving 8 supercharges with $N_f$ flavors becomes  $Sp(N_f/2)$ or $SO(N_f)$ instead of the original $U(N_f)$ symmetry.  We will reach the same conclusion in Chapter 3 while studying F-theory and making use of Tate's algorithm.  A construction involving orthogonal $O_{p+4}$-plane (breaking $\mathcal N=2$ to $\mathcal N=1$ susy) was analyzed in Gimon-Polchinski \cite{gimpol} and was embedded in F-theory in \cite{DasguptaSW}. 

Systems of parallel NS5-branes with $Dp$-branes stretched between them e.g D4-branes in type IIA, D3-branes in type IIB, also preserve 8 supercharges.  Their physics completely capture that of $\mathcal N=2$ $d=(3+1)$ $U(N_c)$ SYM field theory \cite{WittenMT} and $\mathcal N=2$ $U(N_c)$ SYM $d=2+1$ \cite{HananywittenHW} respectively.  Adding flavor branes -say D6-branes or D5-branes respectively, does not break further supersymmetry.  In their proper limit of validity, these rich branes constructions capture all the moduli of these field theories \cite{GiveonSR}.

It is possible to further break susy from $\mathcal N=2$ to $\mathcal N=1$ by considering system in which some branes span orthogonal directions.  The resulting brane construction can preserve 4 supercharges instead of 8.  A famous example is that of two orthogonal NS5-branes with $N_c$ ``color'' D3-branes stretched between them in the presence of ``flavor'' D5-branes .  This leads to Seiberg duality of $U(N_c)$ $\mathcal N=1$ SQCD in $2+1$ dimensions with $N_f$ flavors \cite{ElitzurFH}.  The relative angle between the branes govern the couplings between the different complex adjoint scalars and fundamental hypermultiplets into play.  For example, rotating one NS5 in Witten's $\mathcal N=2$ SQCD type IIA model corresponds to giving a mass to the adjoint scalar and explicitly breaking $\mathcal N=2$ to an $\mathcal N=1$ system of orthogonal NS5-branes with D4-branes stretched between them in the presence of D6-branes.  These couplings are captured in a superpotential of the form 
\begin{equation}
W\equiv \text{tr}\tilde Q X\tilde Q+\frac{\mu}{2}\text{tr}X^2,
\end{equation}
were $Q,\tilde Q$ are the fundamental hypermultiplets, $X$ the adjoint scalar and $\mu$ the mass term for the scalar.  Susy is enhanced to 8 supercharges when the angle goes to zero, see  \cite{GiveonSR} for details.  We will not elaborate on these systems here, saving discussion for relevant models to be analyze in Chapter 4.

\section{M-theory}

Type IIA and type IIB defined above are valid at small $g_s$, where perturbation theory applies.  We have already seen what happens to type IIB in the large $g_s$ regime namely, S-duality maps type IIB to type IIB so we now focus on type IIA.  When $g_s$ becomes large and is outside the regime of perturbative string theory, type IIA grows an eleventh dimension (parametrized by $x^{10}$) in the form of a circle of size $g_sl_s$.    The new 11-dimensional quantum theory which emerges is called M-theory.  At low energy, and in flat $1+10$ dimensional Minkowski vacuum, the latter is approximateted by 11-dimensional supergravity: a classical field theory.   The only parameter of M-theory is the eleven dimensional Plank scale $l_p$: physics is strongly coupled at scales smaller than $l_p$ and well approximated by weakly coupled semiclassical supergravity for scales much larger than $l_p$ \cite{GiveonSR}.  The spectrum of M-theory includes a 3-form potential $A_{MNP}$ with $M,\ N,\ P=0,1,\cdots,10$ which sources electrically an M-theory membrane called M2-branes while magnetically coupling to an M-theory fivebrane denoted M5-brane.  The tension of $Mp$-branes is given by the $p+1$ power of the Planck's length \cite{GiveonSR}:
\begin{equation}
T_p=1/l_{p}^{p+1}.
\end{equation}
These $Mp$-branes preserve half of the thirty two supercharges of the theory.  In particular, an $Mp$-brane stretched in the hyperplane $(x^1,\cdots,x^p)$ with $p=2,5$ preserve the supercharges $\epsilon Q$ with 
\begin{equation}
\Gamma^0\Gamma^1\cdots \Gamma^p\epsilon=\epsilon.
\end{equation}
As mentioned previously, type IIA with a finite string coupling $g_s$ can be thought of as M-theory compactified on $R^{1,9}\times S^1$ (where the radius of $S^1$ is denoted by $R^{10}$).  In the limit $g_s\to0$ and $S^1\to 0$ we recover ten dimensional type IIA.  The relations below between the M-theory compactification radius $R_{10}$, $l_p$, and the type IIA parameters $g_s,\ l_s$ clarify that the strong coupling limit of type IIA $g_s\to \infty$ ($R_{10}/l_p\to\infty$) is given by the $1+10$ dimensional Minkowski vacuum of M-theory. 
\begin{eqnarray}
\frac{R_{10}}{l^3_p}&=&\frac{1}{l_{s}^2},\\
R_{10}&=&g_sl_s.
\end{eqnarray}

The type IIA branes that were described the previous sections all have interpretations in M-theory.  We review in Table \ref{table 1} some examples of correspondences that will be useful for our discussion in Chapter 3 and 4.  
\begin{table}[ht]
\centering
\begin{tabular}{ccc}
Type IIA&M-theory&Charged under\\
\hline\\
Fundamental string&M2-brane&$B_{\mu 1}=A_{10\mu 1}$\\
 $x^1$&$x^{1,10}$ &\\
&&\\
D4-brane&M5-brane&$\tilde{A}_{10\mu_1\mu_2\cdots\mu_5}$\\
 $x^{01236}$&$x^{0,1,2,3,6,10}$& ($d\tilde A=*dA$)\\
&&\\
NS5-brane&M5-brane&$\tilde A_{\mu_1\cdots\mu_6}$\\
 $x^{012345}$& $x^{012345}$ &\\
&&\\
D6-branes&KK monopole&$A_{\mu}=G_{\mu10}$\\
 $x^{0123456}$&$x^{0123456}$&\\
&&\\
&&
%$NS5 \to M5$ pointlike in $x^{10}$
\end{tabular}
\caption{IIA M-theory correspondence. Directions along which the branes are stretched is indicated.}
\label{table 1}
\end{table}
From the table below, one notices that type IIA D4-branes and NS5-branes both emerges from the same object in M-theory, namely M5-branes.  D4-branes in $R^{1,9}$ corresponds to M5-brane which wraps the $S^1$ direction in $R^{1,9}\times S^1$ whereas the NS5-brane on $R^{1,9}$ becomes a transverse M5 brane on $R^{1,9}\times S^1$ whose world volume is pointlike on $S^1$.  Thus, configurations of parallel NS5-branes with $N_c$ D4-branes stretched between them and generating $\mathcal N=2$ $U(N_c)$ SYM in four dimensions are described by a single fivebrane in M-theory.  The worldvolume of the M5-brane is $R^{1,3}\times\Sigma$ with $\Sigma$ embedded in the four-manifold $Q\cong R^3\times S^1$.  $Q$ is parametrized by the complex coordinates $v=x^4+ix^5$ and $s=x^6+ix^{10}$.  Imposing $\mathcal N=2$ susy means giving $Q$ a complex structure in which $v$ and $s$ are holomorphic.  As a consequence, $\Sigma$ is a complex smooth Riemann surface with genus $g$ equal to the rank of the SYM gauge group.  The latter surface is described by a hyperbolic curve which can take the form of a Seiberg-Witten curve \cite{WittenMT}  or a spectral curves (n-sheeted cover of the genus-g Riemann surface) -  usually used in integrable systems \cite{dw}. 
%\begin{equation}
%\frac{1}{g^2_{SYM}}=\frac{L}{g_sl_s}
%\end{equation}
%where $L$ is the distance between the two parallel NS5-branes. 

%CURVE and Riemann surfaces

M-theory is of course much richer than what we have explored so far.  For, it is believed that all ten dimensional string theories can arise as asymptotic expansions around different vacua of M-theory; dualities connecting the different 10-dimensional theories together.  We will not elaborate on this here.  We will however present in the next section our first example of a gauge theory that we will embed in string theory in Chapter 3.   We thus turn to the presentation of the $\mathcal N=2$ supersymmetric gauge theory in four dimensions called Seiberg-Witten theory.  

\section{Seiberg-Witten theory}

We start by discussing some generic features of four-dimensional $\mathcal N=2$ supersymmetric gauge theory  preserving 8 supercharges.  Later, we will focus on a specific theory named  Seiberg-Witten theory with gauge group $SU(2)$ and 4 flavors in the fundamental representation of the gauge group.   

Four-dimensional $\mathcal N=2$ supersymmetry means that there are 2 generators of supersymmetry  $Q^{A}_\alpha$ with spinor index $\alpha$ and $A=1,2$ with algebra: 
\begin{eqnarray}
\{Q^A_\alpha,\bar Q^B_\beta\}&=&-2\delta^{AB}P_\mu\Gamma^\mu_{\alpha\beta},\\
\{P^\mu,Q^A_\alpha\}&=&0.
\end{eqnarray}
$Q^1, Q^2$ are two Majorana spinors each containing 4 linearly independent supercharges satisfying a reality condition.  One can also write them as 2 Weyl spinors with two complex components each.  We will use the Majorana notation.  $P^\mu$ is the spacetime momentum and $\bar Q\equiv Q^\dagger\Gamma^0$ from Majorana properties \cite{polchinski2}.  $\mathcal N=2$ susy theories in $d=4$ thus have 8 supercharges 
transforming as 2 copies of $\mathbf{ 2+\bar 2}$ of $Spin(1,3)$ which is roughly speaking isomorphic to $SL(2,\mathbb C)$.  Every $\mathcal N=2$ susy theories have an $SU(2)_R$ global symmetry acting on the 2 supercharges. In addition, conformal theories have an extra $U(1)_R$ global symmetry under which chiral supercharges have charge $\pm 1$.  $\mathcal N=2$ theories have 3 massless multiplets: the vector multiplet, the hypermultiplet, and the supergravity multiplet. We will be interested in the first two. \\
The vector multiplet contains the following fields: 
\begin{equation}\label{vectormult}
\begin{array}{ccc}
&A_\mu& \\
\lambda_\alpha&&\psi_\alpha\\
&\phi&
\end{array}
\end{equation}
Where $A_\mu$ is a gauge field, $\lambda_\alpha,\psi_\alpha$ are Weyl fermions and $\phi$ is a complex scalar.  The diamond shape indicates how each line in (\ref{vectormult}) transforms under the global $SU(2)_R$ symmetry: the gauge field and the complex scalar are both singlets under $SU(2)_R$ while the fermions $\lambda,\psi$ form a doublet transforming in the $\mathbf 2$ of $SU(2)_R$.  All the fields forming the vector multiplet transform in the adjoint representation of the gauge group.  Using $\mathcal N=1$ supersymmetry language, the $\mathcal N=2$ vector multiplet decomposes into a $\mathcal N=1$ vector multiplet and a $\mathcal N=1$ chiral multiplet.  The $\mathcal N=1$ vector superfield (also called $\mathcal N=1$ vector multiplet) is given, in the notation of \cite{WessbaggerWB}, by: 
\begin{equation}\label{vavector}
V=-\theta\sigma^\mu\bar \theta A_\mu-i\bar \theta^2(\theta\lambda)+i\theta^2(\bar\theta\bar\lambda)+\frac{1}{2}\theta^2\bar\theta^2D,
\end{equation}  
and the associate gauge covariant field strength is given by:
\begin{equation}
W_\alpha=\bar{\mathcal D}^2(e^{2V}\mathcal D_\alpha e^{-2V}),
\end{equation}
where $V=V_aT^a$, $a=1,\cdots,\text{dim G}$.  In the equation above, $V_a$ is the $N=1$ vector multiplet \eqref{vavector} with $a$ the index labelling the adjoint representation of the gauge group and $T^a$ are the generators of the gauge group $G$ in the representation $R$.  Also, $\mathcal D$ was defined in (\ref{covariantderiv}) where the scalars are now $\phi$.  
The $\mathcal N=1$ chiral superfield is given by:
\begin{equation}
\Phi=\phi+\sqrt 2\theta\psi+\theta^2F.
\end{equation}
The low energy Lagrangian describing the $\mathcal N=2$ vector multiplet can be written in terms of $\mathcal N=1$ superspace as \cite{GiveonSR}: 
\begin{equation}\label{veclag}
\mathcal L_{vec}=\text{Im Tr}\left[\tau\left(\int{d^4\theta\Phi^\dagger e^{-2V}\Phi}+\int{d^2\theta W_\alpha W^\alpha}\right)\right],
\end{equation}
where the complex gauge coupling $\tau$ is defined as 
\begin{equation}
\tau=\frac{\theta}{2\pi}+\frac{i}{g^2_{SYM}},
\end{equation}
and the trace in (\ref{veclag}) is over the gauge group.
%$\mathcal L_{vec}$ is invariant under $U(1)_R$ global symmetry as a consequence of classical conformal invariance.  $\Phi$ has charge 2.  $\mathcal N=2$ in 4d can be obtained from $\mathcal N=1$ in $6d$.
The bosonic part of (\ref{veclag}) decomposes as a kinetic term given by (\ref{iib1}) and a potential of the form 
\begin{equation}
V\sim\text{Tr}[\phi^\dagger,\phi]^2.
\end{equation}
%(\textbf{check this-add discussion on moduli space of vacua here})\\
On the other hand, the $\mathcal N=2$ hypermultiplet are made of 
\begin{equation}\label{hypermult}
\begin{array}{ccc}
&\psi_q& \\
q &&\tilde q^\dagger \\
&\psi^\dagger_{\tilde q}&
\end{array}
\end{equation}
two Weyl fermions $\psi_q$ and $\psi^\dagger_{\tilde q}$ as well as two $2\text{dim}R$ complex scalar $q$, $\tilde q^\dagger$.  In $\mathcal N=1$ superspace language, the $\mathcal N=2$ hypermultiplet decomposes into two $\mathcal N=1$ chiral superfields (multiplets) denoted $Q,\tilde Q$ transforming in the representation $R,\bar R$ respectively of the gauge group $G$.  Again, the diamond shape of (\ref{hypermult}) reminds us of how the fields transform under $SU(2)_R$.  The fermions are singlets under $SU(2)_R$ and carry $U(1)_R$ charge 1 while the scalar components of $Q,\tilde Q$ transform as a doublet under $SU(2)_R$ and carry no charge under $U(1)_R$.   In $\mathcal N=1$ superspace language, the low energy Lagrangian describing the hypermultiplet is given by \cite{GiveonSR}: 
\begin{equation}
\mathcal L_{hyper}=\int{d^4\theta\left(Q^\dagger e^{-2V}Q+\tilde Q^\dagger e^{-2V}\tilde Q\right)}+\int{d^2\theta\tilde Q\Phi Q}+c.c.
\end{equation}
When formulating the complete theory
\begin{equation}
\mathcal L=\mathcal L_{vec}+\mathcal L_{hyper},
\end{equation}
it has a Coulomb branch parametrized by the matrices $\phi$ satisfying $V=0$, e.g $[\phi,\phi^\dagger]=0$.  $\phi$ in the Cartan subalgebra of the gauge group $\phi=\sum_{i=1}^r{\phi_iT^i}$ generates $r=\text{rank}\  G$ complex moduli parametrizing the Coulomb branch leading to a $U(r)$ gauge group.  When given a VEV, the gauge group Higgs to $U(1)^r$.  In the presence of matter in the fundamental representation of the gauge group, the complex scalars in the hypermultiplet parametrize the Higgs branch. 
%The expectation value of the adjoint scalar in $\mathcal N=2$ vectormultiplet parametrizes the Coulomb branch of the theory. 
%The expectation values of the scalars in the  $\mathcal N=2$ hypermultiplet parametrize the Higgs branch of $\mathcal N=2$ susy theories in 4 dimensions (\textbf{check this- why 2 scalars param. Higgs branch }). 
$\mathcal N=2$ susy ensures that the moduli space of vacua is not lifted by quantum effects.  However, the metric is modified.

We now turn to a particular $\mathcal N=2$ susy gauge theory with group $SU(2)$.  We will not put fundamental matter just yet for simplicity.  
It was shown by Seiberg and Witten \cite{sw1} that $\mathcal N=2$ SYM theories in four dimensions with at most two derivatives and four fermions can be solved exactly.  By exactly we mean here that while capturing all non-perturbative effects, they still found exact formulas for the metric on the moduli space of vacua as well as for electrons and dyons masses.
This surprising property of Seiberg-Witten theory comes from the fact that the theory's dynamics is governed by holomorphic quantities. 
%(\textbf{Add discussion here on holomorphic  quantities-Seiberg}). 
In fact, Wilsonian 
%(\textbf{check this}) 
effective action with higher derivative terms are not governed by such holomorphic quantities and would not lead to exact solutions. 

The holomorphicity of the quantity $\mathcal F$, a function of the moduli space called the prepotential, allowed Seiberg and Witten to express the $U(1)$ gauge theory completely in terms of the following action \cite{sw1}: 
\begin{equation}\label{u1metric}
\mathcal L_{vec}=\text{Im Tr}\left[\int{d^4\theta\frac{\partial\mathcal F(\Phi)}{\partial \Phi_i}\bar\Phi_i}+\frac{1}{2}\int{d^2\theta \frac{\partial^2\mathcal F(\Phi)}{\partial\Phi_i\partial\Phi_j}W^i_\alpha W^\alpha_j}\right],
\end{equation}
where $\Phi$ is the $\mathcal N=1$ chiral multiplet with scalar component $\phi$ inside the $\mathcal N=2$ vector multiplet while $W^\alpha$ is the $\mathcal N=1$ vector multiplet inside the $\mathcal N=2$ vector multiplet in $\mathcal N=1$ superspace language.
Since the quantum corrections on the moduli space prevent the $SU(2)=Sp(2)$ classical gauge group from enhancing,  the gauge group is $U(1)$ everywhere on the quantum moduli space and we don't need to go beyond the $U(1)$ action shown above. Gauge symmetry group enhancement point on the classical moduli space correspond to two points where the monopole and dyon are massless. The physical meaning of these singularities can be understood as follows: imagine you are at an energy scale $\Lambda$, some massive states exist at that energy scale so you integrate them out.  You then flow to the low energy effective action.  If you see singularities there, it means that you've integrated out massive states at energy $\Lambda$ which became massless at low energy.  Singularities in the low energy effective Wilsonian action thus mean that massless states were integrated out.  Underlying monodromies on the moduli space tell us what states have been integrated out.  This information is captured by an elliptic curve non-trivially fibered over the moduli space.  More on this in a bit.
Coming back to the prepotential function $\mathcal F$, the latter defines the low energy $U(1)^r$ gauge coupling matrix $\tau_{ij}$ which itself parametrizes the metric on the moduli space
\begin{equation}
\tau_{ij}=\frac{\partial^2\mathcal F}{\partial \phi_i\partial \phi_j},
\end{equation}
\begin{equation}
ds^2=\text{Im }\tau_{ij}\ d\phi_id\bar\phi_j=\text{Im }\mathcal F''(\phi)d\phi d\bar\phi. 
\end{equation}
Demanding $\text{Im } \tau(\phi)>0$ guaranties the existence of a well-defined positive definite metric everywhere on the moduli space.  This was one of the crucial ingredient leading to exact solution in Seiberg-Witten theory. The other essential piece of physics to Seiberg-Witten's result is that the holomorphic prepotential $\mathcal F$ is itself constrained by the weakly coupled limit of the $\mathcal N=2$ theory. 
This can be understood as follows: if one compares the $U(1)$ Wilsonian effective action of $\mathcal N=2$ SYM (\ref{u1metric}) to the $\mathcal N=2$ vector multiplet lagrangian (\ref{veclag}), one sees that classically, the prepotential is given by the following quadratic function \cite{GiveonSR,sw1}:
\begin{eqnarray} 
\mathcal F_0=\frac{1}{2}\tau_{0}\Phi_i\Phi^i=\frac{1}{2}\tau_0\mathcal A^2,
\end{eqnarray}
where $\tau_{0}$ is the bare coupling constant.
After adding the 1-loop correction - in the absence of fundamental matter- Seiberg showed \cite{SeibergNS} that the form of the prepotential is \cite{GiveonSR,sw1}:
\begin{eqnarray}
\mathcal F_1&=&\frac{i}{4\pi}\sum_{\vec \alpha>0}{(\vec \alpha\cdot \vec \Phi)^2\text{log}\frac{(\vec\alpha\cdot\vec\Phi)^2}{\Lambda^2}}=\frac{i}{2\pi}\mathcal A^2\text{ln}\frac{\mathcal A^2}{\Lambda^2},
\end{eqnarray}
where $\vec\alpha$ are the positive root of the Lie algebra of the gauge group $G$, $\mathcal A$ is the $\mathcal N=2$ vector multiplet and $\Lambda$ is the dynamically generated scale.
The logarithm breaks $U(1)_R$ symmetry and is related to the 1-loop beta function.  A non-renormalization theorem assures that higher order  perturbative corrections are absent.  However, there exists an infinite series of non-perturbative corrections coming from instantons corrections.   This series falls off algebraically at large $\Phi$ but is important at small $\Phi$.  
%(\textbf{check this: discuss how Seiberg-Witten showed that instantons corrections are resumable})
The full prepotential function is thus given by 
\begin{equation}
\mathcal F=\frac{i}{2\pi}\mathcal A^2\text{ln}\frac{\mathcal A^2}{\Lambda}+\sum_{k=1}^\infty{\mathcal F_k}\left(\frac{\Lambda}{\mathcal A}\right)^{4k}\mathcal A^2,
\end{equation}
where the last term is the instanton contribution and where $k$ instantons contribute to the $k'$th term.
Note that for cases where the beta-function vanishes such as in $\mathcal N=4$ SYM theories, the classical prepotential is exact e.g has no perturbative or non-perturbative corrections.  The last building block on which Seiberg-Witten theory lies on is the generalization of Montonen-Olive $SL(2,\mathbb Z)$ duality \cite{MontonenMO} in $\mathcal N=4$ superconformal field theories to $\mathcal N=2$ supersymmetric ones.  Seiberg and Witten showed in \cite{sw1} that there also exists an $SL(2,\mathbb Z)$ symmetry governing $\mathcal N=2$ SYM theories which interchanges strongly coupled gauge theory for weakly coupled one, provided on interchanges the electrically stable states for magneticallyy charged ones.  This duality exchanges the $\mathcal N=2$ $U(1)$ action for its dual Lagrangian given by
\begin{equation}
\mathcal L^D_{vec}=\text{Im Tr}\left[\int{d^4\theta \mathcal F'_D(\Phi_D)\bar\Phi_{Di}}+\frac{1}{2}\int{d^2\theta}\mathcal F''_D(\Phi_D)W^i_{\alpha D}W_{jD}^{\alpha}\right],
\end{equation}  
where the subscript $D$ denotes electric-magnetic dual variable and where we used the following relation: 
 \begin{equation}\label{dualparam}
 \phi^D_i=\frac{\partial \mathcal F}{\partial \phi^i}.
 \end{equation}
Accordingly, the metric on moduli space can thus be written in the following compact form:
\begin{equation}
ds^2=\text{Im }(d\phi^D d\bar \phi).
\end{equation}
Lastly, the prepotential determines the mass of BPS states of the theory.  For  BPS saturated states with electric charges $e_i$ and magnetic charges $m^i$ with $i=1,\cdots,r$ under the $r$ unbroken $U(1)$ gauge fields, the supersymmetry algebra yields the following mass
\begin{equation}
M=\sqrt 2|Z|,
\end{equation}
with the central charge $Z$ given by
\begin{equation}
Z=\phi^ie_i+\phi_i^Dm^i,
\end{equation}
which can be written in term of the prepotential by using (\ref{dualparam}).
Therefore, by determining exactly the prepotential $\mathcal F$ for the four-dimensional $\mathcal N=2$ SYM with gauge group $SU(2)$ both with and without matter, Seiberg and Witten completly solved the theory. In addition, they found that $\tau_{ij}$ is the period matrix of a Riemann surface with genus one; showing in the process that the moduli space of vacua of this theory is parametrized by the complex structure of an auxiliary two dimensional Riemann surface.  The elliptic curve mentioned previously which captured the singularities on the moduli space parametrize the aforementioned Riemann surface.  We will see the appearance of such a Riemann surface again when discussing about F-theory.  Before we show how F-theory provides a geometric understanding of all the features of Seiberg-Witten $\mathcal N=2$ SYM theory in 4 dimensions, we turn to a more complicated field theory with the higher rank gauge group $SU(3)$ exhibiting a duality called Argyres-Seiberg duality.  We will later see how to embed both Seiberg-Witten theory and its higher rank generalizations in F-theory. 
%(\textbf{add comments on the matter sector with global symmetry and the modifications therein. Kahler metric on Coulomb branch, Hyperkahler metric on Higgs branch})

\section{Argyres-Seiberg duality}

We would like to now discuss about a more complicated superconformal gauge theory: $\mathcal N=2$ SYM theory with gauge group $SU(3)$ with 6 fundamental flavors.  This theory leads to a duality called Argyres-Seiberg duality \cite{ArgyresSeiberg}.  This is an extension of S-duality (strong-weak duality) of $\mathcal N=4$ supersymmetric gauge theories (also called Olive-Montonen duality \cite{MontonenMO}) to the larger class of $\mathcal N=2$ superconformal gauge theories and will be the corner stone of Chapter 3. 

S-duality in $\mathcal N=4$ superconformal field theories answers the following question: what happens when the gauge coupling constant $g$ inside the complex coupling $\tau=\theta/(2\pi)+(4\pi i)/g^2$ becomes infinite?
The answer for four-dimensional $\mathcal N=4$ SYM theories is that the theory turns into a weakly coupled gauge theory, not necessarily with the same gauge group though.  For simply-laced gauge groups, where theory is self dual, this duality is expressed as an equivalence between the theory at different couplings $\tau\cong-1/\tau$.  The periodicity of $\theta\to \theta +2\pi$ leads to further identification, namely $\tau\cong \tau+1$ \cite{ArgyresSeiberg}.  
\begin{figure}[htb]
        \begin{center}
\includegraphics[height=6cm]{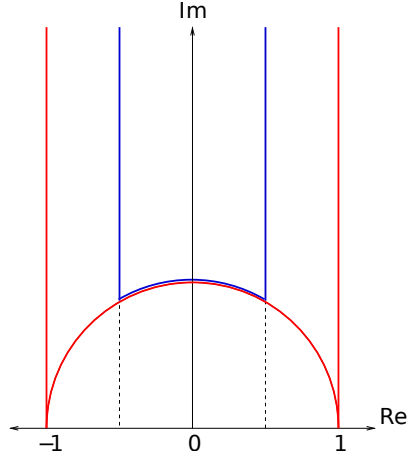}
        \end{center}
        \caption{Fundamental domain in $\tau$ for $SL(2,\mathbb Z)$ (blue) and in $\tilde \tau\equiv 2\tau$ for $\Gamma^0(2)$ (red).  Edges of the domains are identified under reflection by $\text{Re}\tau=0$.  Figure extracted from \cite{ArgyresSeiberg} and modified.} \label{ASfig}
        \end{figure}
        
The two symmetries of the complex coupling $\tau$ generate an $SL(2,\mathbb Z)$ group of identifications whose fundamental domain in the space of couplings is bounded away from infinite coupling given by $\text{Im}\tau=0$ see figure (\ref{ASfig}).
 
The case of infinite coupling in four dimensional scale-invariant $\mathcal N=2$ SYM theories is more subtle.  For $\mathcal N=2$ $SU(2)$ SQCD with four massless hypermultiplets in the fundamental representation of the gauge group, it was shown in \cite{sw1} that Olive-Montonen duality goes through and that there is an $SL(2,\mathbb Z)$ S-duality.  However, this is not the case for higher rank gauge group.  In particular,  for the $\mathcal N=2$ SYM theory with $SU(3)$ gauge group with 6 massless fundamental hypermultiplets it was shown in \cite{plesser} that the S-duality group is $\Gamma^0(2)\subset SL(2,\mathbb Z)$, generated by $\tilde \tau \cong \tilde \tau+2$ and $\tilde \tau \cong -1/\tilde \tau $ where $\tilde \tau\equiv 2\tau$.  As seen in figure (\ref{ASfig}), $\text{Im}\tilde \tau$ can equal to zero, meaning that the theory contains points of infinite coupling in its moduli space.          

Argyres and Seiberg studied the physics at the infinite coupling points of the superconformal $\mathcal N=2$ theory with gauge group $SU(3)$ and provided an M-theory description of the  phenomenon.  Since we will want to embed this duality as well as its extension - called Gaiotto duality - in F-theory, we will not present here the M-theory description they found.  What we will present however is the field theory answer that Argyres and Seiberg found to the following question: what happens when we take the marginal gauge coupling $g$ to infinity in the four-dimensional $\mathcal N=2$ SYM theory with gauge group $SU(3)$ in the presence of 6 massless fundamental hypermultiplets. 

The answer is \cite{ArgyresSeiberg}:  
\bg\label{ASdual}
SU(3)\text{w/}6\cdot({\bf 3\oplus\bar 3})=\rm{SU(2)}\text{w/}(\rm{2}\cdot \bf{2}\oplus\text{SCFT}_{E_6}).
\nd
It reads as follows: 
the gauge theory with gauge group $SU(3)$ coupled to 6 massless hypermultiplets in the fundamental representation of the gauge group with a coupling $f$ is equivalent to an $SU(2)$ gauge theory (with coupling $\tilde f$) with one massless fundamental hypermultiplet.  The $SU(2)$ is also coupled to an isolated rank 1 SCFT with flavor symmetry $E_6$ \cite{ArgyresSeiberg}\footnote{What is meant here by ``isolated SCFT" is an SCFT with no marginal coupling on its own.  The rank of an isolated $\mathcal N=2$ SCFT is equal to the complex dimension of its Coulomb branch.}.  The way the $SU(2)$  gauge group is coupled to the SCFT is by gauging the $SU(2)$ inside the maximal subgroup $SU(2)\times SU(6)\subset E_6$.  Thus, effectively, the flavor symmetry seen by the $SU(2)$ gauge group is given by $U(6)=U(1)\times SU(6)$ where the $U(1)$ comes from the massless hypermultiplet and the $SU(6)$ from what's left of the $E_6$ flavor symmetry after the $SU(2)$ is gauged.   Only in the limit of zero coupling (when $SU(2)$ decouples from the rank 1 SCFT) is the $E_6$ the full flavor symmetry of the theory.  Argyres-Seiberg duality maps an infinite coupling in $f$ to zero coupling in $\tilde f$ where $f\sim e^{i\pi\tilde \tau}\to 0$ at weak coupling and $f\to 1$ at infinite coupling.
A quick check of the duality (\ref{ASdual}) is through the matching of the ranks and flavor groups on both sides \cite{ArgyresSeiberg}.  The rank - real dimension of the Coulomb branch - of the $SU(3)$ gauge group of the left hand side of (\ref{ASdual}) is equal to 2.  On the other hand, both the $SU(2)$ gauge group and the $E_6$ SCFT
on the right hand side of (\ref{ASdual}) have rank 1 matching that of $SU(3)$.  The 6 massless hypermultiplets on the left hand side of (\ref{ASdual}) contribute to an $U(6)$ flavor group.  We have already establish that the $E_6$ SCFT contributes an $SU(6)$ flavor symmetry from the way the $SU(2)$ is gauged inside the maximal subgroup of $E_6$.  Adding the $U(1)$ flavor from the hyper on the right hand side, this matches the flavor symmetry of the left hand side, thus successfully checking the duality.

Since the topics of Chapter 3 include the embedding of Argyres-Seiberg duality and its generalization  in F-theory, we begin the next section by introducing F-theory.  

\section{F-theory}

As we have seen a couple of times now, dualities both in field theories and string theory have allowed us to probe the strong coupling regime of many corners of string theory.  M-theory with its strong-weak duality to type IIA has played a significant role in this endeavour.  However, there are theories, such as type IIB which have a less natural interpretation in M-theory.  Surely, one can understand the $SL(2,\mathbb Z)$ invariance of 10d type IIB by first compactifying the theory to 9d and then compare it with a $T^2$ compactification of 11d M-theory.  One then finds that the $SL(2, \mathbb Z)$ is interpreted as a symmetry of the torus but one recovers 10d type IIB only in the limit where the $T^2$ has zero area- see \cite{GiveonSR} for more details.  Wanting to associate a geometric meaning of the $SL(2, \mathbb Z)$ invariance of type IIB and desiring a strongly coupled dual theory to type IIB (in the same way  M-theory is the strongly coupled dual to type IIA), Vafa proposed in 1996 the F-theory \cite{vafaF}. 

F-theory is a 12-dimensional theory which, once compactified on a 4-dimensional K3 manifolds, gives rise to an 8 dimensional susy field theory preserving half of the supersymmetry.  Vafa argued that the uncompactified 8-dimensions correspond to the worldvolume of 7-branes.  

By definition, an elliptically fibered manifold (a manifolds that admits elliptic fibration) is a manifold $M$ that has the structure of a fiber bundle whose fiber is a two dimensional torus at every points of a base which is some manifold $B$.  F-theory compactified on a manifold $M$ corresponds to type IIB compactified on the manilfold $B$.  More precisely, F-theory compactified on an elliptically fibered K3 manifold corresponds to type IIB compactified on $S^2$ (where the $S^2$ was obtained after compactifying $CP^1$ by putting on it 24 $D7$-branes) \cite{senF}.  Under 10 dimensional $SL(2, \mathbb Z)$ strong-weak duality, the type IIB gauge coupling $\tau=C_0+ie^{-\phi}$ transforms in the same way as the modulus of the torus.  In fact, the complex structure moduli of the elliptic fiber in F-theory is dynamical and under this map from F-theory to type IIB, it corresponds to the gauge coupling (axion-dilaton modulus) of type IIB which captures the aforementioned dynamic by depending holomorphically on the complex coordinates parametrizing $CP^1$.  
%mathematically, $\tau$ is a section of an $SL(2,Z)$ bundle.  $SL(2,Z)$ is the modular symmetry group of a  torus, a section of $SL(2,Z)$ corresponds to the modular parameter of a torus. [read Terning 13.4, 13.5]
If $z$ and $\bar z$ are the coordinates of $S^2$ then $\tau$ only depends on $z$ or $\bar z$.  (Here, holomorphicity comes from looking at vacuum solution of type IIB and demanding that the solution of the low energy Lagrangian preserve 1/2 susy). Since the antisymmetric NS-NS and RR tensors are interchanged under $SL(2,Z)$ transformation (and thus not invariant), there are set to zero in this discussion (i.e set to zero when solving for vacuum solution of type IIB) \cite{vafaF}.  To understand how $\tau$ depends on $z$, one can start by writing how the torus depends on $z$.  The equation for the torus as a function of $z$ is given by the following elliptic curve, also called Weierstrass equation \cite{senF,vafaF}:
\begin{equation}\label{ellipticcurve}
y^2=x^3+f^8(z)x+g^{12}(z),
\end{equation}
where $f^n,g^m$ are degree $n,m$ polynomial in $z$ while $x,y,z\in \mathbb CP^1$.  The above equation defines an elliptically fibered K3 surface where there is a torus at each point on $\mathbb CP^1$ parametrized by the coordinate $z$.  $\tau(z)$, the modular parameter of the torus is given by the ratio:
\begin{equation}\label{jfunc}
j(\tau(z))=\frac{4\cdot(24f)^3}{27g^2+4f^3},
\end{equation}
with 
\begin{equation}
j(\tau)=\frac{(\theta^8_1(\tau)+\theta^8_2(\tau)+\theta^8_3(\tau))^3}{\eta(\tau)^{24}},
\end{equation}
where the theta functions satisfying Jacobi's identity were defined in \cite{sw2} to be
\begin{eqnarray}
\theta_1(\tau)&=&\sum_{n\in\mathbb Z}{q^{\frac{1}{2}\left(n+\frac{1}{2}\right)^2}}\\
\theta_2(\tau)&=&\sum_{n\in \mathbb Z}{(-1)^nq^{\frac{1}{2} n^2}}\\
\theta_3(\tau)&=&q^{\frac{1}{2} n^2},
\end{eqnarray}
with $q=e^{2\pi i\tau}$ and $\eta(\tau)$ is the Dedekind eta function
\begin{equation}
\eta(\tau) = e^{\frac{\pi \rm{i} \tau}{12}} \prod_{n=1}^{\infty} (1-q^{n}). 
\end{equation}
Positions where the torus degenerates correspond to points where the discriminant of the above equation vanishes
\begin{equation}\label{Delta}
\Delta\equiv 4f^3+27g^2.
\end{equation}
From the type IIB perspective, $\tau=C_0(z)+ie^{-\phi(z)}$ and the 7-branes of type IIB transverse to $CP^1$ are located at the zeroes of the above discriminant.  Since generically there are 24 zeroes of $\Delta$, there are 24 of these 7-branes on $CP^1$.
Let $z_i$ be a zero of $\Delta$, then for $z$ near $z_i$, (\ref{jfunc}) and (\ref{Delta}) lead to \cite{senF}:
\begin{equation}
j(\tau(z))\sim\frac{1}{z-z_i}.
\end{equation}
Thus, up to $SL(2,\mathbb Z)$ transformation, the torus modular parameter $\tau(z)$ reads \cite{senF}:
\begin{equation}
\tau(z)\sim\frac{1}{2\pi i}\text{ln}(z-z_i).
\end{equation}
We now observe the following: there are 24 $(p,q)$ 7-branes each carrying some RR charges on a compact $S^2$ manifold.  The flux of the branes charge has no where to go.  A legitimate question to ask is if this picture is inconsistent with Gauss' law?  The answer to this puzzle, provided by Sen in \cite{senF}, is as follows: in the weak string coupling, 16 of the 7-branes are $D7$-branes with charge $(1,0)$.  They are in the presence of 4 $O_7$-planes carrying $-4$ charges of $D_7$ branes (\ref{opcharge}) and thus cancelling the total RR charges on $S^2$.  At strong coupling, these O7-plane split into 2 $(p,q)$ 7-branes each: 1 with charge $(0,1)$ and the other with charge $(1,-1)$.  Thus, in total, the 24 original $(p,q)$ 7-branes are split into 16 $D7$-branes, 4 $(0,1)$ 7-branes and 4 $(1,-1)$ 7-branes\footnote{$(p,q)$ seven-branes are related to each other by $SL(2,\mathbb Z)$ transformations.  This explains why the dyon, in the literature, is sometimes written with charge $(2,1)$ instead of $(1,-1)$.}.  As we will see next, Sen's understanding of F-theory is going to play a crucial role in embedding four-dimensional Seiberg-Witten's $\mathcal N=2$  SYM theory in F-theory.  
%It was also argued that this theory is dual to heterotic string compactification on $T^2$.
%Furthermore, M-theory on a elliptically fibered manifold $K$ is equivalent to compactifying F-theory on $K\times S^1$.  
 \section{F-theory embedding of $\mathcal N=2$ SUSY $d=4$}

It was shown by Sen \cite{senF} and later rendered even more precise by Banks-Douglas-Seiberg \cite{bds} that a single D3-brane in the background of an orientifold 7-plane (with 4 $D7$ branes) reproduces Seiberg-Witten theory (with fundamental matter) \cite{sw1,sw2}.

\begin{figure}[htb]
        \begin{center}
\includegraphics[height=6cm]{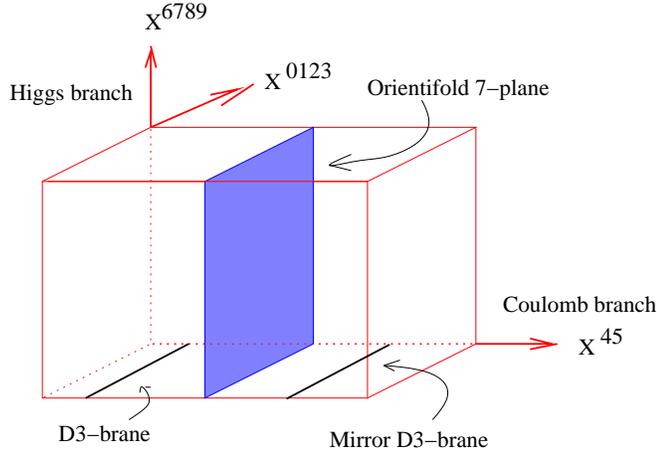}
        \end{center}
        \caption{In F theory
        a single D3-brane probing an orientifold seven-plane background maps to the
classical picture of $Sp(2)$ Seiberg-Witten theory.} \label{fth1}
        \end{figure}

In Seiberg-Witten field theory without matter content\footnote{In Seiberg Witten theory with matter, the moduli space of vacua is six real dimensional and parametrised by one complex scalar in the $N=2$ vector multiplet and two complex scalars in the $N=2$ hypermultipet.  In Seiberg Witten theory without matter, there is no hypermultiplet so the moduli space of vacua is the Coulomb branch.}, the Coulomb branch-loosely speaking referred to as the moduli space of vacua- is a 2 real dimensional plane parametrized by a gauge invariant modulus called $u$.  The $u$-plane can also be written in terms of adjoint complex scalar and thus its complex dimension corresponds to the rank of the gauge group.  Since $G=SU(2)\cong Sp(2)$ has rank 1, recall $SU(r+1)$ has rank r, the $\text{dim}_\mathbb C$(Coulomb branch)$=1$.    
We will embed this field theory in string theory by considering type IIB on $\mathbb R^2/\mathbb Z_2\times \mathbb R^4\times \mathbb R^{0123}$.  The moduli space of vacua, parametrized by $\mathbb R^2/\mathbb Z_2\times \mathbb R^4$ contains 
%is parametrized by $\mathbb R^2/\mathbb Z_2$.  
many branches.  Amongst others, there is the one complex dimensional Coulomb branch parametrized by the $v\equiv x^4+ix^5$ direction with geometry $\mathbb R^2/\mathbb Z_2$ and a Higgs branch along $x^{6789}$ with geometry $\mathbb R^4$.
The $\mathbb R^{0123}$ spans Minkowski space.
When adding sufficient number of 7-branes- in total 24- the moduli space $\mathbb R^2/\mathbb Z_2$ compactifies to $\mathbb T^2/\mathbb Z_2\cong \mathbb {CP}^1$.  This leads to F-theory on an elliptically fibered $K_3\times \mathbb R^4\times R^{0123}$.  Recall that in the latter compact case, the RR fluxes of 7-branes have not enough non-compact transverse directions to escape.  To cancel the RR charge, we consider, classically, 1 $O_7$-plane with 4 parallel $D7$-branes and their mirror. 
The type IIB geometry we will be concerned with at the moment is   $\mathbb R^2/\mathbb Z_2\times \mathbb R^4\times \mathbb R^{0123}$ with branes spanned along the following spacetime directions: 
\begin{eqnarray}\label{seniib1}
D3:&& 0123\\
O_7:&&01236789\\\label{seniib2}
D7:&& 01236789.
\end{eqnarray}
The RR charges, given in (\ref{opcharge}) $Q_{O_7}=-8Q_{D_7}$ thus cancel since the $O_7$ plane carry $-4$ charges of $D7$-branes (-8 charge of $D7$-brane and their mirror).
The full classical configuration generating Seiberg-Witten theory is that of $1\ D3-1\ D3_m,1\ O_7-4\ D7-4\ D7_m$. 
Recall that in the presence of the $O_7$-plane, the D3-brane also gets a mirror D3-brane denoted here by $D3_m$. 
Modulo the $D7$-branes, this classical setup is depicted in figure (\ref{fth1}).
The neutral gauge boson corresponds to the ground state of an open string with both ends on the D3-brane.  On the other hand, charged gauge bosons correspond to the ground states of open strings stretched between the D3-brane and its mirror.  The effective mass of the matter $i^{th}$ quark ($i=1,\cdots, 4$) is given by the relative distance between the $i^{th}$ $D7$-brane and the D3-brane.  Similarly for the mirror $D7$-branes and the mirror D3-brane.
We can make this discussion more precise since all the objects in (\ref{seniib1})-(\ref{seniib2}) are pointlike along the Coulomb branch.  Let the $O_7$-plane be at the origin of the Coulomb branch $v=0$, put the four $D7$-branes and their mirrors at $m_i,\ -m_i$ and let the $D3$-brane and its image be at $v,\ -v$.  The $D7$-branes give rise to $N_f=4$ fundamental hypermultiplets $Q_i,\tilde Q_i$.  The $D7$-branes positions $m_i$ correspond to the bare masses of quarks.  The effective mass of the quarks are given by $m_i-v$ and $m_i+v$ \cite{GiveonSR}. 
The mass of the charged $W^{\pm}$ is $2|v|$ in string units.  When the $D3$-brane and its mirror are coinciding with the orientifold plane, i.e. when $v=0$, the charged gauge boson becomes massless and the gauge group is classically enhanced to $Sp(2)\cong SU(2)$.  When the $D3$-brane and its mirror are away from $v=0$, the gauge boson picks up a mass which higgses the gauge group to $U(1)$.  When $D7$-branes and their mirrors coincide with the $O_7$-plane, quarks become massless and  the gauge symmetry on the 7-branes is enhanced to $SO(8)$.  When the $D7$-branes are away from the $O_7$-plane, the symmetry group is broken to $U(1)^4$.  From the point of view of $1+3$ dimensional physics on the $D3$-branes, the aforementioned $SO(8)$ symmetry is a global symmetry.  As discussed previously, the low energy worldvolume dynamic on the $1+3$ dimensional $D3$-branes in the presence of $D7$-branes and an $O_7$ plane preserve 8 supercharges, classically leading to an $\mathcal N=2$ SYM theory with gauge group $SU(2)$ and global symmetry $SO(8)$ which is exactly Seiberg-Witten theory. 
%\textbf{ make the connection between gauge symmetry and global symm wrt observer more precise}. 
The position of the $D3$-brane along the Coulomb branch is given by the expectation value of the complex scalar in the adjoint representation of the gauge group in the $\mathcal N=2$ $SU(2)$ vectormultiplet: 
\begin{eqnarray}
\langle \phi\rangle=
\left(\begin{array}{cc}
v&0\\
0&-v
\end{array}\right).
\end{eqnarray}
The $u$-plane parametrizing the classical moduli space is written as
\begin{equation}\label{uparam}
u=\frac{1}{2}\text{Tr }{\phi}^2=v^2.
\end{equation}
Similarly, the position of the $D3$-brane along the Higgs branch - which is the direction along the $O_7$ and $D7$-branes, is parametrized by the expectation value of the two complex scalars in the $\mathcal N=2$ hypermultiplet. 
The classical curve corresponding to the dynamic of the classical $Sp(2)$ gauge group is given by
\begin{equation}
y^2=(x^2-u)^2.
\end{equation}
The zeroes of this curve, located at $u=x^2$ indicate the location of the singular region on the classical moduli space where the point of enhanced gauge symmetry is (let $x=0$). 
Since the $u$-parameter corresponds to the location of the $D3$-brane, the aforementioned singular point of enhanced gauge symmetry is consistently locates where the $D3$-brane coincide with the $O_7$ plane at $v=0$. 
The classical moduli space can thus be understood as a probe $D3$-brane in the background of $D7$-branes and a $O_7$-plane.  The complex gauge coupling on the $D3$-brane describing the $\mathcal N=2$ SYM physics is given by the type IIB complex dilaton:
\begin{equation}
\tau=a+\frac{i}{g_s},
\end{equation}
where $a$ here denotes the axion.  Since the $D7$-branes and $O_7$-plane carry charges under the complex dilaton, the presence of these objects in the background of the $D3$-brane modify the value of the complex dilaton and consequently, the value of the complex gauge coupling \cite{GiveonSR}.  In particular, when the $D3$-brane goes once around a $D7$-brane, the complex gauge coupling picks up a monodromy, transforming as $\tau\to \tau+1$.
Since there is $+1$ unit of 7-brane charge where $v=\pm m_i$ and $-8$ unit of 7-brane charge at $v=0$, the gauge coupling, far from the point where the $D3$-brane coincide with either the $D7$-brane or the Orientifold plane,  is given by:
\begin{equation}
\tau(v)=\tau_0+\frac{1}{2\pi i}\left[\sum^4_{i=1}{\left(\text{log} (v-m_i)+\text{log}(v+m_i)\right)-8 \text{log}\ v}\right].
\end{equation}
This can also be rewritten in terms of the gauge invariant modulus $u$  (\ref{uparam}) as: 
\begin{equation}\label{classtau}
\tau(u)=\tau_0+\frac{1}{2\pi i}\left[\sum^4_{i=1}{\text{log}(u-m_i^2)-4\text{log}\ u}\right],
\end{equation}
(the coefficient 4 in the third term is due to the fact that $W^\pm$ bosons carry twice the electric charges of the quarks).  The presence of the logarithmic terms in the equations above indicate that this is a semiclassical result, corresponding to a prepotential at 1-loop.  Since $\text {Im} \tau$ is large and negative for small values of $u$, this expression for the complex gauge coupling can not be exact as it does not satisfies the condition $\text{Im} \tau\geq 0$ everywhere on the moduli space.

\begin{figure}[htb]
        \begin{center}
\includegraphics[height=6cm]{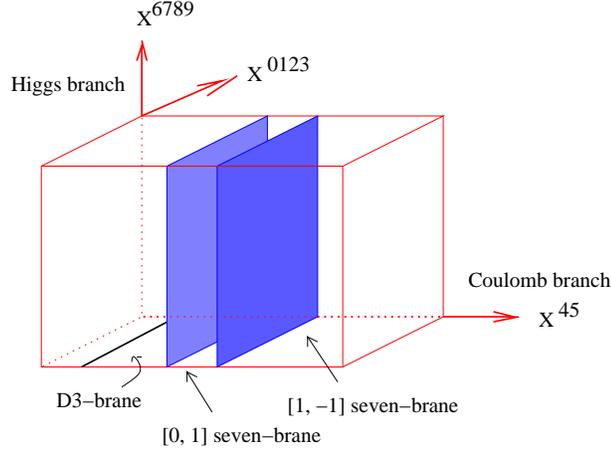}
        \end{center}
        \caption{The quantum corrections in the $Sp(2)$ theory maps to the splitting of the orientifold plane
into two ($p, q$) seven-branes in F-theory.} \label{fth2}
        \end{figure}

When including non-perturbative corrections, the exact effective coupling is a modular parameter $\tau(u)$ of a torus described by the elliptic curve of the form (\ref{ellipticcurve}) where $u\equiv z$.
Under non-perturbative corrections the $O_7$ splits into 2 $(p,q)$ 7-branes of charge $(1,0)$ and $(1,-1)$ corresponding respectively to a monopole and a dyon.  This is shown in figure (\ref{fth2}).  
The $(p,q)$ 7-branes are at a distance of order $exp(2\pi i \tau_0)$ from the original point $u=0$.  The parameter $exp(i\pi\tau_0)\equiv \Lambda$ is the analogue of the QCD scale of the theory.  In the limit $\tau_0\to i\infty$ ($g_s\to 0$) and $|u|>>|\Lambda|^2$, the two singularities coincide and we recover the semiclassical picture (\ref{classtau}) \cite{senF,GiveonSR}.  
Under the splitting of the orientifold plane, the $D7$ and $D3$-mirror branes no longer exists. 
Since there are no longer charged $W^{\pm}$ bosons in the picture, the gauge group on the $D3$-brane can no longer enhance and is $U(1)$ everywhere on the moduli space.  When the $D3$-brane coincide with either the ``dyonic", ``monopole" or $D7$-brane, the $(1,-1)$, $(0,1)$ or $(1,0)$ string between them becomes arbitrarily light i.e the quark becomes massless, corresponding the massless hypermultiplet in the fundamental representation of the gauge group.  The curve capturing the dynamic of the quantum moduli space is:
\begin{equation}\label{quantcurve}
y^2=(x^2-u)^2-\Lambda^4.
\end{equation}
The zeroes located at $u=x^2-\Lambda^2$ and $u=x^2+\Lambda^2$ (let $x=0$) correspond to the two points ($u=\pm \Lambda^2$) where a monopole hypermultiplet and a dyon hypermultiplet become massless respectively.  These two points correspond to the positions where the $D3$-brane coincide with the two ($p,q$) 7-branes respectively.
The final picture of the quantum moduli space has $1\ D3-2\ (p,q)\ 7$-branes, $4\ D7$ where the $D7$-branes are free to move in the $u$-plane but the $(0,1)$ and $(2,1)$ are stuck at $u=\pm \Lambda^2$.  Therefore, the final quantum picture leading to Seiberg-Witten theory has 6 7-branes in type IIB.  Remembering that F-theory is obtained by putting at most 24 7-branes transverse to the type IIB moduli space ($u$-plane), one sees that the full moduli space of F-theory on $\mathbb K_3\times \mathbb R^4\times \mathbb R^{0123}$ with 24 7-branes  captures 4 copies of $\mathcal N=2$ Seiberg-Witten theory with $N_f=4$.
In F-theory language, the zeroes of the discriminant of (\ref{quantcurve}) correspond to points on the moduli space where the fibered torus $T^2$ degenerates on the $u$-plane \cite{senF,bds,GiveonSR}. %[\textbf{mention Tate algorithm}.] [\textbf{how does brane capture confinement, chiral symmetry breaking...]}
%n type IIB, the $u$-plane is $\mathbb R^2/\mathbb Z_2$ while it is $\mathbb {CP}^1$ in F-theory.  D3-brane, $O_7$-plane and 7-branes are all pointlike on the $u$-plane.  Given the type IIB geometry $\mathbb R^2/\mathbb Z_2\times \mathbb R^4\times \mathbb R^{0123}$, we can associate $\mathbb R^2/\mathbb Z_2$ as parametrizing the Coulomb branch ($x^{45}$), $\mathbb R^4$ as being the directions spanned by the Higgs branch ($x^{6789}$) and $\mathbb R^{0123}$ as Minkowski space.
%\section{Dualities}

\section{Possible background deformations}

We have seen so far that for less then 24 7-branes, one can use type IIB on $\mathbb R^2/\mathbb Z_2\times \mathbb R^4\times \mathbb R^{0123}$ to capture the information contained in $\mathcal N=2$ SYM theory in 4 dimensions. On the other hand, if we put 24 7-branes transverse to the type IIB moduli space $\mathbb R^2/\mathbb Z_2$, the latter compactifies to $\mathbb {CP}^1\cong \mathbb T^2/\mathbb Z_2$ and we can make use of the IIB/F-theory duality
\begin{equation}
\text{Type IIB: }\mathbb T^2/\mathbb Z_2\times \mathbb R^4\times \mathbb R^{0123} = \text{F-theory: } K_3\times \mathbb R^4\times R^{0123},
\end{equation}
where the RR charges of the $D7$-branes is cancelled by having 1 $O_7$-plane per bunch of 4 $D7$-branes, leading to 4 copies of Seiberg-Witten theory.  
%Although we will not dwell on this here, we need to mention that studying the non-compact $\mathbb R^2/\mathbb Z_2\times \mathbb R^4\times R^{0123}$ with $7$-branes$<24$ leads to interesting physics with lots of open questions.  [Douglas-Lowe-Schwarz] first analyze the cases where there would be $r$ D3-brane probe [textbf{specify in which limit is a d3-brane a probe brane}] on this background.  Since the moduli space of the latter has two running directions, (\textbf{check how many running directions a brane need for no charge cancellation problem}) there is no problem of charge cancellation for any brane.  The corresponding field theory in that of pure $\mathcal N=2$ SYM with gauge group $Sp(2r)$.  The physics in F-theory is given by a Riemann surface of genus $r$ fibered over the moduli space and parametrized by a hyperelliptic curve.  One of the challenge in studying this theory is to project the information contained in the curve - spectrum of monopoles/dyons, maximal Argyres-Douglas points, etc- unto the 2 real dimensional type IIB moduli-space since the curve does not decompose as $r$ copies of the elliptic one as one would naively expect.  Elaborate discussions and attempts can be found in [DSW]: we  do not elaborate on this here since it is not part of the models we wish to understand. 
%(\textbf{check dimensions of moduli space for Seiberg-Witten and this-see douglas/lowe/schwarz}).  
Recall the direction spanned by certain branched of moduli space of the type IIB theory dual to F-theory: 
\begin{equation}\label{branchesdeform}
\text{IIB:}\ \underbrace{\frac{\mathbb T^2}{\mathbb Z_2}}_{\text{Coulomb branch}}\times\underbrace{\mathbb R^4}_{\text{Higgs branch}}\times\underbrace{\mathbb R^{0123}}_{\text{Minkowski}},
\end{equation}
staring at this long enough, one realized that certain deformations of this background are allowed and might lead to new and interesting physics.  These are summarized in the table below \cite{DasguptaSW}:

\begin{table}[htb]
 \begin{center}
\begin{tabular}{|cccc|}\hline
&&&\\
 &$IIB:\ \mathbb T^2/\mathbb Z_2\times \mathbb R^4\times \mathbb R^{0123}$&&\\
 &&&\\ \hline\hline
 &&&\\
$\mathbb R^4\to\mathbb R^4/\mathbb Z_k$ & $\text{ALE}_{k>1}$ & non-compact&$\mathcal N=2$ \\ 
&&&\\\hline
&&&\\
$\mathbb R^4\to\mathbb T^4/\mathbb Z_2$  & K3& compact &\ \ \ $\mathcal N=2,1$ \\ 
&&&\\ \hline
&&&\\
$\ \ \ \ \ \ \ \mathbb R^4\to\mathbb T^2/\mathbb Z_2\times \mathbb R^2$  &&&$\mathcal N=1$ \\ 
&&&\\\hline
  \end{tabular}
\end{center}
  \caption{Possible background deformations of the F-theory geometry}
  \label{bckdeform}
\end{table}
From (\ref{branchesdeform}), we see that each deformation in Table (\ref{bckdeform}) corresponds to a different compactification of the Higgs branch. 
Here, we will discuss each of them briefly.  The main content of this thesis will focus exclusively on the first deformation and on the new physics that emerges from it.  A detailed analysis of the other cases can be found in \cite{DasguptaSW}.  \\
\begin{enumerate}
\item Chapter 3 will deal with the first deformation consists in taking:
\begin{equation} \label{deformation1}
\mathbb R^4\to \mathbb R^4/\mathbb Z_k,
\end{equation}
under type IIB/F-theory duality, it leads to:
\begin{equation}
\text{IIB: } \mathbb T^2/\mathbb Z_2\times \mathbb R^4/\mathbb Z_k\times \mathbb R^{0123}\to \text{F-theory: }K3\times \mathbb R^4/\mathbb Z_k\times \mathbb R^{0123}.
\end{equation}
Where $\mathbb R^4/\mathbb Z_k\cong TN_k$ with $TN_k$ a $k$-centered Taub-NUT space.  The new type IIB/F-theory obtained preserve $\mathcal N=2$ supersymmetry in 4 dimensions.  We will analyze the physics of multiple $D3$-branes probing F-theory on $K3\times TN_k$ with 7-branes wrapped on the $TN_k$.  We will study this model: 
\begin{itemize}
%\item in the presence of multiple probe $D3$-branes
\item in the conformal and non-conformal limit
\item see how it reproduces Gaiotto-model, Benini-Benvenuti-Tachikawa-model, and Gaiotto duality
\item discuss about the supergravity dual of the conformal regime 
\item see how it leads to $\mathcal N=2$ cascade in the non-conformal limit
\end{itemize}
%\item An alternative way to study Gaiotto-type models using fractional branes to probe the F-theory
%geometry with multi Taub-NUT spaces.
%\item A study of a class of Gaiotto dualities via chiral anomaly cancellations, anti-GSO projections and
%brane transmutations in a ${\rm TN}_n$ background with wrapped seven-branes.
%\item A study of the ultra-violet and infra-red geometries using branes anti-branes system, including an
%analysis of the holographic duals.
%\item A study of many conformal examples in Gaiotto models and a map to the brane network picture. Explicit
%examples of %Gaiotto
%dualities in these models.
%\item A study of many new non-conformal models directly from F-theory, including an interesting
%example of non-conformal cascading ${\cal N} = 2$ theories. Examples of the mapping to cascading ${\cal N} = 1$ theories.
\ \\
\item The second deformation
\begin{equation}
\mathbb R^4\to \mathbb T^4/\mathbb Z_2,
\end{equation}
was first studied by \cite{DRS} where the authors show that it preserved both $\mathcal N=2$ and $\mathcal N=1$ susy.
Under type IIB/F-theory duality, they obtained: 
\begin{equation}
\text{IIB: }\mathbb T^2/\mathbb Z_2\times K3\times \mathbb R^{0123}\to\text{F-theory: }K3\times K3\times \mathbb R^{0123},
\end{equation}
where $\mathbb T^4/\mathbb Z_2\cong K3$. 
\cite{DasguptaSW} revisited this model and studied the following: 
\begin{itemize}
\item supergravity solutions for $D3-\bar D3$ probing $K3\times K3$ in F-theory with $G$-fluxes
\item observed duality between abelian instantons and $G$-fluxes in M-theory
\item connected to non-trivial M(atrix) theory on $K3\times K3$ with fluxes
\end{itemize}
\item The last deformation consists in
\begin{equation}
\mathbb R^4\to\mathbb T^2/\mathbb Z_2\times \mathbb R^2.
\end{equation}
This model was first studied in \cite{gimpol} where it was shown to lead to $\mathcal N=1$ susy theory in 4-dimensions.  The associate background is given by  
\begin{eqnarray}
&&\text{IIB: }\left({{\mathbb T}^2\over \Omega\cdot (-1)^{F_L}\cdot {\cal I}_{45}} \times 
 {{\mathbb T}^2\over \Omega\cdot (-1)^{F_L}\cdot {\cal I}_{89}} \right) \times {\mathbb R}^2 \times {\mathbb R}^{0123}\\
 &&\\
&&\text{F-theory: }\left({\rm CY}_3\right) \times {\mathbb R}^2 \times {\mathbb R}^{0123}.
\end{eqnarray}
Where the new $\mathbb Z_2$ orbifold was written explicitly so that we don't confused it with the existing orbifold where $\mathcal I_{89}:x^{8,9}\to-x^{8,.9}$. When lifting to F-theory, the moduli space of type IIB becomes an elliptically fibered Calabi-Yau 3-fold.
In this model, \cite{DasguptaSW} studied the physics of multiple $D3$-branes probing intersecting $7$-branes and $O_7$-planes background in type IIB or analogously, F-theory on a $CY_3$.  
They analyzed 
\begin{itemize}
\item a dual map to the heterotic theory on a non-Kahler $K3$ manifold that is not a conformally Calabi-Yau manifold
\item new examples of type IIB and M-theory compactifications on non-Kahler manifolds
\end{itemize}
\end{enumerate}
As mentioned previously, Chapter 3 will address the first background deformation and explore in depth its consequences.  Chapter 4 will try to clarify one of the new physical phenomenon raised in Chapter 3.  \newpage
%\section{Omission}
%\begin{itemize}
%\item Bosonic part of the 11-dimensional supergravity action : equation 12.1.1 Polchinski vol.2
%\item Type IIA action: equation 12.1.10a Polchinski vol. 2
%\item Type IIB action: equation 12.1.26a Polchinski vo.2
% getting non-abelian gauge theories on $Dp$-branes: Becker-Becker-Schwatz 195-200.
%low energy effective action: p.300-p.301 Becker-Becker-Schwarz.
% T-duality p.188 Becker-Becker-Schwatz
% Superspace and superfields: Polchinski 2, p.103
% various coupling in low energy action: Polchinski 2, p.109.
%\end{itemize}
%\newpage

\clearpage

\chapter{F-theory embedding of new supersymmetric gauge theories}
%%%%%%%%%%%%%%%%
%%%%%%%%%%%%%%%%%%%%%%%%%%%%%%%%%%%%%%%%%%%%%%%%%%%%%%%%%%%%%%%%%%%%%%%%%%%%%%%%%%%%%%%%%%%%%%%%%%NEW MODEL 2
\section{Introduction} 

Inspired by S-duality and in particular by Argyres-Seiberg duality, Gaiotto \cite{gaiotto} asked if the latter strong-weak duality could hold for generic gauge groups.  He explored the strongly coupled limit of various $\mathcal N=2$ superconformal gauge theories in four dimensions and found that an Argyres-Seiberg-like duality does hold for $\mathcal N=2$ SCFT with gauge groups of the form $\prod_{i=1}^{n} SU(N_i)$ and appropriate amount of global symmetry to make the theory conformal.   In particular, he found that in all these conformal cases, a dual weakly coupled theory emerges from the very strongly coupled regions in the moduli space.  The weakly coupled gauge theory is coupled to interacting $\mathcal N=2$ SCFT with no exactly marginal deformations called $T_N$.  By rearranging these building blocks, he was able to generate a wide class of new $\mathcal N=2$ generalized quivers.  Providing a Seiberg-Witten curve for all these generalized quivers, Gaiotto gave a brane description of these gauge theories as $N$ M5-branes wrapped on a Riemann surface \cite{gaiotto}.  This brane construction provided him with a recipe for constructing four dimensional gauge theories as a compactification of six dimensional $(2,0)$ SCFT of $A_{N-1}$ type. 

\section{Multiple $D3$-branes probing seven-branes on a Taub-NUT background}\label{model2section}

Based on our understanding of Sen \cite{senF} and Banks-Douglas-Seiberg's \cite{bds} embedding of Seiberg-Witten theory \cite{sw1,sw2} in F-theory \cite{vafaF}, it is natural to ask if it is possible to capture Gaiotto's $\mathcal N=2$ $d=4$ superconformal gauge theories and their associated dualities \cite{gaiotto} in F-theory.  The aim is to obtain an even more geometric picture than what M-theory already provided and see if the embedding could shed some light on new physics.  We will see in this section that our construction succeeds in doing just that. 

\begin{figure}[htb]
        \begin{center}
\includegraphics[height=6cm]{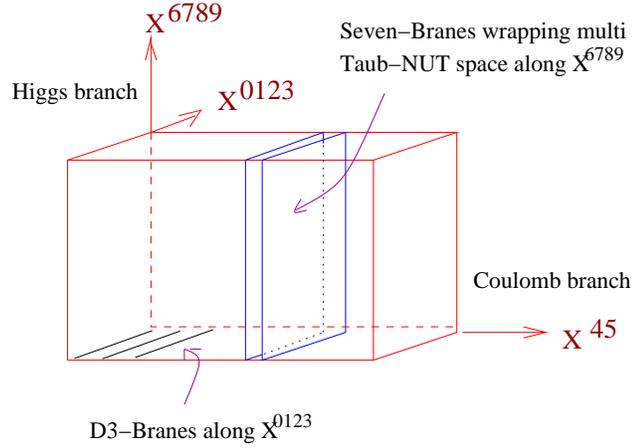}         \end{center}
          \caption{Multiple $D3$-branes probing seven-branes on a multi Taub-NUT geometry.}          \label{taubNUTbranefig}
        \end{figure}

This is achieved by first deforming the background of the original type IIB $D3/D7$ system by replacing the ${\mathbb R}^4$ for a more non-trivial
four-dimensional space. The simplest non-compact example is an ALE space ${{{\mathbb R}^4}\over {\mathbb Z}_2} $, or more locally, a Taub-NUT space- see discussion around \eqref{deformation1}.  Asymptotically, $\mathbb R^4/\mathbb Z_2$ is $\mathbb R^3\times S^1$.  If the radius of the circle is of finite size, we call the geometry an ALF (asymptotically locally flat) space whereas it is called an ALE (asymptotically locally Euclidean) space if $R\to\infty$.  Although we will refer to $\mathbb R^4/\mathbb Z_2$ or more generally to $\mathbb R^4/\mathbb Z_k$ as ALE space throughout this text, we will alway take the constant radius limit of the asymptotic circle of $\mathbb R^4/\mathbb Z_k$.   
We thus obtain:
\bg\label{joba2}
&&{\rm Type~ IIB ~on} ~~{{\bf T}^2\over \Omega\cdot (-1)^{F_L}\cdot {\cal I}_{45}}
\times {{{\mathbb R}^4}\over {\mathbb Z}_2} \times {\mathbb R}^{0123}
~ = ~ \nonumber \\
&&{\rm F~ Theory ~on}~~ {\rm K3} \times {{{\mathbb R}^4}\over {\mathbb Z}_2} \times {\mathbb R}^{0123}
\nd
probed by a single $D3$-brane.  
As we will discuss below, once we increase the number of $D3$-branes, we can also
make the Taub-NUT space multi-centered without breaking further supersymmetries. The brane configuration is
given by {figure \ref{taubNUTbranefig}} and {table \ref{directionstable}}.
\noindent In our configuration the $D3$-branes are oriented along the spacetime $x^{0,1,2,3}$ directions. The seven-branes
(and seven-planes) are parallel to the $D3$-branes and also wrap multi-centered Taub-NUT space oriented along
$x^{6,7,8,9}$. Therefore as before, the Coulomb branch will be the complex $u \equiv x^4 + ix^5$ plane, whereas the
Higgs branch will be along the Taub-NUT space.  The supersymmetry of this configuration still remains ${\cal N} = 2$ as one can incorporate a Taub-NUT space in a $D3/D7$ system without further breaking susy.  A simple trick to prove this statement is to T-dualize our construction\footnote{At low energy i.e at far IR-this will be refined latter} along the $x^6$ direction - since T-duality does not break susy.  There, one finds that our brane picture is given by Table \ref{dswtdual} which the famous Witten's construction \cite{WittenMT} preserving eight supercharges. 
\begin{table}[htb]\label{dswtdual}
 \begin{center}
\begin{tabular}{ccc}
Type IIB\ \ (DSW$_{IR}$) &$\underrightarrow{T_{x^6}}$ &Type IIA\ \ (Witten) \\ 
&&\\\hline
D3\ \ 0123\ .\ .\ .\ .\ .\ . & &D4\ \ 0123\ .\ .\ 6\ .\ .\ \ \ \ \ \ \\
D7\ \ 0123\ .\ .\ 6789\ \  & &D6\ \ 0123\ .\ .\ .\ 789\ \ \ \ \  \\
TN\ \ .\ .\ .\ .\ .\ .\ 6789\ \ \ & &$NS_5$\ 012345\ .\ .\ .\ .\ \ \ \ \ \ \ \\ 
&&\\\hline
  \end{tabular}
   \end{center}
  \caption{Naive link between our model in the IR and type IIA existing construction} 
  \label{TdualIR}
\end{table} 
\newpage  
\section{Brane anti-brane on a Taub-NUT background \label{subsecDDbarTN}}

\begin{figure}[htb]
        \begin{center}
\includegraphics[height=6cm]{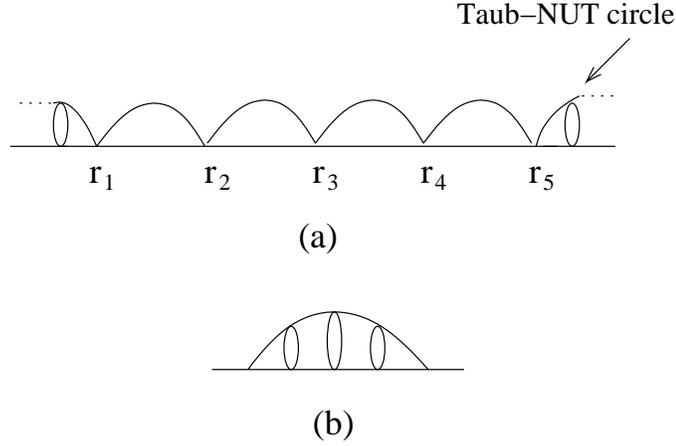}        \end{center}
        \caption{In fig (a) the singularities of Taub-NUT space are shown. The compact direction is the
Taub-NUT circle that is fibered over the base. The various points at which the circle (which is along $x^6$ direction
in the text) degenerate are the singular points. %These singular points form a
Between two singular points form a
two-cycle, as shown in fig (b), once we assume that the $x^6$ circle is degenerating along a line parametrized by
the $x^7$ coordinate. Thus the ${\bf P}^1$'s are labelled by $x^{6, 7}$ coordinates in the text.}
\label{tnfrac}\end{figure}

Let us start with a singular ${\mathbb Z}_2$ ALE space
along directions
$x^{6,7,8,9}$. The node is really a 5-plane filling the remaining
directions. Close to the singular point $x^{6,7,8,9}=0$, the space can be replaced by
a 2-centre (separated in $x^6$) Taub-NUT metric with coincident (in $x^{0,1,2,3}$) centres see figure (\ref{tnfrac}). This
is equivalent to saying that we have two coincident Kaluza-Klein
monopoles. We also know \cite{aspinwall} that the ${\mathbb Z}_2$ orbifold hides
half a unit of $B_{\rm NS}$ flux through the shrunk 2-cycle
$\Sigma$. The four moduli associated to this ALE space are three
geometrical parameters, which can be thought of as the blowup of the
ALE to form a smooth Eguchi-Hanson metric, and the $B_{\rm NS}$ flux \cite{aspinwall}.

Take a $D3$-brane transverse to the ALE space, filling the
directions $x^{0,1,2,3}$. (More generally we start with $r$ such
$D3$-branes.) When the ALE space is singular, the
world-volume theory of the 3-brane has two branches: a Higgs branch,
when the brane is separated from the singularity along
$x^{6,7,8,9}$, and a Coulomb branch when the brane hits the
singularity and dissociates into a pair of fractional branes which
can move around only in the $x^{4,5}$ directions. However, if the ALE
space is blown up, then the Coulomb branch gets disconnected from the Higgs branch because the 3-brane cannot
dissociate supersymmetrically into pair of fractional branes.

The fractional $D3$-brane is interpreted as D5-brane wrapped on a ${\bf P}^1$ with fluxes or
$\overline{\rm D5}$-brane wrapped on a ${\bf P}^1$ with different choice of fluxes (see details
below). Therefore an integer $D3$-brane would be a pair of five-branes whose
D5-brane charges cancel, hence they are really a ${\rm D5}$-$\overline{\rm D5}$ (D5-brane --
anti-D5-brane) pair (see also \cite{mukdas}, \cite{PolchinskiMX, poln3, aharonyn2}
where a somewhat similar model has been discussed).
However, they carry $D3$-brane charge proportional to the relative difference of five-brane fluxes by virtue of
the Chern-Simons coupling on D5-branes. Denoting the world-volume gauge field
strength on the D5-brane by $F_1$, we have the coupling
\bg\label{cscoupling}
\int (B_{\rm NS} - F_1)\wedge C_4
\nd
where $C_4$ is the self-dual 4-form potential in the type IIB
string. At the orbifold point we have $\int_\Sigma B_{\rm NS} = {1\over 2}$ and
hence half a unit of $D3$-brane charge. The $\overline{\rm D5}$ (anti-D5-brane) (whose world-volume gauge field strength is
denoted by $F_2$) will have a coupling
\bg\label{antics}
-\int (B_{\rm NS} - F_2)\wedge C_4.
\nd
Now let us also turn on a
world-volume gauge field strength $F_2$ on the anti-D5-brane and give it
a flux of $+1$ unit through the vanishing 2-cycle $\Sigma$ (more
generally, we assign unit flux to the relative gauge field $F_- = F_2
- F_1$). In this configuration, the ${\rm D5}$-$\overline{\rm D5}$ pair has
in total $D3$-brane charge equal to 1, or more generally:
\bg\label{charge}
\int (F_2 - F_1) \wedge C_4 \equiv \int_{\Sigma \times {\mathbb R}^{0123}} F_- \wedge C_4.
\nd

In a slightly more generalized setting with multi Taub-NUT space the situation is somewhat similar. To see this, let
us consider a Taub-NUT space with $m$ singularities as shown in {figure \ref{tnfrac}}. Once we bring the $D3$-branes
near the Taub-NUT singularities, they decompose as $m$ copies of D5-${\overline{\rm D5}}$ wrapping the various two-cycles
of the Taub-NUT space. Each of the wrapped $k$'th
D5's can be assumed to create a fractional $D3$-brane on its world-volume via
the world-volume $F_{1,k}$ fluxes by normalising the total integral of $F_1$ over all the two-cycles to equal the number
of integer $D3$-branes.
The ${\overline{\rm D5}}$ branes, on the other hand, are used only to cancel the
D5 charges as their world-volume fluxes are taken to be zero. These fractional $D3$-branes can now move along the
Coulomb branch as expected. T-dualising this configuration gives us D4-branes between the NS5-branes which may be
broken and moved along the Coulomb branch, see table (\ref{dswtdual}). The above way of understanding the fractional branes has two immediate
advantages:

\begin{itemize}
\item Since every ${\bf P}^1$ of the multi Taub-NUT space is wrapped by D5-${\overline{\rm D5}}$, 
and the system is symmetrical, one is restricted to
switching on same gauge fluxes on each of the ${\bf P}^1$'s. However for non-compact ${\bf P}^1$'s  %(\textbf{explain what is meant by non-compact $P^1$})
this restriction doesn't hold as the wrapped branes give rise to flavors and not colors, and one may switch on different fluxes.
This will be used to understand the Hanany-Witten
brane creation process later in the text.
\item If we are interested in non-conformal scenarios ($\beta=N_f-2N_c\neq 0$), one way to break the balance between the number of color branes $N_c$ and the number of flavor branes $N_f$ is to manually wrap {\it additional} D5-branes on each of the ${\bf P}^1$.  Since for the conformal case,  the number of D5-${\overline{\rm D5}}$ on each of the ${\bf P}^1$'s are the same, the presence of additional D5-branes  will break conformal invariance leading to
cascading theories\footnote{Instead, if we wrap
additional ${\overline{\rm D5}}$'s on the ${\bf P}^1$'s, they will break supersymmetry. Also the number of
D5-${\overline{\rm D5}}$'s on each ${\bf P}^1$ should remain same so as to cancel the tachyons across each
wrapped ${\bf P}^1$'s.}. 
\end{itemize}
In the above analysis, we assumed that quantum corrections were taken into account and that the $O_7$-planes dissociated into $(p,q)$ 7-branes. 
%In the above analysis of fluxes, we have been ignoring one subtlety related to the
%orientifold action $\Omega$ discussed at the beginning of this section. Due to the orientifold
%action we expect all $B_{\rm NS}$ and $B_{\rm RR}$ fluxes that do not have one component along the
%$x^{4,5}$ directions to be projected out. Therefore it would seem that due to the orientifold action
%the $D3$-branes cannot apparently dissociate into pairs of ${\rm D5}$-$\overline{\rm D5}$ branes. However as is well known,
%quantum corrections in F-theory can take us away from the orientifold point, and we can study the theory
%completely in terms of {\it local} and {\it non-local} (\textbf{understand this})seven-branes without resorting to orientifold planes.  Therefore we will 
Henceforth we will assume that the F-theory background probed by the $r$ $D3$-branes is described
completely in terms of the seven-branes wrapped on multi Taub-NUT space which, alternatively, would also mean
that the moduli space is a ${\bf P}^1$ with 24 transverse seven-branes (not to be confused with the $P^1$ of $TN_k$), i.e:
 \bg\label{t2def}
{{\bf T}^2\over \Omega\cdot (-1)^{F_L}\cdot {\cal I}_{45}} ~ \longrightarrow ~ {\bf P}^1.
\nd
Thus going to the Coulomb branch, by tuning all of $x^{6,7,8,9}$ to 0,
the picture is somewhat different. At this point a $D3$-brane splits
into a pair of fractional branes which can move independently along
$x^{4,5}$. 
%(The geometric orbifold singularity now cannot be blown up any more. \textbf{needed?})
On the T-dual side, one understands this as regular D4-branes, stretched along 2 NS5-branes on a compactified $x_6$ direction.  The D4-branes split into two partially wrapped pieces on each side of the $x^6$ circle (now fractional D4-branes), moving independently along the Coulomb branch $x^{4,5}$ i.e. along the two NS5-branes.  Of course, this is simply the classical picture, in the correct non-perturbative scenario (reproducing our F-theory picture) the D4-branes and NS5-branes form thin M5-branes tubes. . 

%This is easy to see on the T-dual type IIA side, where the
%D4-brane splits into two pieces that stretch along the two intervals
%between the two NS 5-branes (one from each side of the $x^6$ circle).
%These partially wrapped 4-branes can move independently along the
%NS 5-branes, namely in the $x^{4,5}$ directions. 

To summarize, the type IIB picture on the Coulomb branch is that the
relative world-volume gauge field strength $F_-$ on the ${\rm D5}$-$\overline{\rm D5}$ pair must be turned on over the 2-cycle $\Sigma$ and
gives rise to a 3-brane in the space transverse to that cycle. The
spacetime $B_{\rm NS}$ flux over $\Sigma$ changes the relative tensions of
the wrapped D5-brane and anti-D5-brane keeping the total constant. The directions of
various branes and fluxes in out set-up therefore is given in table \ref{directionstable}. Supersymmetry is preserved because the
${\rm D5}$-$\overline{\rm D5}$ pairs wrap vanishing 2-cycles of the multi Taub-NUT space, in addition to the conditions mentioned
earlier. There are also additional % Ramond-Ramond (just for clarity and pedagogy, we could add this?)
background fluxes like axion-dilaton, two-forms and four-forms. The metric of the
Taub-NUT space will be deformed due to the backreactions of the branes and fluxes, that we will discuss later.
All these effects conspire together to preserve ${\cal N} = 2$ supersymmetry on the fractional probe $D3$-branes. Note that
if the Taub-NUT cycles are blown-up then, in the presence of the seven-branes, supersymmetry will be broken.

\begin{table}[h!]
 \begin{center}
\begin{tabular}{|c||c|c|c|c|c|c|c|c|c|c|}\hline Directions & 0 & 1 & 2
& 3 & 4 & 5 & 6 & 7 & 8 & 9 \\ \hline
D5 & --  & --   & --   & --  & $\ast$  & $\ast$ & -- & -- & $\ast$
& $\ast$\\  \hline
$\overline{\rm D5}$ & --  & --   & --   & --  & $\ast$  & $\ast$  & --  & -- &
$\ast$  & $\ast$\\  \hline
D7 & --  & --   & --   & --  & $\ast$  & $\ast$ & -- & -- & --  & --
\\  \hline
Taub-NUT & $\ast$  & $\ast$  & $\ast$  & $\ast$ & $\ast$  &
$\ast$  & --  & -- & --  & -- \\  \hline
Fluxes & $\ast$  & $\ast$  & $\ast$  & $\ast$ & $\ast$  & $\ast$
& -- & -- & $\ast$ & $\ast$ \\  \hline
  \end{tabular}
\end{center}
  \caption{The orientations of various branes and fluxes in out set-up. The dashed lines for the
branes are the directions parallel
to the world-volume of the branes; and for the fluxes and Taub-NUT space are the directions along which we have
non-trivial fluxes and metric respectively. The stars denote orthogonal spaces. Supersymmetry is preserved in the
presence of vanishing cycles and background fluxes.}% XXX Note that `f' in the first column stands for fractional brane.}
  \label{directionstable}
\end{table}

\section{Anomaly inflow, anti-GSO projection and brane transmutation \label{subsecTimeVary}}

There is an interesting subtle phenomenon that
happens to our system when we switch on a time-varying vector potential $A_\mu(t)$ along the Taub-NUT space. 
%(\textbf{check this: describe T-dual $=$ moving $D7$-branes across NS5? })%% Mention something about Euler number and b^1  somewhere for completeness and friendliness?
However before we go about discussing this in detail, we want to point out
an important property of the
underlying Taub-NUT space, namely, the existence of a normalizable harmonic two form $\Omega$. For $m$-centered
Taub-NUT there would be equivalently $m$ normalizable harmonic forms $\Omega_i, i = 1, 2, \cdots , m$. The existence
of these harmonic forms are crucial in analyzing the phenomena that we want to discuss. %% For consistency, use $m$ for number of centers.

To see what happens when we switch on time-varying Wilson line, note first that
the seven-brane wrapping the Taub-NUT space will give rise to a $D3$-brane
bound to it. The charge of the $D3$-brane is given by the non-trivial
$B_{\rm NS}$ background on the Taub-NUT. To see this, consider some of
the couplings on the world volume of the $D7$-brane (we are neglecting
constant factors in front of each terms):
\bg\label{coupdseven}
\int *C_0 ~+~ \int C_4 \wedge F \wedge B_{\rm NS} ~+~
\int  C_4 \wedge F\wedge F ~+~ \cdots.
\nd
These couplings are derived from the Wess-Zumino coupling $\int C\wedge
e^{B-F}$, where $C$ is the formal sum of the RR potentials.  The first
term $\int *C_0$ gives the charge of the $D7$-brane.

Coming back to our phenomena, notice that we cannot turn on a flat
connection on this space. Instead, a self-dual connection can be
turned on. This self-dual connection is of the form:
\bg\label{selfdual}
F= dA = \Omega,
\nd
where $\Omega$ is the unique normalizable harmonic two-form on the
Taub-NUT space. This harmonic two form, being normalizable, goes to
zero at infinity, hence we have a flat connection there. Recall that at infinity, the multi-centered Taub-NUT space asymptotes to $\mathbb R^3\times S^1$ therefore, at infinity, the flat connection corresponds to a Wilson line on that $S^1$.  
%there is an $S^1$ and therefore the flat connection corresponds to
%a Wilson line
%\footnote{Multi-center Taub-NUT also known as asymptotically locally flat (ALF) space,
%locally asymptotes at infinity to ${\mathbb R}^3 \times {\bf S}^1$. When the size of the ${\bf S}^1$ goes to
%infinity, we get the asymptotically locally Euclidean (ALE) space.}.

The above choice of background \eqref{selfdual} however doesn't take the fluctuations of gauge fields into account.
A more appropriate choice for our case is to decompose the field strength $F$ as
\bg\label{fdecom}
F = \Omega + F_1.
\nd
instead of just \eqref{selfdual}.
Now $F_1$ will appear as a gauge field on the $D7$ (or Taub-NUT plane).
Inserting \eqref{fdecom} in \eqref{coupdseven} and integrating out $\Omega$,
we get the required $D3$-brane charge (see also \cite{mukdas} for more details).
This confirms that a bound state of a $D3$ with the $D7$-brane appears once we
switch on a self-dual connection (which is of course the Wilson line for our case).

However the situation at hand demands a {\it time-varying} gauge field on the
world volume of the $D7$-brane. A typical time-varying gauge field $A_\mu$ can be
constructed from $F_1$ in \eqref{fdecom} by making it time-dependent. Such a time-varying gauge
field creates a chiral anomaly along the $S^1$ at the
asymptotic region of the Taub-NUT space. This $1+1$ dimensional anomaly is of the form \cite{BDGreen}
\bg\label{anomaly}
\int d^2x ~\omega \epsilon^{ab} \partial_a A_b,
\nd
 where $\omega$ is the gauge transformation parameter. Another way to see this anomaly is
to dualize the $D7$-brane and the Taub-NUT space into a D6/D4 system oriented along
$x^{0,1,2,3,4,5,6}$ and $x^{0,6,7,8,9}$ respectively\footnote{Use the
following set of dualities to go from one picture to another:
T-dualities along $x^{6, 1, 2, 3}$ then a S-duality followed by another
T-duality along $x^6$.}. The chiral anomaly is along the
$x^6$ direction.

It is suggested in \cite{mukdas} that the term cancelling the aforementioned anomaly is given by:
\bg\label{anomterm}
S = \int\, G_5\wedge A\wedge F
\nd
on the world volume of the seven-brane. Here $G_5 = dC_4$, the pullback of the
background four-form, in the absence of any source. The cancellation
takes place via anomaly inflow. We have a coupling, \eqref{anomterm}, in
$(7+1)d$ spacetime. Along a $(1+1)d$ subspace of this, chiral fermions
propagate and give rise to the anomaly \eqref{anomaly}.
Since $D3$-brane is
the source for $G_5$, we find that changing the Wilson line
produces a change of flux of $G_5$. In other words, a gauge
transformation $\delta A = d\omega$ on the world-volume will vary
\eqref{anomterm} by:
\bg\label{vary}
-\int dG_5 \wedge (\omega F)
\nd
Since $dG_5 \ne 0$ in the presence of a source of $G_5$ flux\footnote{Since $G_5$ is self-dual, this switches on
a $D3$-brane with orientations along $x^{0,1,2,3}$ directions.},
we end
up with:
\bg\label{anomcancel}
\delta S = - \int d^2 x~ \omega\, \epsilon^{ab}\del_a A_b
\nd
resulting in the inflow which cancels the anomaly \eqref{anomaly} by creating a $D3$-brane.
%\footnote{Another way to
%see this is the following. Switching on a $D3$-brane amounts to switching on an instanton action of the form
%$${\cal L}_{\rm eff} = \int d^4y~\theta ~\epsilon_{abcd} F^{ab}F^{cd}$$ along the Taub-NUT world volume. Here
%$\theta$ is the remnant of the $D3$-brane term in the CS coupling of the $D7$-brane or the more popular, $\theta$-term
%of gauge theory.
%Now as shown by \cite{callan} a gauge transformation will effectively give us
%$$\delta_\omega {\cal L}_{\rm eff} = -\int d^2y ~\omega \epsilon^{ab} F_{ab}$$ with the correct minus sign to
%cancel the anomaly. Incidentally if the above action becomes non-abelian, exactly similar computation will again
%cancel the underlying anomaly. Therefore the upshot is that, switching on a $D3$-brane
%oriented along $x^{0,1,2,3}$ directions will cancel the gauge anomaly in the system.}.

The story is however {not} complete. There is an {\it additional} phenomena that happens
simultaneously that actually reduces the number of $D3$-branes instead of increasing it (as we might
have expected from the above discussion). This additional phenomena relies on the dissociation
of the $D3$-branes into ${\rm D5}$-$\overline{\rm D5}$ pairs discussed in the previous subsection. Recall that the
tachyon between the D5 and the $\overline{\rm D5}$ is cancelled for $F_- \equiv F_2-F_1=\pm 1$. Here we set
\bg\label{f2f1}
F_2 = 1, \qquad F_1 = 0,
\nd
which also implies that $F_- = 1$ in \eqref{charge}, giving rise to a unit $D3$-brane charge.
\begin{figure}[htb]
        \begin{center}
\includegraphics[height=4cm]{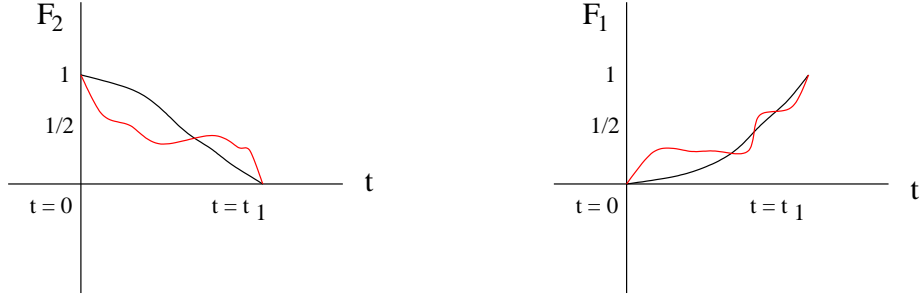}        \end{center}
        \caption{The gauge fluxes on the D5 and the $\overline{\rm D5}$ branes are changed with time. The red and the
black lines denote two different changes with the same initial and final values.} \label{fluxchange}\end{figure}

\begin{figure}[htb]
     \begin{center}
\includegraphics[height=4cm]{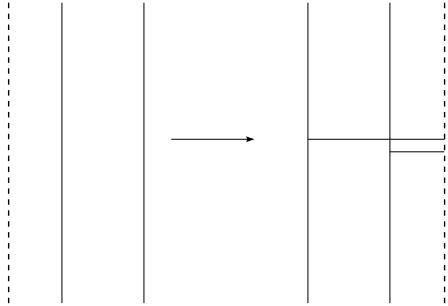}   
     \end{center}
        \caption{HW brane creation: solid vertical lines are NS5-branes (23456), dashed vertical line is D5-brane (56789), solid horizontal line is a $D3$-brane along (156) .} \label{HW}
        \end{figure}

\begin{figure}[htb]
        \begin{center}
\includegraphics[height=4cm]{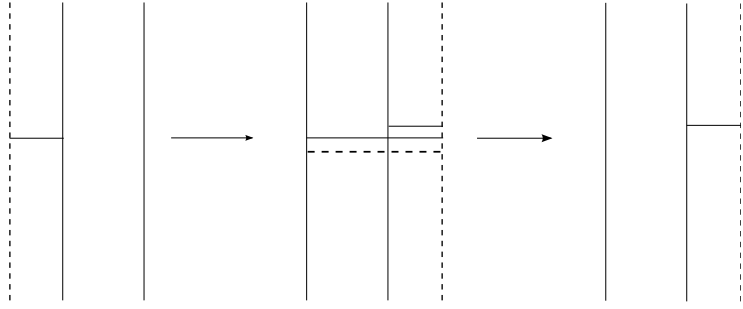}   
     \end{center}
        \caption{HW brane creation: solid vertical lines are NS5-branes (023456), dashed vertical line is D5-brane (056789), solid horizontal line is a $D3$-brane along (0156), horizontal dashed line is an anti-$D3$-brane (0156)} \label{hwwtfluxe}\end{figure}

Now imagine that we change the world volume fluxes as described in figure \ref{fluxchange}, starting with (\ref{f2f1}): 
\begin{table}[htb]
 \begin{center}
\begin{tabular}{ccc}
$t=0$&$F_1=0,$&$F_2=1$\\
&&\\
$t=t'$&$F_1=\frac{1}{2},$&$F_2=\frac{1}{2}$\\
&&\\
$t=t_1$&$F_1=1,$&$F_2=0$\\
&&
  \end{tabular}
\end{center}
  \caption{Variation of world-volume fluxes with time} 
  \label{Table3}
\end{table}
\ \\
where $(t=0)<t'<(t=1)$.  Since the $D3$-brane charge is given by  \eqref{charge} , we see that at $t=t'$, the $D3$-brane charge vanishes.  On the other hand, when $t=t_1$, the relative difference in the fluxes $F_-=F_2-F_1$ is negative, indicating that the ${\rm D5}$-$\overline{\rm D5}$ are generating an $\overline{\rm D3}$.  Note that since the norm of $F_-$ is still equal to 1, the tachyon between the D5 and the $\overline{\rm D5}$ is still massless and the supersymmetry is not broken.  We have just seen that variations in the flux of ${\rm D5}$-$\overline{\rm D5}$ transmute a $D3$-branes into a $\overline{\rm D3}$-brane.
%In the beginning, $(F_1, F_2)=(0,1)$ at $t=0$, and then at some time they both become $\frac{1}{2}$, finally at $t=t_1$ $(F_1, F_2)=(1,0)$.
%As seen in the figure, when
%both the $F_i$ fluxes approach the mid value $F_i = {1\over 2}$ the D3-brane charge in \eqref{charge} vanishes.
%However at $t = t_1$ when
%\bg\label{fcha}
%F_2 ~ = ~ 1~ \to ~ {1\over 2} ~ \to~ 0, \qquad F_1~ = ~ 0~ \to ~ {1\over 2} ~ \to~ 1
%\nd
%then one may easily check that the ${\rm D5}$-$\overline{\rm D5}$ pair give rise to an anti-D3-brane ($\overline{\rm D3}$). However since the {\it difference}
%between the fluxes have still remained 1, the tachyon between the D5 and the $\overline{\rm D5}$ continue to remain massless
%and the supersymmetry doesn't get broken by this process. Therefore this process transmutes a D3-brane into an
%$\overline{\rm D3}$! 
Thus switching on a time-varying Wilson line has the following two effects:
\begin{itemize}
\item Chiral anomaly cancellation via anomaly inflow and creation of a new $D3$-brane.
\item $D3$-brane transmutation to an $\overline{\rm D3}$ brane via flux change.
\end{itemize}
\noindent Together these two effects would remove one of the existing $D3$-brane in the system. Therefore the color degree of freedom would change via this process. If we do this multiple times, we can reduce the number of $D3$-branes in the model. One way I like to think about the above two phenomena is in the brane language of Hanany-Witten \cite{HananywittenHW}. 
In was shown in \cite{BDGreen} that turning on a time varying gauge field $A_1(t)$ on intersecting D5-branes along (12345) and (16789) is T-dual to the relative motion of orthogonal D4-branes along the transverse direction $x^1(t)$, creating a fundamental string every time the fourbranes cross each other.  \cite{BDGreen} showed that their setup is T-dual to that of Hanany-Witten's.  Since we know that our brane construction is also T-dual to Hanany-Witten's model- see table \ref{dswtdual}, one can think of the chiral anomaly cancellation via anomaly inflow and creation of a $D3$-brane simply as brane creation in Hanany-Witten figure \ref{HW}.  
If we combine this phenomenon with flux change which transmute a $D3$-brane into an anti-$D3$ brane, we obtain brane annihilation as shown in figure (\ref{hwwtfluxe}).
 %The variation of the time dependent gauge field $A_6(t)$ on the worldvolume of a $D7$-brane is T-dual to the variation of the position of $D6$-brane along the transverse $x^6$ direction [cite: Bachas, Douglas, Green, hep-th/9705074].  In the presence of TN space in type IIB, the one can think of turning on a time varying gauge field on the worldvolume of the $D7$-brane wrapped on a 2-cycle of the Taub-NUT space as moving a 
We will say a few more words on the Hanany-Witten mechanism when we will discuss about the brane networks.  
%Once we go to the brane network model, we will show that the above phenomena are related to the brane annihilation and the Hanany-Witten effect.
%
\noindent Of course
in the {\it absence} of the Taub-NUT space, none of the above arguments would work, and
so there would be no brane creation/annihilation. This is perfectly consistent with our
expectation.

Coming back to the system under study, imagine now that we switch on gauge fluxes $F = n \Omega$. This would imply that we have
two new sources of the form:
\bg\label{nsour}
{n^2\over 2} \int C_4, \qquad {\rm and} \qquad n\int C_4 \wedge (B_{\rm NS} - F_1).
\nd
The latter doesn't break supersymmetry as was explained in \cite{mukdas}. In fact overall the
supersymmetry will never be broken if we take $r$ $D3$-branes and we simultaneously consider
$m$-centered Taub-NUT. Therefore the $r$ $D3$-branes wrap $m$ different vanishing 2-cycles.

From the above discussions we see that we have two models that are {\it dual} to each other while preserving
supersymmetry. The duality criteria for our case can be presented in the following way:
\begin{itemize}
\item $r$ $D3$-branes probing seven-branes wrapping a $m$-centered Taub-NUT space. The $D3$-branes dissociate
as $m$ copies of ${\rm D5}$-$\overline{\rm D5}$
pairs that move along the Coulomb branch as depicted in
{ figure \ref{wrappedTN}}.
The seven-branes could
be arranged to allow for any global symmetries and axion-dilaton moduli, including the conformal cases.
\item Time-varying
self-dual connections on the seven-branes along the Taub-NUT direction that create {\it and} transmute
${\rm D5}$-$\overline{\rm D5}$ sources by changing the $F_i$ in \eqref{cscoupling} and \eqref{antics}. Due to this
some $D3$-branes may annihilate,
thereby changing the local and possibly the global symmetries
of the model.
\end{itemize}
Our claim therefore is the following. The above two dual descriptions, coming from chiral anomaly cancellation,
$D3$-brane creation and $D3$-brane transmutation, are related by the recently proposed Gaiotto dualities. In the
next subsection we will supply more evidences for this conjecture.

Note that the total moduli in both the models are exactly similar, although both color and flavor degrees of freedom
may apparently differ. The seven-branes could be arranged such that we could either have F-theory at constant
couplings {\it a la} \cite{DM1}, or non-constant couplings. However due to the underlying F-theory constraints, the 
flavor degrees of freedom remain below 24 although the color degrees of freedom could be anything arbitrary.
In addition to that there is also a M-theory uplift of our model that is quite different from the M-theory brane
constructions studied by Witten \cite{WittenMT} and Gaiotto \cite{gaiotto}. We will discuss this soon.

\section{Mapping to Gaiotto theories and beyond}
Now that we deformed the background and presented our model, let us try to reproduce some of the Gaiotto gauge theories construction \cite{gaiotto}.
%After having constructed our model, let us try to map to some of the Gaiotto's constructions. 
Our first map will be
to the brane network model studied by \cite{bbt} recently. We will then argue how gravitational duals for
our models, at least in the conformal limit, may be derived. These gravity solutions should be compared to the 
recently proposed gravity duals given in \cite{GM}. We will see that our model can be extended to the non-conformal cases
in two ways: by moving the seven-branes around or by wrapping by hand fractional fivebranes. In fact we will see an interesting class of {\it cascading} ${\cal N} = 2$
models appearing naturally out of our constructions.
Additionally, new states in the theory could appear in the generic cases when the
$D3$-brane probes are connected by string junctions or string networks. In the later part of this section we will give
some details on these issues, extending the scenario further.

\subsection{Mapping to the type IIB brane network models \label{subsecNetwork}}

Recently the authors of \cite{bbt} have given a set of interesting brane network models that may
explain certain conformal constructions of the Gaiotto models, including ways to see how the Gaiotto
dualities occur from the networks. The obvious question now is whether there exist some regime of
parameters in our set-up that could capture the brane network models of \cite{bbt}.

\begin{figure}[htb]
        \begin{center}
\includegraphics[height=6.5cm]{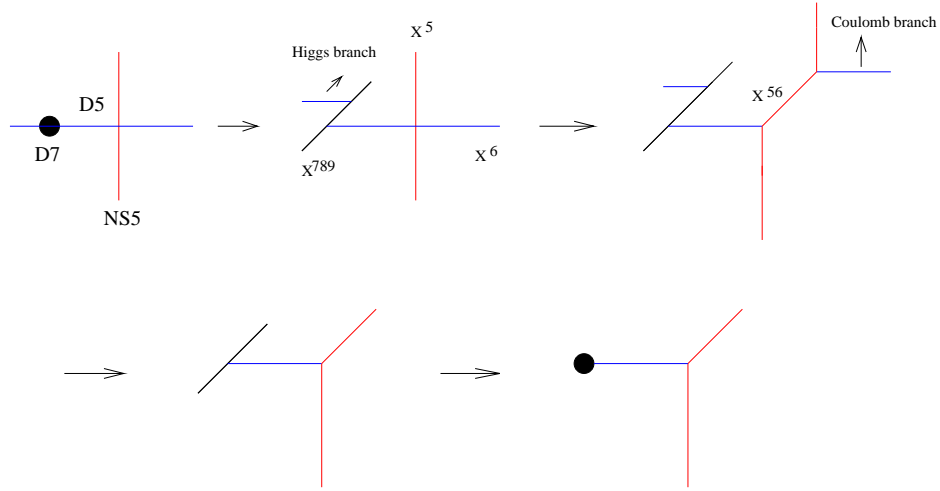}
        \end{center}
        \caption{The simplest brane junction from our Taub-NUT configuration.} \label{TN2I}
        \end{figure}
It turns out the mapping to \cite{bbt} is not straightforward. The orientations of various branes in our
set-up are given in {table \ref{direction2}}. A naive T-duality along $x^4$ and $x^6$
direction will convert the Taub-NUT
space to a NS5-brane oriented along $x^{0,1,2,3,4,5}$ and the $D7$-brane into another $D7$-brane oriented along
$x^{0,1,2,3,4,7,8,9}$. However the ${\rm D5}$-$\overline{\rm D5}$ pairs will continue as ${\rm D5}$-$\overline{\rm D5}$ pairs although with a slightly different
orientation.

%\begin{table}[h!]
% \begin{center}
%\begin{tabular}{|c||c|c|c|c|c|c|c|c|c|c|}\hline Directions & 0 & 1 & 2
%& 3 & 4 & 5 & 6 & 7 & 8 & 9 \\ \hline
%f D5 & --  & --   & --   & --  & --  & $\ast$ & $\ast$ & -- & $\ast$
%& $\ast$\\  \hline
%f $\overline{\rm D5}$ & --  & --   & --   & --  & --  & $\ast$  & $\ast$ & -- &
%$\ast$  & $\ast$\\  \hline
%D5 & --  & --   & --   & --  & --  & $\ast$ & -- & $\ast$ & $\ast$
%& $\ast$\\  \hline
%$D7$ & --  & --   & --   & --  & --  & $\ast$ & $\ast$ & -- & --  & --
%\\  \hline
%NS5 & --  & --  & -- & -- & --  &
%--  & $\ast$ & $\ast$ & $\ast$  & $\ast$ \\  \hline
%  \end{tabular}
%\end{center}
%  \caption{T-duality along $x^4$ direction from {table \ref{D4table}}.}
%  \label{T46table}
%\end{table}

\begin{table}[h!]
 \begin{center}
\begin{tabular}{|c||c|c|c|c|c|c|c|c|c|c|}\hline Directions & 0 & 1 & 2
& 3 & 4 & 5 & 6 & 7 & 8 & 9 \\ \hline
D5 & --  & --   & --   & --  & $\ast$  & $\ast$ & -- & -- & $\ast$
& $\ast$\\  \hline
$\overline{\rm D5}$ & --  & --   & --   & --  & $\ast$  & $\ast$  & --  & -- &
$\ast$  & $\ast$\\  \hline
$D7$ & --  & --   & --   & --  & $\ast$  & $\ast$ & -- & -- & --  & --
\\  \hline
Taub-NUT & $\ast$  & $\ast$  & $\ast$  & $\ast$ & $\ast$  &
$\ast$  & --  & -- & --  & -- \\  \hline
  \end{tabular}
\end{center}
  \caption{The orientations of various branes in out set-up. Same as
the earlier table but now the flux informations are not shown.}
  \label{direction2}
  \end{table}

\begin{figure}[htb]
        \begin{center}
\includegraphics[height=5cm]{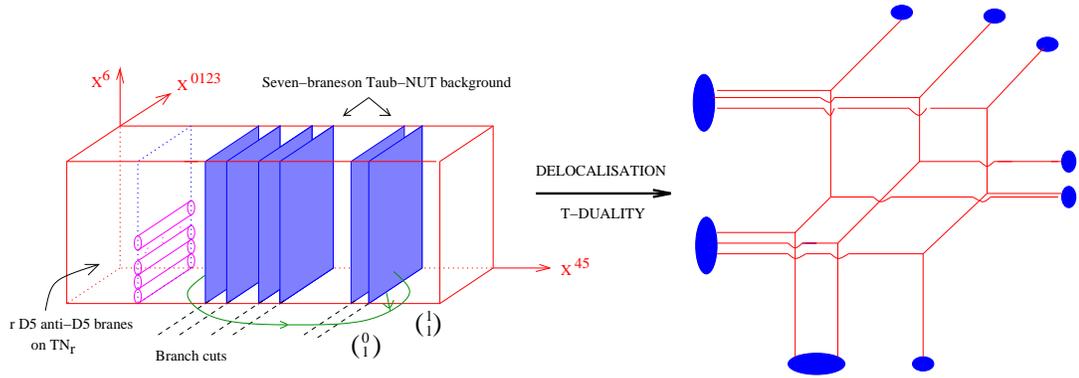}
        \end{center}
        \caption{Under special arrangement of the seven-branes, delocalization and T-dualities map our model
%may map
to the brane network studied in \cite{bbt}. For other configurations there are no simple
map to the brane networks. The blue patches on both sides represent the seven-branes. The $r$ ${\rm D5}$-$\overline{\rm D5}$ pairs
are wrapped on vanishing 2-cycles of a multi Taub-NUT space with $m$ centers. %% r and m??
} \label{bbt}
        \end{figure}

This is not what we would have expected for the model of \cite{bbt}. Furthermore, because of the fluxes as well
as other fields that have non-trivial dependences along the $x^{4,6}$ directions T-dualities along these directions
are not possible. Therefore there is no simple map to the brane network model of \cite{bbt}. However we
can go to a corner of the moduli space of solutions where:
\begin{itemize}
\item[(a)] We are at low energies i.e at far IR, so the ${\rm D5}$-$\overline{\rm D5}$ pairs behave as fractional D3-branes,
\item[(b)] We have delocalized completely along the two T-duality directions $x^{6,4}$.
\end{itemize}
Under these two special cases together, we can T-dualize along $x^{6,4}$ directions to convert our
configuration to the brane network model of \cite{bbt} as depicted in {figures \ref{TN2I}} and { \ref{bbt}}.
In { figure \ref{TN2I}} the configuration in { table \ref{direction2}} is T-dualized following the above criteria
to get to the brane intersection model in the top left of the figure. Motion in the Coulomb branch is precisely the
decomposition of the $D3$-brane into ${\rm D5}$-$\overline{\rm D5}$ pair, such that each of them support a fractional $D3$-brane.
Once we have the fractional $D3$-branes we can move one of them along the $x^{4,5}$ direction\footnote{Of course this is
the generic case. But for wrapped ${\rm D5}$-$\overline{\rm D5}$-branes there could be
situations where in the T-dual set-up the D5-brane may
terminate on NS5-brane (much like the one in \cite{mukdas}).}.
On the other hand we also
need to break the D5-brane on the seven-brane and move this
along the Higgs branch, as depicted in { figure \ref{TN2I}}. This is achieved by expressing the fractional
$D3$-brane (on the D5-brane) as an {\it instanton} on the seven-brane and then further decomposing the instanton
as fractional instantons on the seven-brane. Moving one set of fractional branes along the Higgs branch will
eventually give us the brane junction studied by \cite{bbt} as shown in { figure \ref{TN2I}}.
\cite{KlebanovHB}
Clearly this T-dual mapping works most efficiently with fractional $D3$-branes and ignoring their
${\rm D5}$-$\overline{\rm D5}$ origins. As we saw before, this dissociation is crucial in the
presence of multi Taub-NUT space and therefore the mapping to \cite{bbt} only works under special
circumstances. It also means that once we map our model to \cite{bbt} we may lose many informations
of our model. In particular all the high energy informations, like the presence of ${\rm D5}$-$\overline{\rm D5}$ pairs, fluxes
and massless tachyons are completely lost on the other side. But certain low energy informations do map
from our model to \cite{bbt}. For example a crucial ingredient of \cite{bbt} is the Hanany-Witten
brane creation process that occurs when we move the $D7$-brane across the NS5-brane. The $D7$-brane is located at
$x^6_{(1)}$ and the NS5-brane is located at $x^6_{(2)}$. The relative motion of the $D7$-brane will induce
following T-duality map:
\bg\label{jogini}
\int d^2y~\left[{\partial x^6_{(1)} \over \partial t} - {\partial x^6_{(2)} \over \partial t}\right]
~\longrightarrow~
\int d^2y~\epsilon^{06}\partial_0 A_6,
\nd
which is of course one term of the chiral anomaly $\int \omega \epsilon^{ab}\partial_a A_b$ as we saw before. A
cancellation of the chiral anomaly therefore maps to the brane creation picture of \cite{bbt}, although the
brane transmutation in our model (that relies on the dissociation of $D3$-brane into ${\rm D5}$-$\overline{\rm D5}$ pair) cannot be
seen directly from the T-dual model (although there may exist some equivalent picture).

Another interesting ingredient of \cite{bbt} is the so-called s-rule that preserves supersymmetry. In this
configuration the D5-branes ending on same $D7$-branes must end on {\it different} NS5-branes, i.e not more than one D5-brane may end on a
given pair of NS5-brane and $D7$-brane, otherwise supersymmetry will be broken. At low energy we saw that T-duality can map our model to
\cite{bbt}. The $m$-centered Taub-NUT space can map to the multiple configuration of the NS5-branes. Similarly, ($p, q$) five branes can be understood as explained above. The ${\rm D5}$-$\overline{\rm D5}$ pairs wrap the vanishing cycles of the
multi Taub-NUT geometry and we may keep $r$ pairs of ${\rm D5}$-$\overline{\rm D5}$ with $m$-centered Taub-NUT space. This means that there
may not be a simple map of the s-rule of \cite{bbt} to our set-up. This is understandable
because making a single T-duality to type IIA
along $x^6$, and removing the seven-branes, give us the NS5/D4 configuration where multiple D4-branes can end on
NS5-branes. However in this mapping all information of the {\it non-local} seven-branes are completely lost including
informations about exceptional global symmetries etc. Thus our F-theory model including number of seven-branes captures additional information.

\subsection{The UV/IR picture and gravity duals \label{subsecUVIRgrav}}

In the limit when we take the number of ${\rm D5}$-$\overline{\rm D5}$ pairs to be very large, we expect the near horizon
geometry to give us the gravity duals of the associated theories. One may arrange the seven-branes in such
a way that the axion-dilaton coupling doesn't run. In that case the corresponding theories should be conformal
at least both at UV and IR. Recently Gaiotto and Maldacena \cite{GM} have studied the gravity duals of
some of the Gaiotto models and have provided explicit expressions for the IR pictures. In this subsection we will
provide some discussions on this using our set-up. More detailed derivations will be provided in the sequel to this
paper.

\begin{figure}[htb]
        \begin{center}
\includegraphics[height=6cm]{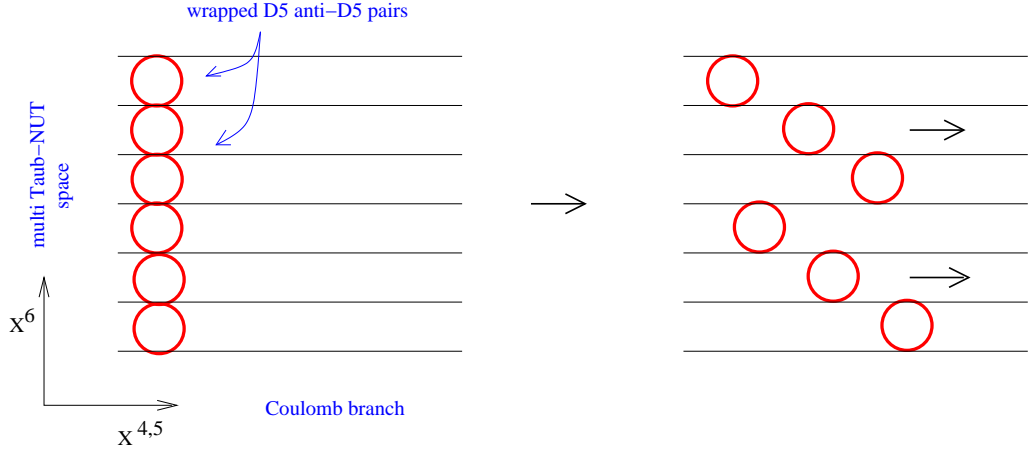}
         \end{center}
       \caption{${\rm D5}$-$\overline{\rm D5}$ branes wrapped on 2-cycles of a multi Taub-NUT space. When each of these
$N$ set of
${\rm D5}$-$\overline{\rm D5}$ pairs wrap vanishing cycles of the multi Taub-NUT space, they can give rise to
$[SU(N)]^m$ gauge groups where $m$ denote the number of vanishing Taub-NUT 2-cycles. In the next figure
various such pairs are broken and moved along the Coulomb branch.}  \label{wrappedTN}
        \end{figure}

One aspect of the gravity dual should be clear from the F-theory model that we present here: the UV of the theory
should be different from the IR. In fact we expect the UV to be a six-dimensional theory whereas the IR should be
four-dimensional. This is bourne out from the following observations. In the UV, when the system is probed using
high energy wavelengths, the complete large $m$ Taub-NUT singularities should be visible. Therefore UV description
should be given by a large $r$ ${\rm D5}$-$\overline{\rm D5}$ pairs wrapped on the vanishing 2-cycles of the muti Taub-NUT space.
Because of the presence of ${\rm D5}$-$\overline{\rm D5}$ pairs we expect the theory should become six-dimensional, and by arranging
the seven-branes appropriately the UV should be a 6d SCFT.

On the other hand the IR is simple. Since at IR we are probing the geometry with large wavelengths, the subtleties
of the geometry will be completely washed out and we will see only a simple Taub-NUT space with no non-trivial cycles.
The ${\rm D5}$-$\overline{\rm D5}$ pairs wrapping this geometry will effectively behave as four-dimensional,
and therefore the IR geometry
should be four-dimensional.

We can make this a bit more precise. The supergravity solution for pairs of ${\rm D5}$-$\overline{\rm D5}$ branes
on a {\it flat}
background can be given in the following way\footnote{It would be interesting to compare the analysis below with the one
done in \cite{poln3} where supergravity solution related to $D7$ and fractional $D3$-branes is studied. Our analysis is
very different from the one in \cite{poln3} as we will be studying the system from its D5-${\overline{\rm D5}}$
perspective, and not from its $D3$ perspective,
so as to capture the UV and IR behaviors.} \cite{townsend1, townsend2, bk}:
\bg\label{bakkarch}
ds^2 =&&-V_1^{-1} V_2^{-1/2}(dx_0 - kdx_7)^2 + V_2^{-1/2}dx^2_2 + V_2^{-1/2} dx_1^2
+ V_2^{-1/2} dx_3^2 \nonumber\\
&& \qquad + V_2^{1/2} (V_1^{-1}dx_6^2 + dx_7^2) + V_2^{1/2} (dx_4^2 + dx_5^2 + dx_8^2 + dx_9^2),
\nd
where stability and supersymmetry requires us to switch on an electric field $F_{0i}$
with $\vert F_{0i}\vert^2 < 1$ and a magnetic field $F_{67}$
with opposite signs on the ${\rm D5}$-$\overline{\rm D5}$ pairs. This is slightly different choice of the
world-volume fluxes compared
to the ones that we took in the previous subsections and in appendix E of \cite{DasguptaSW}. One may however easily
verify that both the
choices result in identical physics\footnote{In this framework
one might worry about the fundamental string oriented parallel to the
seven-branes i.e along $x^6$ direction. This can be dissolved in one of the $D7$-brane and then moved away in the
$u$-plane, so that local ${\cal N} = 2$ supersymmetry remains unaffected. This is equivalent to the statement that
we can go to a frame where only world-volume magnetic field is turned on and the electric field is zero.
In the framework studied in the earlier
sub-section there are only fractional $D3$-branes and no fundamental strings.}.

Due to the existence of an electric field (with say $i = 6$) there would be bound fundamental strings, and due to the
magnetic fields $F_{67}$ there would be bound $D3$-branes. The $D3$-brane charge is then typically given by
$F^{(2)}_{67} - F^{(1)}_{67}$ as we saw before. If we keep the five-branes at the same point in the $u = x^4 + i x^5$
plane but separate the anti five-branes very slightly along the $r = \sqrt{(x^8)^2 + (x^9)^2}$ directions, then
\bg\label{dista}
V_i = 1 + \alpha_i\left[{1\over \vert u\vert^2 + r^2}+  {1\over \vert u\vert^2 + (r- \epsilon)^2}\right], ~
k = \beta\left[{1\over \vert u\vert^2 + r^2} - {1\over \vert u\vert^2 + (r- \epsilon)^2}\right], \nonumber\\
\nd
where $\alpha_{1,2}, \beta$ are functions of $g_s, l_s$; and the number of
fundamental strings, $D3$-branes and five-branes respectively.
Note that in $k$ the two terms come with a relative minus sign, so that when $\epsilon \to 0$, $k$
vanishes.

The above picture is not complete as we haven't yet accounted for the multi Taub-NUT space and seven-branes. Let us
first consider the multi Taub-NUT space oriented along $x^{6,7,8,9}$ directions. The multi Taub-NUT
space modifies the $x^{6,7,8,9}$ directions in the following way:
\bg\label{dhosta}
ds^2_{\rm TN} = \left(1 + \sum_{\sigma}{1\over \vert \vec{w}-\vec{w_\sigma}\vert}\right) d\vec{w}^2 +
\left(1 + \sum_{\sigma}{1\over \vert \vec{w}-\vec{w_\sigma}\vert}\right)^{-1} \left(dx^6 + \sum_\sigma F^\sigma_i
dx^i\right)^2, \nonumber\\
\nd
where $\sigma$ denotes %$m$
Taub-NUT singularities and $\vert \vec{w}\vert = \sqrt{\vert r\vert^2 + (x^7)^2}$
denotes the distance along the Taub-NUT space.

In addition to the Taub-NUT space, we also have the seven-branes distributed in some way to give rise to the
global symmetries in the theory. For generic distribution of the seven-branes the resulting gauge theory is not
conformal. The metric orthogonal to the seven-branes along the $u$-plane is given by the following
expression (see for example \cite{gsvy}):
\bg\label{ortho}
ds^2_u = \tau_2(u) \Bigg\vert \eta^2(\tau(u))\prod_{i=1}^{24} {du\over (u - u_i)^{1/12}}\Bigg\vert^2,
\nd
where $\tau_2(u)$ is the imaginary part of $\tau(u)$ on the $u$-plane and $\eta(\tau)$ is the $\eta$-function (we are
using the notations of \cite{gsvy}).

Now combining \eqref{ortho}, \eqref{dhosta} and \eqref{bakkarch} we obtain the total
background metric. The $dx^6$ component of \eqref{bakkarch} should be replaced by the
$U(1)$ fibration metric of \eqref{dhosta} and the ($dx^4, dx^5$) part of \eqref{bakkarch} should be replaced by the
backreaction from the seven-branes, i.e \eqref{ortho}. Together, the final picture would be pretty involved, and
will take the following form:
\bg\label{totla}
ds^2 =&&-f_1V_1^{-1} V_2^{-1/2}(dx_0 - kdx_7)^2 + f_2V_2^{-1/2}dx^2_2 + f_3V_2^{-1/2} dx_1^2
+ f_4V_2^{-1/2} dx_3^2 \nonumber\\
&& + f_5V_1^{-1}V_2^{1/2} \left(dx^6 + \sum_\sigma F^\sigma_i dx^i\right)^2
+ f_6V_2^{1/2} \tau_2(u) \Bigg\vert \eta^2(\tau(u))\prod_{i=1}^{24} {du\over (u - u_i)^{1/12}}\Bigg\vert^2 \nonumber\\
&& + f_7 V_2^{1/2} dx_7^2 + f_8V_2^{1/2}(dx_8^2 + dx_9^2),
\nd
where we expect ($f_5, f_7, f_8$) to be functions of ($\vert u\vert, \vec{w}$) so that informations about the
multi Taub-NUT space can be captured\footnote{Note that the multi Taub-NUT geometry is deformed due to the
backreactions of branes and fluxes in the background.}.
The other $f_i$ would definitely be functions of $\vert u\vert$ but could
have dependences on other coordinates too. The $V_i$'s now specify the harmonic functions for all the wrapped
D5-$\overline{\rm D5}$ pairs on the multi Taub-NUT two-cycles.

In the above metric we can go to either the conformal or the non-conformal limits. The conformal limits will be
given by some special re-arrangements of the 24 seven-branes whose individual contributions appear in \eqref{totla}.
The non-conformal limits are of course any generic distributions of the seven-branes in \eqref{totla}. In addition, this limit can be probed by adding fractional D5-branes as will be explained in the next section. Each of the
two limits would also have their individual UV and IR behaviors. The IR behavior for both the conformal as well as the
non-conformal limits shouldn't be too difficult to determine from the above form of the metric \eqref{totla}.

At IR we expect that all informations about the multi Taub-NUT space will be washed out (because we are probing
the system with wavelengths larger than the resolutions of the Taub-NUT singularities). This means at IR the
system is probed by $D3$-branes. We also expect
\bg\label{ko}
k ~ \equiv ~ \beta\left[{1\over \vert u\vert^2 + r^2} - {1\over \vert u\vert^2 + (r- \epsilon)^2}\right] ~ = ~ 0
\nd
in \eqref{totla}, as the ${\rm D5}$-$\overline{\rm D5}$ pairs would effectively overlap, and so there would be no $dx_0dx_1$ cross-terms
in the metric. The metric then takes the following form:
\bg\label{dhongsho}
ds^2 &&=  {1\over \sqrt{V_2}}\Big(-f_1 V_1^{-1} dx_0^2 + f_3 dx_1^2 +
f_2 dx_2^2 + f_4 dx_3^2\Big)\\
&&+ \sqrt{V_2}\Bigg[f_5V_1^{-1}\left(dx^6 + \sum_\sigma F^\sigma_i dx^i\right)^2 \nonumber\\
&&+ f_7 dx_7^2 + f_8 (dx_8^2 + dx_9^2)
+ f_6 \tau_2(u) \Bigg\vert \eta^2(\tau(u))\prod_{i=1}^{24} {du\over (u - u_i)^{1/12}}\Bigg\vert^2 \Bigg],
\nd
which is very suggestive of the multi $D3$-brane metric provided certain conditions are imposed on
($f_1, \cdots,  f_4$) at IR. The condition that we want for our case would be the following obvious one:
\bg\label{obvious}
f_1 V_1^{-1} ~ \approx ~ f_i
\nd
with $i = 2, 3, 4$. This is not too difficult to show. In the far IR, as we discussed above, the system is described
by $D3$-branes probing the geometry instead of the ${\rm D5}$-$\overline{\rm D5}$ pairs. This means that the
Taub-NUT geometry is essentially decoupled from the $D3$-brane geometry implying that the metric seen along
the $D3$-brane directions is given by the first line of \eqref{bakkarch} with $k = 0$ i.e $f_1 = f_i$. Additionally the
other warp factor $V_1$ is defined in terms of the fundamental strings, $\alpha_1$, as shown in \eqref{dista}. We can
always make a Lorentz transformation to go to a frame of reference where only world-volume magnetic fields,
$F_{67}^{(1, 2)}$, are turned on
and the electric field, $F_{0i}$, is zero (see also footnote 19). Thus $\alpha_1 \to 0$ and $V_1 \approx  1$, so that
\eqref{obvious} is satisfied.
Therefore our {\it ansatze} for the IR metric will take the following form:
\bg\label{irgeometry}
ds^2_{\rm IR} = && {\cal F}_1^{-1/2} ds^2_{0123} + {\cal F}_1^{1/2} ds^2_\perp =
{\cal F}_1^{-1/2} (-dx_0^2 + dx_1^2 + dx_2^2 + dx_3^2) + {\cal F}_2 \vert d\vec{w}\vert^2 \nonumber\\
&& + {\cal F}_3 \left(dx^6 + \sum_\sigma F^\sigma_i dx^i\right)^2 + {\cal F}_4
\tau_2(u) \Bigg\vert \eta^2(\tau(u))\prod_{i=1}^{24} {du\over (u - u_i)^{1/12}}\Bigg\vert^2,
\nd
where ${\cal F}_i$ are related to each other by supergravity EOMs. Thus the non-conformal IR limit does not have any
immediate simplification. But if we go to the special arrangements of the seven-branes where we expect
constant coupling scenarios \cite{senF, DM1} then the EOMs connecting ${\cal F}_i$ should simplify to give us
the near-horizon $AdS_5$ geometry.

On the other hand, our background \eqref{totla} could also tell us about the UV geometry. At UV we cannot ignore the
Taub-NUT singularities and therefore the ${\rm D5}$-$\overline{\rm D5}$ pairs would wrap various vanishing cycles of the multi
Taub-NUT geometry. The D5-brane charges cancel, but as we discussed before, the $D3$-brane charges add up and in fact
the $D3$-branes are {\it delocalized} along the $x^{6,7}$ directions. From the above discussions, we now expect the UV
metric to be given by:
\bg\label{ltubaz}
ds^2_{\rm UV} &&=  {1\over \sqrt{V_2}}\Bigg[-f_1 V_1^{-1} dx_0^2 + f_3 dx_1^2 +
f_2 dx_2^2 + f_4 dx_3^2 + f_7 V_2dx_7^2\nonumber\\
&&+ f_5 V_2 V_1^{-1}\left(dx^6 + \sum_\sigma F^\sigma_i dx^i\right)^2\Bigg] \nonumber\\
&& + \sqrt{V_2}\left[f_8 (dx_8^2
+ dx_9^2)
+ f_6 \tau_2(u) \Bigg\vert \eta^2(\tau(u))\prod_{i=1}^{24} {du\over (u - u_i)^{1/12}}\Bigg\vert^2 \right].
\nd
Therefore the UV physics is now captured not by
a four-dimensional spacetime, but by a six-dimensional spacetime! In the constant coupling scenario of \cite{senF, DM1}
the near-horizon geometry should give us an $AdS_7$ spacetime. It would be interesting to compare the UV and IR
limits with \cite{GM} (and the earlier work of \cite{poln3}).

Before moving further, let us make two comments on the IR metric of \eqref{irgeometry}. This will help us to compare
our F-theory constructions with the brane constructions in type IIA \cite{WittenMT} and the brane network in type
IIB \cite{bbt}.

\begin{itemize}
\item The above metric \eqref{irgeometry} {\it cannot} come from a type IIA brane configuration with
NS5, D4 and D6-branes. In fact even in the so-called delocalized limit the form \eqref{irgeometry} cannot be
recovered. In particular it is not possible to see how the second term in the second line of \eqref{irgeometry} could
appear from D6-branes of type IIA\footnote{One might observe that a T-duality along the isometry direction of the
Taub-NUT space i.e along the $x^6$ direction, naively leads to a NS5-brane {\it delocalized} along the
$x^6$ direction. In \cite{ghm} this issue has been addressed in great details and the final
answer reveals an additional dependence of the NS5 harmonic function along the angular $x^6$ direction (see also
\cite{tong}). However similar analysis have not been attempted for the seven-branes, and at this stage it is
not {\it a-priori} clear to us how this T-duality should be taken to allow for localized gravitational solutions.}.
\item As we discussed in the previous subsubsection \ref{subsecNetwork}, a T-duality along $x^4$ or $x^5$ to get the brane
network model of \cite{bbt} is {\it not} possible because the metric \eqref{irgeometry} has non-trivial
dependence along the $u$-plane! If we delocalize along these directions then we can recover the brane network of
\cite{bbt} but will lose all non-trivial information on the $u$-plane. Therefore the F-theory picture captures
more information than the brane network of \cite{bbt}.
\end{itemize}

Thus from the above comments we see that the F-theory models are in some sense better equipped to capture non-trivial
informations of the corresponding gauge theories as the probe branes have direct one-to-one connections to the
corresponding gauge theories. The only restriction that we could see in our models has to do with the {upper-bound}
on the {\it number} of seven-branes. F-theory tells us that the number of seven-branes have to be at most 24 otherwise
the singularities on the $u$-plane will be too drastic to have a good global description \cite{vafaF}. This restriction on the
number of seven-branes (or to the global symmetries of the corresponding gauge theories) should not be too much of
an issue because one may resort to only local F-theory description assuming that the global completions may be done
by introducing anti-branes that would preserve ${\cal N} = 2$ supersymmetry up to certain energy scales (see also
\cite{kleban}).
The energy scale
may be chosen in such a way that all the above discussions may succinctly fit in. The global symmetries in these
theories may then be made arbitrarily large so as to encompass most of the Gaiotto's models. It would of course be
an instructive exercise to explicitly demonstrate a concrete example with a large global symmetry that, in the
Seiberg-Witten sense, remains
{\it integrable}. Once there, the far UV picture of this model should be interesting to unravel from our set-up.

One final thing before we end this subsection is to analyze the background fluxes. At the far IR the six-form
charges should cancel completely but at UV they should appear as dipole charges\footnote{For
${\rm D5}$-$\overline{\rm D5}$ pairs,
the tachyonic behavior emerges at distances of
order $\sqrt{\alpha'}$ or less \cite{dhhk}. (See appendix E in \cite{DasguptaSW}.)
In a tachyon-free system, we expect D5-$\overline{\rm D5}$ to be separated by a distance larger than that, creating a dipole moment in the system.
}.
The four-form charges should be
quantized and should be proportional to the number of ${\rm D5}$-$\overline{\rm D5}$ pairs. In addition to that there would be a
background axion-dilaton $\tau(u)$ that is a function on the $u$-plane, and  NS and RR two-forms field with the
required three-form field strengths. For the conformal cases we expect $\tau(u)$ to take one of the values given in
\cite{senF, DM1}. These fluxes and branes deform both the Taub-NUT and the seven-brane geometries and together
they preserve the required supersymmetry for our case.

\subsection{Mapping to the conformal cases \label{ConfGMap}}

The ${\rm D5}$-$\overline{\rm D5}$-brane pairs at the Taub-NUT singularities also tell us what the UV gauge symmetry should be for our case.
Imagine we have a $m$ multi-centered Taub-NUT geometry, then the $N$ ${\rm D5}$-$\overline{\rm D5}$ brane pairs wrapped around the
$m$ vanishing cycles lead to $mN$ fractional $D3$-branes where each of the $N$ fractional $D3$-branes carry a total RR
charge of $N/m$ in appropriate units.
Since there are $m$ copies of this, there is a total charge of $N$ $D3$-branes, leading us to
speculate the UV gauge symmetry to be $m$ copies of $SU(N)$, i.e:
\bg\label{uvgs}
SU(N) \times SU(N) \times SU(N) \times \cdots \times SU(N)  .
\nd
Once the wrapped ${\rm D5}$-$\overline{\rm D5}$ pairs are decomposed in terms of fractional $D3$-branes\footnote{Recall that there are no
D5-brane charges in the background.},
these fractional branes can now freely move along the F-theory $u$-plane, i.e
the Coulomb branch of the theory\footnote{Recall that if these $D3$-branes move along the Taub-NUT directions (i.e
the Higgs branch) they become fractional instantons.}.
This is illustrated in {figure \ref{wrappedTN}}. However even the individual
set of $N$ fractional branes may separate by further Higgsing to $U(1)^N$. In that case the individual fractional
$D3$-brane carry a net RR charge of ${1/m}$ in appropriate units.

It is now interesting to see how supersymmetry
and global symmetries would constrain the underlying picture. Since the
D5-brane charges cancel, the model only has fractional $D3$-branes and therefore the fractional-$D3$ and seven-branes
preserve the required supersymmetry as we discussed before. However the global symmetries are crucial. So we
should look for various arrangements of the seven-branes that allow $\tau(u) = $ constant in the $u$-plane. These
arrangements should be related to the models studied by Gaiotto \cite{gaiotto}.
\begin{figure}[htb]
        \begin{center}
\includegraphics[height=3cm]{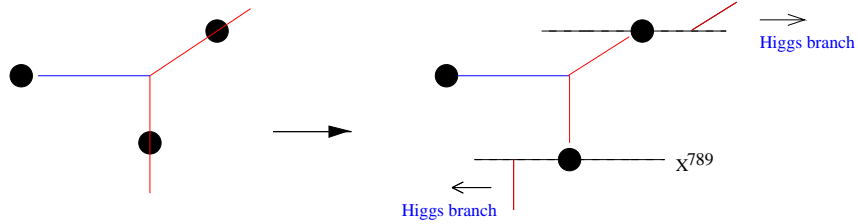}
        \end{center}
        \caption{The brane junction that we get in the left of the figure may further be truncated by
breaking the NS5 and the ($p, q$) five-brane and moving along the Higgs branch. However as we argue below,
this may not be required to see the underlying dualities.} \label{bbtreq}
        \end{figure}
One interesting example is related to the construction that we had in {figure \ref{TN2I}} wherein we showed how
the simplest T-dual brane network may come out from our scenario. The D5-brane ends on the seven-brane, and so would
any horizontal (i.e along $x^6$) D5-branes in this scenario. However the NS5-branes and the ($p, q$) five-branes
have to intersect the seven-branes as shown in the left of {figure \ref{bbtreq}}. From \cite{bbt} we might
expect the figure on the right where parts of the NS5 and the ($p, q$) five-branes have been moved away along the
Higgs branch. This configuration is in principle rather non-trivial to get from the Taub-NUT scenario, but
there is no reason for an exact one-to-one correspondence \cite{bbt} as we argued earlier.
The ${\cal N} = 2$ dualities should in principle be seen
as long as we have the D5-brane configurations right.

\begin{figure}[htb]
        \begin{center}
\includegraphics[height=3cm]{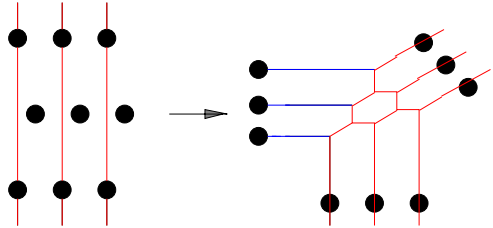}
        \end{center}
        \caption{On the left is a non-susy configuration that appears from naive T-duality of the
Taub-NUT model, as the branch cuts of the seven-brane (shown as solid black circles) would modify the
parallel NS5-branes configuration from the local axion-dilaton charges. This is similar to the deformation of a
string-network from background axion as shown in \cite{DMstring}. The arrangement of the seven-branes along
$x^6$ direction come from the original framework of the wrapped seven-branes with $A_6$ switched on. T-duality convert
$A_6$ to $x^6$ shown on the left.
On the right is the susy configuration by moving the seven-branes across the
``bent'' NS5-branes.} \label{asmodel}
        \end{figure}

To see further how this is implemented let us consider
our UV configuration of a three
set of three seven-branes wrapping three-centered Taub-NUT manifold with no fractional $D3$-branes. One immediate
advantage of this is that, since there are no fractional $D3$-branes to start with, a T-dual map to \cite{bbt} will be
easier.
The Weierstrass
equation governing the background at a given point is given by:
\bg\label{weq1}
y^2 ~ = ~x^3 + x (c_0 + c_1 z) + b_0 + b_1z + b_2 z^2
\nd
where we are choosing the {\it split} case of the Tate algorithm \cite{tate} where the choices of ($c_i, b_i$)
can be read off from eq. (4.7) of \cite{bikmsv}. This means that the discriminant locally is of the form
\bg\label{disloc}
\Delta ~ \sim ~ z^3
\nd
and so if we have three copies of this on the $u$-plane, we are
guaranteed that we will have {\it no} gauge symmetry but only a global symmetry of
\bg\label{glon}
 SU(3) \times SU(3) \times SU(3).
\nd
In one set of three seven-branes we can first switch on constant $A_6$ fields so that a T-duality along $x^6$
may lead to an arrangement of the seven-branes shown in the LHS of {figure \ref{asmodel}}. Note however that
due to the background axion-dilaton, the T-dual NS5-branes will {\it not} remain straight. The background axion-dilaton
will affect the NS5-branes and they will in turn get bent. This phenomena is exactly what we see for string
networks. In \cite{DMstring} it was shown how a network of ($p, q$) strings get bent in the presence of axion-dilaton.

Once this is taken care of, we can
switch on a time-varying gauge field on the same set of seven-branes in exactly the similar way we discussed
earlier. This would create fractional $D3$-branes to cancel the gauge anomalies which, in the T-dual framework, is
given by the RHS of {figure \ref{asmodel}}. The other two sets of three seven-branes\footnote{Clearly not all the
seven-branes are $D7$-branes, as we would need [$p, q$] seven-branes for consistency with Gauss' law.}can be arranged to
intersect the NS5 and the ($p, q$) five-branes. This is also exactly the configuration studied in \cite{bbt} (with
mild differences).

Note that the above configuration is in principle {\it different} from the configuration of
three fractional $D3$-branes probing seven-branes background where the seven-branes wrap
multi-centered Taub-NUT geometry. The T-dual of the $m$-center Taub-NUT space would be $m$ parallel NS5-branes as above.
The UV gauge group will be determined as \eqref{uvgs} but we may only consider
the low energy limit where the Taub-NUT singularities are not prominent. This however doesn't mean that we have
recovered the above model because there would still be a remnant gauge symmetry in the model even at far IR. We may
play the same game of removing fractional $D3$-branes by
switching on time-varying gauge field on each of the Taub-NUT cycles but the model will not be similar to our earlier
case and the gauge theory dynamics will be different.

Coming back to our model, we can now rearrange the seven-branes using F-theory Weierstarss equation to go to another
limit with a different global symmetry. This time the Weierstrass equation can be changed from \eqref{weq1}
to the following local form:
\bg\label{weq2}
y^2 ~ = ~ x^3 + z^4,
\nd
implying a global $E_6$ symmetry on the gauge theory side. From F-theory side the discriminant locus and the
underlying four-fold ${\cal M}_8$ will become respectively:
\bg\label{disffold}
\Delta ~ \sim ~ z^8, \qquad {\cal M}_8 ~ = ~ {\mathbb R}^4/{\bf Z}_3 ~ \times ~ {\bf TN}_3.
\nd
Observe that globally the K3 manifold has degenerated to its ${\bf Z}_3$ orbifold limit with a full global symmetry
of $E_6^3$ \cite{DM1}. For this global symmetry we are indeed at the constant coupling
point \cite{DM1} (see also \cite{minahan1, minahan2}).

In our Taub-NUT picture we have now redistributed the seven-branes now as three sets with {\it eight} seven-branes
in each set. We can move the two set of sixteen seven-branes in the $u$-plane so that we only allow a global symmetry
of $E_6$. Our picture can also be supported by the T-dual brane network of \cite{bbt}. This then would realize the
Argyres-Seiberg duality \cite{ArgyresSeiberg}.

Yet another example to consider would be to view the $SU(3)$ global symmetry to come from an $SU(4)$ symmetry by
Higgsing the ${\bf 4}$. This means we are bringing in another set of seven-branes so that the overall configurations wrap
a Taub-NUT with four singularities. The local Weierstrass equation now will be:
\bg\label{weq3}
y^2 ~ = ~ x^3 + x(c_0 + c_1 z) + b_0 + b_1z + b_2z^2 + b_3z^3
\nd
with special relations between ($c_i, b_i$) such that we are at the {\it split} ${\bf A}_3$ case \cite{tate}.
These relations are
worked out in \cite{bikmsv} which the readers may look up for more details. The discriminant locus is as expected:
\bg\label{dlo}
\Delta ~ \sim ~ z^4,
\nd
so that we have a global $SU(4)$ symmetry. As before if we make three copies of this we will have the required
global symmetry of $SU(4)^3$. This configuration maps directly to the brane network studied in \cite{bbt} so we
don't have to go through the details. It suffices to point out that the rearranged seven-branes may now give a
global symmetry of (see also \cite{DM1})
\bg\label{rags}
E_7 ~\times ~E_7 ~\times ~ SO(8),
\nd
so that the axion-dilaton remains constant throughout the $u$-plane. Locally near one of the $E_7$ singularity the
F-theory manifold is typically an orbifold of the form:
\bg\label{orbF}
{\mathbb R}^4/{\mathbb Z}_4 ~ \times ~ {\bf TN}_4,
\nd
 which means that our K3 has become a ${\mathbb Z}_4$ orbifold of the four-torus.

The above decomposition of the underlying K3 manifold into its various orbifold limits give us a hint what the next
configuration would be. This would be the ${\mathbb Z}_6$ orbifold of the four-torus so that the conformal global symmetry
should be \cite{DM1}
\bg\label{cgssb}
E_8 ~\times ~ E_6 ~\times ~ SO(8).
\nd
Now since the ${\mathbb Z}_6$ orbifold creates a deficit angle\footnote{Recall that for a singularity to be of an
orbifold type the deficit angle has to be $2\pi \left(1 - {1\over n}\right)$ for a fixed point of order $n$.}
of at most ${5\pi \over 3}$ we know that this is an orbifold with a fixed point
of order 6. Therefore our starting point would be to put three copies of six seven-branes wrapping a Taub-NUT
with six-singularities leading to an $SU(6)^3$ global symmetry. This then clearly enhances to \eqref{cgssb} with the
local F-theory four-fold given by:
\bg\label{orbF}
{\mathbb R}^4/{\mathbb Z}_6 ~ \times ~ {\bf TN}_6.
\nd
The above set of configurations were studied without incorporating any ${\rm D5}$-$\overline{\rm D5}$-branes in the background. Once we
introduce the probes we will not only have global symmetry, but also gauge symmetry. A special rearrangement of the
seven-branes may help us to study the conformal theories leading to other Gaiotto dualities. We will discuss a
more detailed mappings to these cases in the sequel.

\subsection{Beyond the conformal cases \label{subsecNonConf}}

\begin{figure}[htb]
        \begin{center}
\includegraphics[height=6cm]{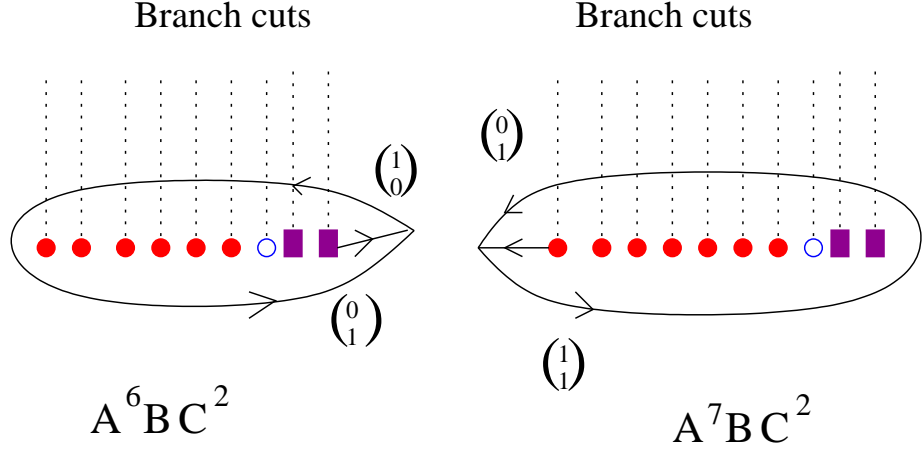}
        \end{center}
        \caption{A simple non-conformal deformation of the theory where the
multiple ${\rm D5}$-$\overline{\rm D5}$ branes are probing a $E_7 \times E_8$ singularity. The
global symmetry will have additional $U(1)$ that are not shown in the figure.
In the language of \cite{gaberdiel} the filled circles are $A$-branes, the
empty circles are $B$-branes and the filled squares are $C$-branes.} \label{nonconformal}
        \end{figure}

Since our model is a direct construction in F-theory, all informations of the type IIB background
under non-perturbative corrections are transferred directly to the $D3$-brane probes. This in particular means that
arrangements of the seven-branes that lead to non-trivial axion-dilaton backgrounds would also be transferred to the
$D3$-brane probes, except now they would appear as non-conformal theories on the $D3$-branes. A simple non-conformal
deformation, with our set-up discussed in the previous subsection, is given in {figure \ref{nonconformal}}. This
could be generated from $SO(8)$ in \eqref{rags}
breaking completely to $SU(2)$ by first going to $SO(7)$ and then $SO(7)$ breaking to
$SU(2)\times SU(2)\times SU(2)$. Recall from \cite{gaberdiel} that\footnote{In the following ${\bf A}, {\bf B}$ and
${\bf C}$ are the three monodromy matrices given as:
$${\bf A} ~ \equiv ~ \begin{pmatrix} 1&{1}\\ {0} & 1 \end{pmatrix}, ~~~~
{\bf B} ~ \equiv ~ \begin{pmatrix} ~4&{~9}\\ {-1} & ~2 \end{pmatrix}, ~~~~
{\bf C} ~ \equiv ~ \begin{pmatrix} ~2&{~1}\\ {-1} & ~0 \end{pmatrix}$$
These monodromy matrices are derived from the monodromies around $D7$ and the two ($p, q$) seven-branes in {figure \ref{figure 8}}
respectively. For more details the authors may refer to \cite{gaberdiel}.}
\bg\label{so8}
SO(8) ~ \equiv ~ {\bf A}^4 {\bf B} {\bf C},
\nd
so that the perturbative pieces generate the subgroup $SU(2) \times SU(2)$. Separating the [$0, 1$] and the [$1, -1$]
seven-branes from the bunch of the six seven-branes allow us to achieve this. Once we further break the other $SU(2)$,
we can easily generate the ${\bf 248}$ of $E_8$ from \eqref{rags} via:
\bg\label{gene8}
{\bf 248} ~ = ~ ({\bf 3}, {\bf 1}) ~ \oplus~ ({\bf 1}, {\bf 133}) ~ \oplus~ ({\bf 2}, {\bf 56}) .
\nd
\noindent This is clearly a non-conformal deformation in the Taub-NUT background as the axion-dilaton-dilaton
values are no longer
constant in the $u$-plane. It is interesting to note that if we take other model \eqref{cgssb} then there exist a
limit where the non-conformal deformation in this model is precisely the non-conformal deformation of the earlier case.
This is when $SO(8)$ in \eqref{cgssb} is completely broken to $U(1)$ by moving all the ${\bf A}, {\bf B}$ and
${\bf C}$ branes except one ${\bf A}$ brane. Under this circumstances the ${\bf 56}$ of $E_7$ is easily generated from
\eqref{cgssb} for the $D3$-brane probes to see identical physics as the earlier case:
\bg\label{56e7}
{\bf 56}~ = ~ {\bf 1} ~ \oplus ~{\bf 1} ~ \oplus ~{\bf 27} ~ \oplus ~ {\overline{\bf 27}}.
\nd
Finally, to see similar non-conformal deformation from the first $E_6^3$ model that we studied above, we can go back to
the {\it unenhanced} case for one of the $E_6$ group, namely the $SU(3)^3$ global symmetry with the particular
arrangements of the seven-branes as in {figure \ref{asmodel}}. For this case the ${\bf 56}$ of $E_7$ is
generated as \eqref{56e7}, but the ${\bf 248}$ of $E_8$ is now generated via:
\bg\label{e8e7ks}
{\bf 248} ~ = ~ ({\bf 8}, {\bf 1})~ \oplus ~({\bf 1}, {\bf 78})~ \oplus ~({\bf 3}, {\bf 27})~ \oplus
~({\bar{\bf 3}}, {\overline{\bf 27}}),
\nd
provided of course that the remnant subgroup of $SU(3) \times SU(2) \times U(1)$ is completely broken. Only under this
case the physics seen by the $D3$-brane probes will be identical.

The above non-conformal deformations were extensions of the conformal theories with exceptional global symmetries. They
aren't the simplest non-conformal models that we could study here. There exist simpler models if we introduce, in
addition to the ${\rm D5}$-$\overline{\rm D5}$-brane probes, some {\it additional} D5-branes wrapping vanishing 2-cycles of the Taub-NUT
space.

%below, changed r into m. for number of TN centers.
Let us take a concrete example where we have $k$ ${\rm D5}$-$\overline{\rm D5}$-brane pairs at a point in the Taub-NUT space with $m$
singularities. In addition to these probes, let us also introduce $M_i$ (with $i = 1, \cdots, m$)
D5-branes wrapping the $m$ vanishing 2-cycles
of the Taub-NUT space. It is immediately clear that the gauge symmetry now will change from \eqref{uvgs} to the
following:
\bg\label{gsnwb}
\prod_{i = 1}^m SU(k + M_i) ~ \equiv ~ SU(k + M_1) ~\times ~SU(k + M_2) ~\times ~\cdots~ \times ~SU(k + M_m).
\nd
The above theory is obviously non-conformal as the additional wrapped D5-branes break the conformal invariance
already in the absence of any flavor symmetry. If we take the Taub-NUT and wrap $M$ D5-branes
on the vanishing 2-cycle,
then the gauge group
will be special case of \eqref{gsnwb}, namely
\bg\label{kebot}
SU(k + M) ~ \times ~ SU(k).
\nd
This brings us exactly to the cascading models of \cite{PolchinskiMX, poln3, aharonyn2, benini2, BeniniIR} where the authors
have argued cascading behavior in this model (see \cite{benini2, BeniniIR} for a more recent study)! It is then
clear that our model can have an even more interesting cascading dynamics because there is an option of having a
 much bigger gauge group as can be seen from \eqref{gsnwb}.
Furthermore due to the presence of seven-branes, the cascading model is of the
Ouyang type \cite{ouyang1, ouyang2}. This means
there is a chance that cascades would be {\it slowed} down by the presence of
fundamental flavors, much like the one
studied in \cite{ouyang1, ouyang2}.

The connection to ${\cal N} = 1$ cascade \cite{KlebanovHB} is now clear: we can break the ${\cal N} =2$ supersymmetry
by non-trivially fibering the Taub-NUT over the compactified $u$-plane. Writing $u \equiv z_4 = x^4 + ix^5$, the
fibration is explicitly:
\bg\label{fibex}
z_1^2 ~+~ z_2^2 ~+~ z_3^2 ~= ~-z_4^2 ~ \equiv ~ -u^2,
\nd
which is an ALE space with coordinates ($z_1, z_2, z_3$)
fibered over the $u$-plane. Near the node $x^4 = x^5 = 0$ the geometry is
our familiar Taub-NUT space, and the equation \eqref{fibex} is a conifold geometry.
Once this is achieved the ${\cal N} =1$ cascading behavior can take over with the
termination point governed by {\it either} a confining theory, or a conformal theory depending on the choice of the flavor symmetry. 

This concludes our discussion about the connection between a class of Gaiotto models and F-theory with a multi Taub-NUT space.  We seem to have provided a geometry with all the right ingredients to generate a cascade for $\mathcal N=2$ non-conformal supersymmetric gauge theories in four dimensions.  Limited by the complexity of type IIB/F-theory language, we can not, at this point, push further our analysis of the dynamic of this cascade.  For this reason, we will turn, in the next chapter, to the type IIA/M-theory language where the origin of $\mathcal N=2$ cascade will becomes clarified.

\clearpage
\newpage
\chapter{Type IIA perspective on cascading gauge theories}
\section{Introduction}

The system of $p$ $D3$-branes and $M$ $D5$-branes at the tip of the conifold 
in type IIB string theory \cite{KlebanovHB} exhibits many non-trivial phenomena such as 
confinement, dynamical symmetry breaking and a rich landscape of ground states. 
It is also an important example of gauge/gravity duality, which plays a role in studies 
of string phenomenology and early universe cosmology. 

The low energy effective field theory of this system is an $\mathcal N=1$ supersymmetric 
four dimensional gauge theory with gauge group $SU(M+p)\times SU(p)$ and 
matter in the bifundamental representation \cite{KlebanovHB}. The rich vacuum structure 
of this gauge theory was described in \cite{DymarskyXT}. It can be interpreted in terms of 
a ``duality cascade'' -- a sequence of gauge theories with varying ranks which provide 
a description of the different vacua. Some of these vacua have a regular type IIB  
supergravity description, which has also been extensively investigated. 

The T-dual of the type IIB  construction of \cite{KlebanovHB} is given by a system 
of $NS5$-branes and $D4$-branes in type IIA string theory. This system was 
mentioned in \cite{KlebanovHB} and further studied in \cite{AharonyMI}. One of our goals 
below will be to build on the results of \cite{AharonyMI} and reproduce the results of 
\cite{KlebanovHB, DymarskyXT} using the IIA description. We will also discuss 
some non-supersymmetric aspects of the dynamics. 

We will see that the IIA description provides a nice picture of the supersymmetric 
and non-supersymmetric vacua. As is standard in studying
brane dynamics in string theory, the three descriptions (gauge theory, IIA and IIB) are valid in
different regions in the parameter space of the brane system. This should not matter for the
supersymmetric vacuum structure, and indeed we will reproduce the results of 
\cite{KlebanovHB,DymarskyXT} in the IIA language. Many aspects of the 
non-supersymmetric vacuum structure are also expected to agree, and we will find 
that to be the case.  

Cascading behavior was found in a wide variety of theories, some of which do not exhibit Seiberg 
duality. We will briefly discuss an example of this phenomenon, an $\mathcal N=2$ supersymmetric quiver theory
closely related to $\mathcal N=2$ SQCD, and use the IIA description to identify the origin of the cascade in this theory. 

In gauge theory there are actually two versions of Seiberg duality. The strong version asserts that the electric and magnetic theories of \cite{SeibergPQ} are equivalent in the infrared at the origin of moduli space and in the absence of deformations of the Lagrangian. In general one or both of these (conformal) theories are strongly coupled, and their equivalence has not been proven to date. The weaker version concerns the infrared equivalence of the two theories in the presence of deformations, and/or along moduli spaces of flat directions. In this case, one can often analyze the long distance behavior of both theories precisely and show their equivalence. Examples of this were studied in the original work of \cite{SeibergPQ} and many subsequent papers. A discussion in a context closely related to the cascading gauge theory appears in \cite{GiveonEF}. In the IIA brane description of Seiberg duality \cite{ElitzurFH}, the strong version of Seiberg duality involves exchanging $NS$ and $NS'$-branes connected by $D4$-branes, which involves fivebranes intersecting at a point in the extra dimensions. The statement of duality is that this process is smooth, which is non-trivial and unproven to date. The weak version of the duality involves smooth deformations of the brane system, which obviously do not change the low energy behavior. The discussion below will make it clear that the cascading gauge theory only requires the weak version of the duality. This is why it is manifest in the brane description. It also will clarify  that while the authors of \cite{DymarskyXT} used Seiberg duality to derive the vacuum structure of the model, one should be able to do this without that assumption, and show that the resulting vacuum structure exhibits the correct duality structure. 

The plan of the paper is as follows. In section \ref{section 2} we introduce the classical $\mathcal N=1$ supersymmetric 
gauge theory and IIA brane system that reduces to it at low energies. We review the structure of the 
classical moduli spaces in both languages, and show that they agree. We also discuss some non-supersymmetric 
vacua that appear for non-zero Fayet-Iliopoulos (FI) coupling. 

In section \ref{section 3} we discuss the quantum theory. We show that the quantum moduli space of the brane system
is the same as that of the gauge theory, and in particular exhibits the cascading behavior found in
\cite{KlebanovHB, DymarskyXT}. The brane picture gives a simple description of the cascade and helps 
understand which vacua of theories with different values of $p$ agree, and which do not. In this
picture, the cascade is associated with the fact that for a given value of the UV cutoff the fivebrane in general 
winds around a circle. As one reduces the UV cutoff, the winding number decreases. This 
corresponds in the field theory language to decreasing $p$ by a multiple of $M$. Vacua in which the fivebrane
does not wind (or does not wind enough times) around the circle are not in general the same in theories with 
different values of $p$. 

In section \ref{section 4} we discuss non-supersymmetric vacua of the brane system. We show that for the non-supersymmetric vacua that appear for non-zero FI coupling the quantum brane picture incorporates chiral symmetry breaking, which is expected to occur in the corresponding low energy gauge theory. We also discuss the IIA analog of the metastable states discussed in the IIB language in \cite{KachruGS}. 
We find that these states are present in the brane system, but the barrier that separates them from the supersymmetric states goes to infinity in the field theory limit. Thus, they become stable in that limit. 

In section \ref{section 5} we discuss the $\mathcal N=2$ supersymmetric analog of the cascading gauge theory, and in particular address the question how a theory that does not have Seiberg duality can have a duality cascade. We show that the situation is similar to that in the $\mathcal N=1$ case -- the gauge theory has a rich set of vacua, some of which exhibit cascading behavior.  Even in these vacua, different theories along the cascade differ by abelian factors in the gauge group. 

Section 6 contains a brief discussion of our results; an appendix summarizes some aspects of the IIA description of quantum $\mathcal N=2$ SQCD, which are useful for the discussion in section \ref{section 5}.

\section{Classical theory}\label{section 2}

We start with a brief description of the classical gauge theory and the corresponding type 
IIA brane system. We refer the reader to \cite{DymarskyXT,GiveonSR} for a more detailed 
discussion of the two topics. We will draw heavily on the results described in these papers.

As mentioned in the introduction, we will be studying an $\mathcal N=1$ supersymmetric gauge theory with gauge group 
\bg\label{ggroup}G=SU(N_1)\times SU(N_2)\nd
with 
\bg\label{defnonetwo}N_1=M+p;\qquad N_2=p\nd
The matter consists of chiral superfields $A_{\alpha i}^a$ and $B_{\dot\alpha a}^i$,
where $i=1,\cdots, N_1$, $a=1,\cdots, N_2$, are gauge indices, and $\alpha,\dot\alpha=1,2$
are global symmetry labels. As implied by the notation, the matter fields transform under
$G$ as follows:
\bg\label{transfab}
A_\alpha &\qquad (N_1,\bar N_2)\\
B_{\dot\alpha} &\qquad (\bar N_1, N_2)
\nd
There is also a tree level superpotential,
\bg\label{www}
W_0={h\over2} \epsilon^{\alpha\beta}\epsilon^{\dot\alpha\dot\beta}
A_{\alpha i}^aB_{\dot\alpha b}^iA_{\beta j}^bB_{\dot\beta a}^j=h\left(A_{1 i}^aB_{1 b}^iA_{2 j}^bB_{2 a}^j-A_{1 i}^aB_{2 b}^iA_{2 j}^bB_{1 a}^j\right).
\nd
To compare to standard discussions of $\mathcal N=1$ SQCD, it is useful to note 
that the $SU(N_1)$ factor in the gauge group \eqref{ggroup} ``sees'' $2N_2$ flavors,
and similarly for $SU(N_2)$. Since $N_1>N_2$ \eqref{defnonetwo}, in the quantum 
theory the quartic superpotential \eqref{www} is a relevant perturbation of the IR fixed 
point of the $SU(N_1)$ gauge theory obtained  by turning off the $SU(N_2)$ 
gauge coupling, and an irrelevant perturbation of the corresponding $SU(N_2)$ fixed point. 
 
The global symmetry of the model includes $SU(2)\times SU(2)$, with the
two factors acting on the indices $\alpha$ and $\dot\alpha$, and a $U(1)$ 
symmetry which assigns charges $+1$ to $A$ and $-1$ to $B$; this symmetry
is usually referred to as baryon number. We will denote it by $U(1)_b$
and will mostly consider the  theory in which it is gauged, since this is the case 
in the IIA brane system we will study. It is also useful to consider this case in the  
IIB theory, since it is relevant to the embedding of the conifold geometry in a compact 
Calabi-Yau manifold. 

\begin{figure}[htb]
        \begin{center}
\includegraphics[height=2 cm]{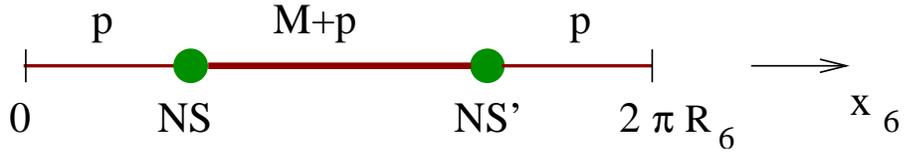}
        \end{center}
        \caption{The IIA brane configuration that realizes the cascading gauge theory.  
The $NS5$-branes are depicted in green, and are connected by $M+p$ 
$D4$-branes on one side of the $x^6$ circle (whose radius is $R_6$) and
by $p$ $D4$-branes on the other} \label{figure 1}
        \end{figure}

The IIA brane system that gives rise to the above gauge theory at low energies is
depicted in figure \ref{figure 1}. The system contains two kinds of branes: $NS5$-branes 
localized on a circle $x^6\sim x^6+2\pi R_6$, represented by green circles in the figure, and stacks of 
$D4$-branes connecting them, represented by red lines. The orientations of the 
different branes in the $9+1$ dimensional spacetime are as follows:
\bg\label{braneor}
NS: \qquad & 012345\\
NS': \qquad & 012389\\
D4: \qquad & 01236
\nd
This configuration preserves $\mathcal N=1$ SUSY in the $3+1$ dimensions common to all
the branes, $(0123)$. 
The low energy effective theory of this brane system contains $U(M+p)\times U(p)$ 
$\mathcal N=1$ SYM, associated with the massless excitations living on the $M+p$ and $p$
$D4$-branes, respectively, and  chiral superfields $A, B$  \eqref{transfab}, which come from
strings connecting the two stacks of $D4$-branes. An overall $U(1)$ in 
$U(M+p)\times U(p)$ is decoupled, and can be ignored, but the relative $U(1)$ is
precisely the $U(1)_b$ discussed above. Hence, the brane construction
gives the theory in which this symmetry is gauged, as mentioned above. 

The classical $U(N_i)$ gauge couplings are determined by the lengths of the corresponding
branes, $L_i$,
\bg\label{gili}
{1\over g_i^2}={L_i\over g_sl_s}
\nd
As is clear from figure \ref{figure 1},
\bg\label{sumlength}
L_1+L_2=2\pi R_6,\qquad i.e. \qquad 
{1\over g_1^2}+{1\over g_2^2}={2\pi R_6\over g_sl_s}
\nd
Our purpose in the remainder of this section is to compare the classical moduli space
of supersymmetric vacua of the gauge theory, studied in \cite{DymarskyXT},  to that of the 
brane system. We will also discuss some non-supersymmetric vacua of the theory.
In the next section we will describe the quantum moduli space.

The D-term equations of this gauge theory can be written as the following matrix equations
for the $(p+M)\times p$ matrices $A_\alpha$, $B_{\dot\alpha}^\dagger$:
\bg\label{dterms}
\sum_\alpha A_\alpha A^\dagger_\alpha-\sum_{\dot\alpha}B^\dagger_{\dot\alpha} B_{\dot\alpha}&=&
{\mathcal U\over p} I_p\\
\sum_\alpha A^\dagger_\alpha A_\alpha-\sum_{\dot\alpha}B_{\dot\alpha} B^\dagger_{\dot\alpha}&=&
{\mathcal U\over M+p} I_{M+p}
\nd
with $I_n$ an $n\times n$ identity matrix, and
\bg\label{defuu1}\mathcal U=\text{Tr}\left(\sum_\alpha A_\alpha A^\dagger_\alpha-
\sum_{\dot\alpha}B^\dagger_{\dot\alpha} B_{\dot\alpha}\right)
\nd
In the theory with gauged $U(1)_b$, one must set $\mathcal U=0$; turning on a 
Fayet-Iliopoulos (FI) term $\xi$ for this $U(1)$, modifies this to 
\bg\label{uxi}\mathcal U=\xi\nd
%%%
Classical supersymmetric vacua correspond to solutions of the D-term equations \eqref{dterms} -- \eqref{uxi} as well
as the F-term conditions for the superpotential \eqref{www}. For general $M$, $p$, and setting $\xi=0$ for now,
the solutions of these equations can be written, up to gauge transformations, in the diagonal form
\bg\label{formab}
A_\alpha=\left(
\begin{array}{ccccccccc}
A_{\alpha1}^1& & & & & & & &\\
 &A_{\alpha2}^2& & & & & & &\\
 & &A_{\alpha3}^3& & & & &\\
 & & &.& & & & &\\
 & & & &. & & & &\\
 & & & & &A_{\alpha p}^p& &
 \end{array}\right);
B^{t}_{\dot\alpha}=\left(
\begin{array}{ccccccccc}
B_{\dot\alpha1}^1& & & & & & & &\\
 &B_{\dot\alpha2}^2& & & & & & &\\
 & &B_{\dot\alpha3}^3& & & & &\\
 & & &.& & & & &\\
 & & & &.& & & &\\
 & & & & &B_{\dot\alpha p}^p&  &
 \end{array}\right)
\nd
The eigenvalues  $A_{\alpha a}^a$ and  $B_{\dot\alpha a}^a$, $a=1,\cdots, p$ satisfy the constraints 
\bg\label{dconst}\sum_\alpha|A_{\alpha a}^a|^2-\sum_{\dot\alpha}|B_{\dot\alpha a}^a|^2=0\nd
For given $a$, the eigenvalues are four complex fields, which satisfy one real constraint  
\eqref{dconst}. Another real field (for each $a$) is removed by the (Higgsed) gauge symmetry. 
Thus, the moduli space is $3p$ (complex) dimensional. It can be described by the $4p$ 
complex coordinates 
\bg\label{zzaa}z^a_{\alpha\dot\alpha}=A_{\alpha a}^aB_{\dot\alpha a}^a\nd
which satisfy the (complex) constraints 
\bg\label{singcon}\det_{\alpha\dot\alpha} z^a_{\alpha\dot\alpha}=0\nd
Together with the symmetry of permutation of the $p$ eigenvalues, we conclude that 
the classical moduli space is a symmetric product of $p$ copies of the singular conifold 
\eqref{singcon},
\bg\label{clasmod}\mathcal M_0={\rm Sym}_p(\mathcal C_0)\nd
At a generic point in the moduli space, the low energy theory consists of an $SU(M)$ $\mathcal N=1$ 
SYM theory\footnote{The unbroken $SU(M)$ is a  subgroup of the $SU(M+p)$ factor in  \eqref{ggroup} 
and  $p$ copies  of $\mathcal N=4$ SYM with gauge group $U(1)$.}
%%%

\begin{figure}[htb]
        \begin{center}
\includegraphics[height=5 cm]{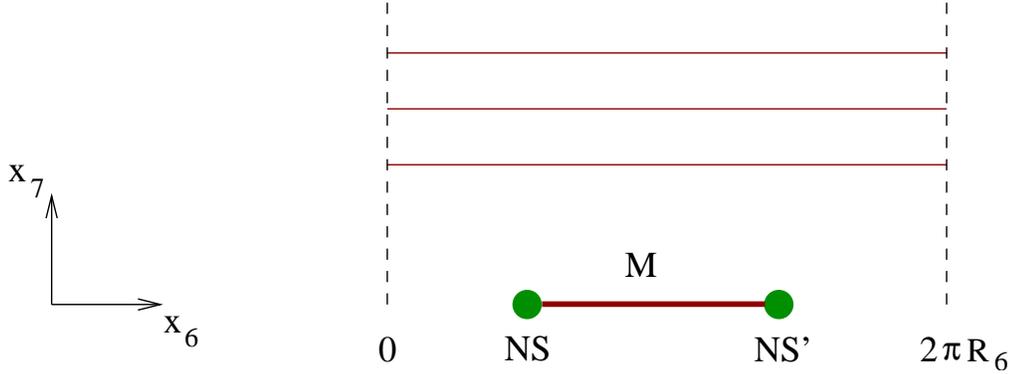}
        \end{center}
        \caption{A generic point in the classical moduli space. $p$ $D4$-branes wrap the $x^6$ circle 
and can move in the transverse space; $M$ are stretched between the fivebranes and give 
rise to $SU(M)$ $\mathcal N=1$ SYM} \label{figure 2}
        \end{figure}

In terms of the brane system of figure \ref{figure 1}, the moduli space described above is obtained by 
noting that the configuration contains $p$ $D4$-branes that wrap the circle, and can thus 
freely move in the $\mathbb R^5$ labeled by $(45789)$; another (compact) dimension 
of moduli space is obtained from a component of the gauge field on the fourbanes, $A^6$. 
A generic point in the moduli space is described in figure \ref{figure 2}.  The $p$ mobile branes support 
$U(1)^p$ $\mathcal N=4$ SYM, while the $M$ localized branes give rise to pure $\mathcal N=1$ SYM with 
gauge group $SU(M)$ (and a decoupled $U(1)$ mentioned above), in agreement with the 
gauge theory analysis. 

The form of the moduli space 
\eqref{clasmod} is very natural from the brane perspective: in the classical gauge theory limit, the 
separation between the fivebranes goes to zero \cite{GiveonSR}, and the $M$ $D4$-branes in figure \ref{figure 2}
\ref{figure 2} can be ignored. The fivebranes are described by the equation  $vw=0$, where 
\bg\label{vvww}v=x^4+i x^5~;\qquad w=x^8+i x^9\nd
This is known to be a dual description of the conifold (obtained by T-duality in $x^6$; see e.g
\cite{UrangaVF,DasguptaSU,McOristIN}). Under this T-duality, the mobile $D4$-branes turn into 
$D3$-branes living on the conifold, in agreement with \eqref{clasmod}. 
 
We next turn to the case where the FI parameter $\xi$ \eqref{uxi} is non-vanishing. In general, supersymmetry 
is then broken, with vacuum energy $V\sim g^2\xi^2$ \cite{DymarskyXT}. The case 
\bg\label{pkm}p=kM\nd
with integer $k$ is special. In that case the gauge theory has an isolated supersymmetric vacuum, 
in which for $\xi>0$ one has $B_{\dot\alpha}=0$, 
\bg\label{susyaone}
A_{\alpha=1}=C\left(
\begin{array}{cccccc}
\sqrt{k}&0&0&.&0&0\\
0&\sqrt{k-1}&0&.&0&0\\
0&0&\sqrt{k-2}&.&0&0\\
.&.&.&.&.&.\\
0&0&0&.&1&
\end{array}\right)
\nd
and 
\bg\label{susyatwo}
A_{\alpha=2}=C\left(
\begin{array}{cccccc}
0&1&0&.&0&0\\
0&0&\sqrt{2}&.&0&0\\
0&0&0&\sqrt{3}&.&0\\
.&.&.&.&.&.\\
0&0&0&.&0&\sqrt{k}
\end{array}\right)\nd
%%%
Each entry in the matrices \eqref{susyaone}, \eqref{susyatwo} is proportional to an $M\times M$ 
unit matrix, and the constant $C$ satisfies \eqref{defuu1}, \eqref{uxi}, $\xi=k(k+1)M|C|^2$. For 
$\xi<0$, one finds a similar vacuum with $A\leftrightarrow B$. The low energy theory 
in the baryonic vacuum is described by an unbroken $SU(M)$ $\mathcal N=1$ SYM, but unlike
the mesonic branch, this $SU(M)$ is embedded non-trivially in both factors of the
gauge group $G$ \eqref{ggroup}. 
 
In the theory with gauge group  $SU((k+1)M)\times SU(kM)$ 
(i.e with ungauged baryon number), \eqref{susyaone}, \eqref{susyatwo} give rise to a one complex
dimensional moduli space of vacua labeled by $C$, which is usually referred to as the
baryonic branch. Our interest is in the theory where $U(1)_b$ is gauged, 
in which (classically) it only appears for non-zero $\xi$ and is isolated. 

\begin{figure}[htb]
        \begin{center}
\includegraphics[height=5.5 cm]{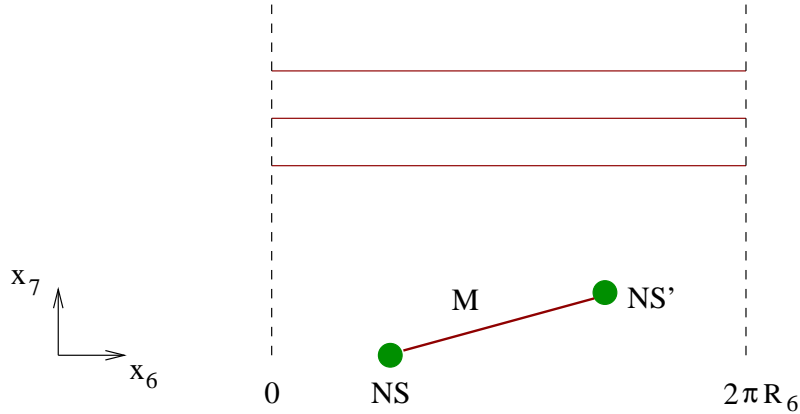}
        \end{center}
        \caption{Turning on an FI term in general leads to branes at an angle and breaks SUSY.} \label{figure 3}
        \end{figure}

In the brane system, the FI coupling $\xi$ corresponds to the relative displacement 
between the fivebranes in $x^7$ \cite{GiveonSR}. It is clear that generically this breaks 
supersymmetry, leading to configurations such as that of figure \ref{figure 3} , in which different 
$D4$-branes  are not mutually BPS. 
 
\begin{figure}[htb]
        \begin{center}
\includegraphics[height=4 cm]{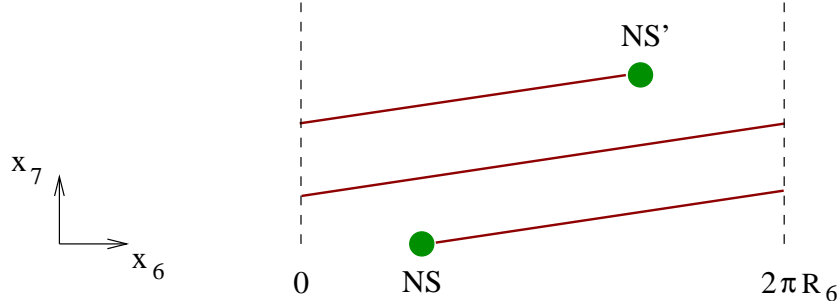}
        \end{center}
        \caption{The baryonic vacuum for $\xi\not=0$ corresponds to a stack of $M$ $D4$-branes
connecting the fivebranes and winding $k$ times around the circle ($k=2$ in the figure). } \label{figure 4}
        \end{figure}

The baryonic vacuum \eqref{susyaone}, \eqref{susyatwo} is described in terms of the branes by the 
configuration of figure \ref{figure 4}. The red line corresponds to a stack of $M$ $D4$-branes, which
connects the $NS$ and $NS'$-branes, in the process winding $k$ times around the circle. 
It is easy to check that all the branes in figure \ref{figure 4}  are mutually BPS and the configuration 
is supersymmetric.  Note that the vacuum of figure \ref{figure 4}  is isolated, as expected from the 
gauge theory analysis. Turning off the FI term, i.e taking the $NS'$-brane in figure \ref{figure 4} to the
 $x^6$ axis, leads to the configuration of  figure \ref{figure 1} , with $p=kM$. Thus, the baryonic vacuum 
 coincides in this case with the origin of the mesonic branch, as in gauge theory.

It is also clear from the figure that the low energy theory in this vacuum is pure $\mathcal N=1$ SYM 
with gauge group $SU(M)$, and that this $SU(M)$ is non-trivially embedded in the full gauge 
symmetry  $SU((k+1)M)\times SU(kM)$. A quick way to find this embedding is to calculate
the gauge coupling of the unbroken $SU(M)$, $g$. Taking $\xi\to 0$, the 
coupling is related to the length of the branes, as in \eqref{gili}:
%%%

\bg\label{couplbar}{1\over g^2}={L_1+2\pi kR_6\over g_s l_s}={k+1\over g_1^2}+{k\over g_2^2}\nd
Thus, the unbroken gauge group is the diagonal $SU(M)$ subgroup of an $SU(M)^{k+1}$ subgroup
in $SU((k+1)M)$ and an $SU(M)^k$ subgroup of $SU(kM)$,  in agreement with the gauge theory. 

Finally, figure \ref{figure 4} makes it clear that by moving the $NS'$-brane in $x^6$, we can 
change the winding number of the spiraling $D4$-branes, and thus $k$, by one 
or more units, without changing the low energy theory \footnote{Of course, if we keep 
the parameters $L$, $R_6$ fixed in the process, the (classical) gauge coupling of 
the $SU(M)$ gauge theory changes, but we can adjust these parameters so that 
it does not.}  This is a classical precursor of  Seiberg duality \cite{SeibergPQ}, which is known to play 
an important role in the quantum dynamics of the cascading gauge theory. The way
 it appears here is reminiscent of the discussion of \cite{ElitzurFH}. We will discuss its
 quantum analog in the next section.

In addition to the supersymmetric vacuum of figure \ref{figure 4} (or eqs \eqref{susyaone}, \eqref{susyatwo} in gauge 
theory) the brane system has a series of non-supersymmetric vacua labeled  by the winding 
number of the $M$ $D4$-branes stretched between the fivebranes, $l=0,1,2,\cdots, k$. The 
vacuum with winding number $l$ contains $M$ $D4$-branes connecting the $NS$ and 
$NS'$-branes while winding $l$ times around the circle, and $p-lM$ mobile $D4$-branes.  
The vacuum with $l=0$ is the one described in figure \ref{figure 3}, while that with $l=k$ corresponds to figure \ref{figure 4} 
(and is supersymmetric for $p=kM$). 

Since the vacua with $l<k$ are not supersymmetric, it is natural to ask what is the potential 
on the $3(p-lM)$ complex dimensional pseudomoduli space. Far along the moduli space (i.e
for large $A$, $B$ in \eqref{dconst}, or $z$ in \eqref{zzaa}) and at weak IIA string coupling, it is clear 
that the leading effect is due to closed string exchange between the mobile fourbranes and those
stretching between the fivebranes. Since these branes are not parallel, the gravitational attraction 
does not precisely cancel the RR repulsion, and there is a net attractive force pulling the mobile
branes towards the localized ones. 

\begin{figure}[htb]
        \begin{center}
\includegraphics[height=9cm]{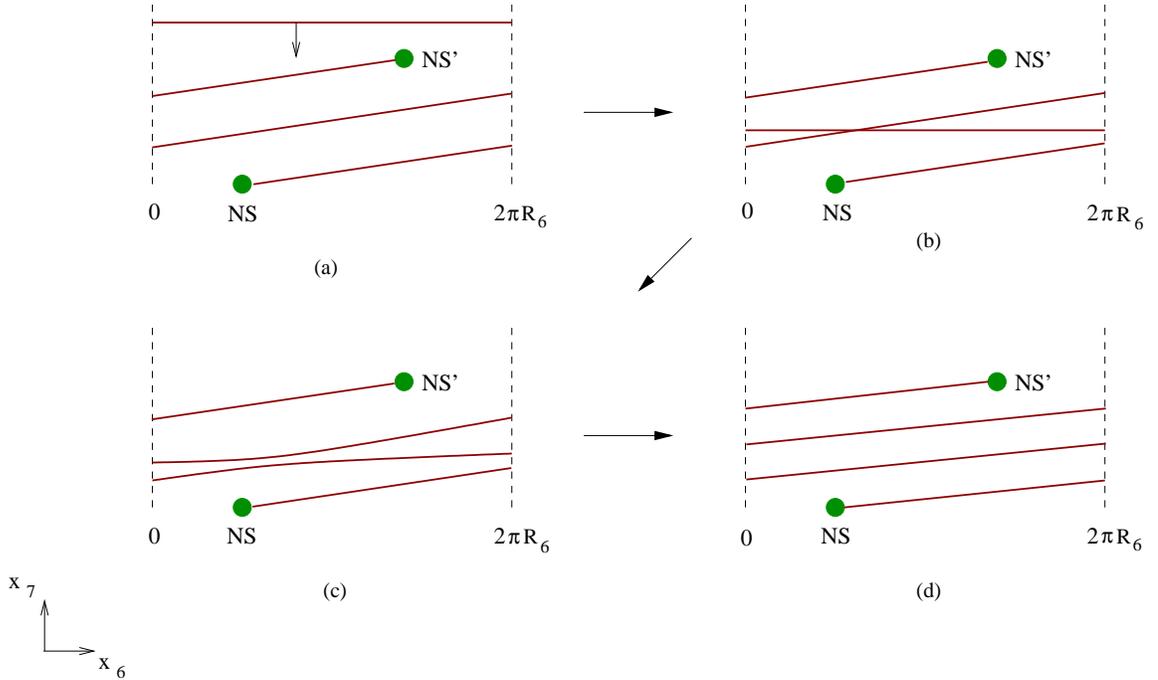}
        \end{center}
        \caption{Open string tachyon condensation connects vacua with different values of $l$.} \label{figure 5}
        \end{figure}

The resulting dynamics facilitates a change in $l$, as demonstrated in figure \ref{figure 5}, in which a stack of 
$M$ mobile $D4$-branes is pulled towards the fivebranes (a); when it intersects them (b), the brane 
configuration has an instability towards reconnection (c), due to the presence of an 
open string tachyon living at the intersection. Condensation of this tachyon leads to the 
configuration (d), in which $l$ has increased by one unit. The endpoint of this process is the 
supersymmetric vacuum, with $l=k$ (figure \ref{figure 4}) \footnote{One might think that there are other non-supersymmetric
but locally stable ground states in which the winding numbers of the different fourbranes connecting the fivebranes 
are different and the $SU(M)$ gauge symmetry is broken, but this is not the case.}  

The above brane discussion has a gauge theory counterpart. The F and D term potential at 
non-zero $\xi$ has a series of non-supersymmetric vacua in which the matrices $A$ and $B$
split into a block of size $(l+1)M\times lM$ in which they look like \eqref{susyaone}, \eqref{susyatwo} with 
$k\to l$, and a block of size $p-lM$, which looks like  \eqref{formab}. The eigenvalues \eqref{dconst} label 
the pseudomoduli space as in the discussion around \eqref{zzaa}. The potential for the pseudomoduli
is classically flat,  however near the origin of pseudomoduli space there is a tachyonic instability 
in a different direction in field space, which takes the system towards the supersymmetric vacuum 
\eqref{susyaone}, \eqref{susyatwo}. 

A natural question is what is the field theory analog of the classical gravitational attraction that in 
the brane description gives a potential on pseudomoduli space and leads to the isolated baryonic 
vacuum of figure \ref{figure 4}. In other closely related brane systems, such as those that appear in the discussion 
of the ISS model, this potential is the Coleman-Weinberg (CW) potential computed in \cite{IntriligatorDD}. 
It is natural to expect the same to happen here; we will leave a detailed analysis to future work. 

The gravitational brane attraction can be described from the field theory point of view in terms of 
a non-canonical Kahler potential for the light fields. As disussed in 
\cite{GiveonFK, GiveonEW, GiveonUR}, this effect is not identical to the CW potential. The two
are dominant in different regions in the parameter space of the brane system,  but tend 
to lead to similar dynamics. 

\begin{figure}[htb]
        \begin{center}
\includegraphics[height=3cm]{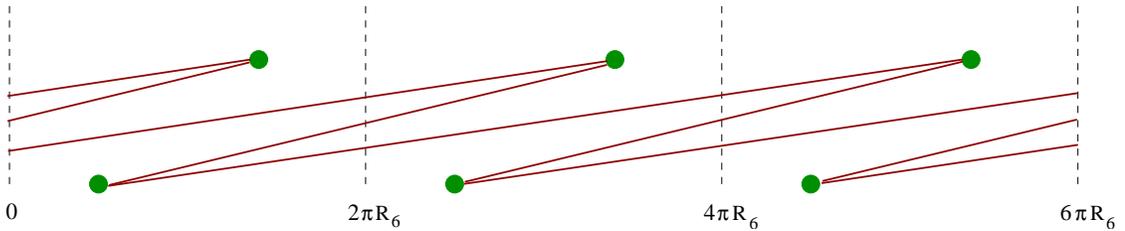}
        \end{center}
        \caption{The ground state of the brane system with non-zero $\xi$, viewed in the covering space of
the $x^6$ circle. $M-\tilde p$ fourbranes have winding $k$, while $\tilde p$ have winding $k+1$.
The specific case exhibited is  $k=\tilde p=1$, $M=2$, $p=3$.} \label{figure 6}
        \end{figure}

So far we focused on the case $p=kM$ \eqref{pkm}, but it is easy to generalize to
\bg\label{ptilde}p=kM+\tilde p;\qquad 1\le \tilde p\le M-1\nd
Most of the discussion of this case is the same as before. After all but $\tilde p$ of the mobile 
$D4$-branes have combined with the localized fourbranes via the process of figure \ref{figure 5}, we are 
left with $\tilde p<M$ branes in the bulk. These branes are also attracted to the spiraling fourbrane
and undergo a process similar to that of figure \ref{figure 5}, except now it affects only $\tilde p$
of the $M$ spiraling fourbranes. This leads to a state in which we have $M-\tilde p$ 
fourbranes which stretch between the fivebranes while winding $k$ times around the 
circle, and $\tilde p$ fourbranes which wind $k+1$ times (see figure \ref{figure 6}). Clearly, this state is not
supersymmetric (for generic $\tilde p$). We see that in this case, turning on an FI term causes
the moduli space to collapse to an isolated non-supersymmetric vacuum.

\bigskip
%%%

The low energy dynamics of the theory with non-zero $\tilde p$ can be read off from figure \ref{figure 6}.  
The unbroken gauge group is $SU(M-\tilde p)\times SU(\tilde p)\times U(1)_b$. 
The embedding of this group in \eqref{ggroup} can be determined in a similar way to the discussion 
around \eqref{couplbar}. The gauge group can be written as 
\bg\label{gaugtildep}SU(M+p)\times SU(p)=SU\left((k+1)(M-\tilde p)+(k+2)\tilde p\right)\times SU(k(M-\tilde p)+
(k+1)\tilde p)\nd
The first factor contains a $\left[SU(M-\tilde p)\right]^{k+1}\times \left[SU(\tilde p)\right]^{k+2}$ subgroup;
the second contains $\left[SU(M-\tilde p)\right]^k\times \left[SU(\tilde p)\right]^{k+1}$. The gauge group 
corresponding to figure \ref{figure 6} involves the diagonal $SU(M-\tilde p)\times SU(\tilde p)$ of all these factors.

There are two kinds of matter fields. One comes from strings both of whose ends lie on the same stack 
of fourbranes. In addition to the gauge fields, these give fermions in the adjoint representation of the gauge
group, which in the absence of the second stack of fourbranes would be the gauginos of an $\mathcal N=1$ 
supersymmetric model. The second comes from open strings stretched between the two stacks, and is 
localized at their intersections. As shown in \cite{gkunpub},  two $D4$-branes ending on an $NS5$-brane at a 
generic angle give rise to a massless Dirac fermion \footnote{The lightest bosonic fields have a mass that 
depends on the angle between the fourbranes and is non-zero unless these branes are parallel.}
Thus, the vacuum of the theory of figure \ref{figure 6} contains fermions in the 
bifundamental of  $SU(M-\tilde p)\times SU(\tilde p)$ charged under $U(1)_b$, i.e the
light matter is similar to that of the original cascading gauge theory, without the scalars. 

So far we discussed the non-supersymmetric vacuum of the theory with generic $\tilde p$ from the point
of view of the IIA brane construction, but it is easy to repeat the discussion in the gauge theory language. 
For $\xi>0$, the vacuum field configuration is obtained by splitting each $M\times M$ block on the diagonal 
in \eqref{susyaone}, \eqref{susyatwo} into 
blocks of size $M-\tilde p$ and $\tilde p$. Looking back at \eqref{gaugtildep} we see that in the blocks of size 
$M-\tilde p$ we should use the ansatz \eqref{susyaone}, \eqref{susyatwo}, with $C=C_{M-\tilde p}$ . In the blocks of
size $\tilde p$ we should use a similar ansatz, with $k\to k+1$ and $C=C_{\tilde p}$. The D-term potential
takes the form (up to an overall constant)
\bg\label{dtermpot}
V_D&\simeq & k(M-\tilde p)\left(|C_{M-\tilde p}|^2(k+1)-{\xi\over p}\right)^2+
(k+1)\tilde p\left(|C_{\tilde p}|^2(k+2)-{\xi\over p}\right)^2\\
&+&  (k+1)(M-\tilde p)\left(|C_{M-\tilde p}|^2k-{\xi\over p+M}\right)^2\\
&+&(k+2)\tilde p\left(|C_{\tilde p}|^2(k+1)-{\xi\over p+M}\right)^2
\nd
Minimizing w.r.t. $C_{\tilde p}$ and $C_{M-\tilde p}$ we find 
\bg\label{cmp}
|C_{M-\tilde p}|^2&=&{\xi\over 2k+1}\left({1\over p}+{1\over M+p}\right),\\
|C_{\tilde p}|^2&=&{\xi\over 2k+3}\left({1\over p}+{1\over M+p}\right).
\nd
For $\tilde p=0,M$ this reduces to the supersymmetric result of \cite{DymarskyXT}. 

To summarize, we found that the classical brane configuration has the same vacuum structure as 
the classical gauge theory. As usual \cite{GiveonSR}, the IIA description provides a simple geometric 
picture of the vacuum structure and low energy dynamics in a certain region of the parameter 
space of brane configurations. In the next section we move on to the quantum theory and compare 
the structure one finds in the gauge theory and brane pictures. 
%%%
\section{Quantum theory}\label{section 3}
In studying quantum effects we start from small values of $p$, and then proceed to larger ones. 

\subsection{$p=0$}

The field theory  described in section \ref{section 2} is in this case $\mathcal N=1$ pure SYM with gauge group $SU(M)$.
This theory generates dynamically a mass gap $\Lambda_1$,  and has $M$ isolated vacua in which 
the  superpotential takes the values 
\bg\label{dynsup}W=M\Lambda_1^3e^{2\pi i r\over M};\qquad r=1,\cdots, M\nd
The index $r$ labels $M$ vacua related by a $Z_{2M}$ R-symmetry (the anomaly free part of a $U(1)_R$
symmetry), which is dynamically broken to $Z_2$.

The brane description leads to a similar structure. The fivebranes and the $D4$-branes
ending on them combine into a smooth curved fivebrane \cite{WittenEP} whose form is given by
\bg\label{quantumfive}vw=\zeta^2;\qquad v=\zeta e^{-z/\lambda_M}\nd
where 
\bg\label{defzz}z=x^6+i x^{11}\nd, 
and $\lambda_M=g_sl_sM=RM$, with $R$ the radius of the M-theory circle, $x^{11}\simeq x^{11}+2\pi R$. We assume that
the IIA string coupling is small but $g_sM$ is large, so that $R=g_sl_s$ is small but the characteristic 
size of the fivebrane \eqref{quantumfive}, which is governed by $\zeta$, $\lambda_M$, is large (in string units).

\begin{figure}[htb]
        \begin{center}
\includegraphics[height=6cm]{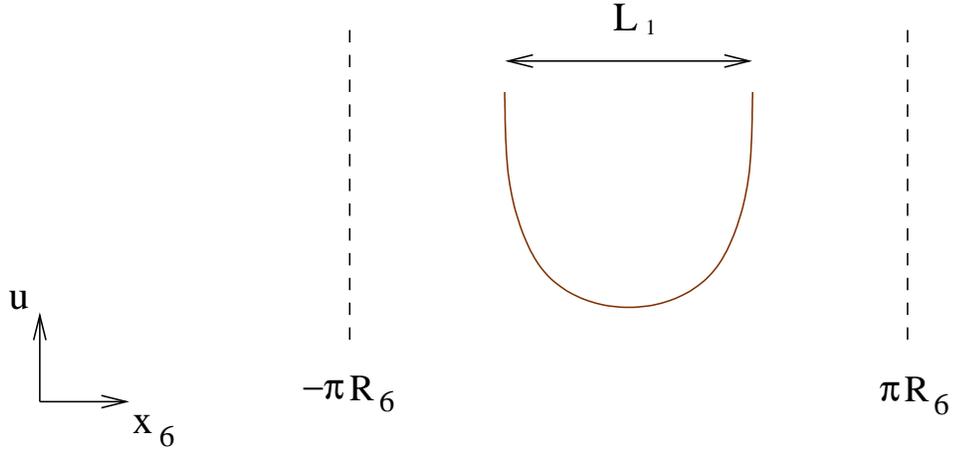}
        \end{center}
        \caption{The quantum ground state of the brane system with $p=0$ is described by the curved fivebrane \eqref{defuu}, \eqref{boundcond}.} \label{figure 7}
        \end{figure}

Since the position of the fivebranes in $x^6$ does not approach a constant value at large $v, w$, we need 
to impose a UV cutoff on the brane configuration. One way to do that \cite{AharonyMI} is to define the radial 
coordinate 
\bg\label{defuu}u^2=|v|^2+|w|^2=2\zeta^2\cosh{2x^6\over\lambda_M}\nd
and take it to be bounded, $u\le u_\infty$. The curved fivebrane \eqref{quantumfive} must satisfy the boundary
condition  
\bg\label{boundcond}\Delta x^6(u_\infty)=L_1\nd
We assume that $L_1<2\pi R_6$, i.e. the distance between the fivebranes at the cutoff
scale is smaller than the size of the circle. The profile of the brane is schematically
exhibited in figure \ref{figure 7} .

\bigskip

%%%
%%%
Note that the quantum theory is defined by specifying the parameters $M$, $\lambda_M$, 
$u_\infty$, $R_6$ and $L_1$. The dynamical scale $\zeta$ is a derived quantity, 
and can be calculated in terms of these parameters by using \eqref{defuu}:
\bg\label{xiuv}\zeta=u_\infty\exp(-L_1/2\lambda_M)=u_\infty\exp(-1/2\lambda_M^{(4)})\nd
where we defined the four dimensional 't Hooft coupling  in the usual way 
\cite{GiveonSR}, $\lambda_M^{(4)}=\lambda_M/L_1$, and assumed that it is small. 
One can think of $\lambda_M^{(4)}$ as the coupling at the UV cutoff scale,
$u_\infty$. The coupling at an arbitrary scale $u$ can be similarly defined by 
replacing $\Delta x^6(u_\infty)=L_1$ by the distance between the two arms of 
the curved fivebrane in figure (\ref{figure 7}), $\Delta x^6(u)$. For large $u$ it takes the form 
\bg\label{largeucurve}{1\over \lambda_M^{(4)}(u)}\simeq 2\ln{u\over\zeta}\simeq 
{1\over \lambda_M^{(4)}(u_\infty)}+ 2\ln{u\over u_\infty}\nd
The preceding discussion is very similar to what happens in gauge theory, where the role of 
$u$ is played by the RG scale, $u_\infty$ is the UV cutoff, and $\lambda_M^{(4)}$ the 't Hooft 
coupling.  The analog of the relation \eqref{xiuv} then gives the QCD scale of the theory, which
we denoted by $\Lambda_1$ in \eqref{dynsup}; the analog of \eqref{largeucurve} governs the RG flow of
the gauge coupling.

As is well known \cite{WittenEP}, the fivebrane \eqref{quantumfive} actually describes a system with $M$ vacua, 
associated with multiplying $\zeta$ by an $M$'th root of unity. These $M$ vacua correspond to
the ones labeled by $r$ in \eqref{dynsup}. 

\subsection{$0<p<M$}

The gauge theory has in this case three scales:
the dynamically generated scales of the two factors in the gauge group \eqref{ggroup}, $\Lambda_1$, 
$\Lambda_2$, and the superpotential coupling $h$ (which has units of inverse energy). Due to holomorphy, the moduli space can be studied for any ratio of these scales. A convenient regime
is one in which the gauge coupling of $SU(N_2)$, $g_2$, and Yukawa coupling $h$, are small at the
scale of $SU(N_1)$, $\Lambda_1$. In that case we can first analyze the $SU(N_1)$ dynamics, and then
add the other interactions. 

Since for $p<M$ the $SU(N_1)$ theory has fewer flavors than colors, we can describe the 
supersymmetric vacua in terms of the $2p\times 2p$ meson matrix 
\bg\label{mesmat}M_{\alpha\dot\alpha b}^a=A_{\alpha i}^aB_{\dot\alpha b}^i\nd
The superpotential for these fields takes the form 
\bg\label{wweff}W_{\rm eff}=W_0+(M-p)\left(\Lambda_1^{3M+p}\over\det M\right)^{1\over M-p}\nd
The F-term constraints of \eqref{wweff} lead to $M$ vacua, which can be thought of as the $M$ vacua 
of the $SU(M)$ pure SYM theory that appears at a generic point in the classical moduli space 
discussed in section \ref{section 2}. 

The mesons \eqref{mesmat} transform in the adjoint ($+$ singlet) representation of $SU(N_2)$. Their $SU(N_2)$ 
dynamics is weakly coupled at low energies. The main effect of this dynamics is to impose the D-term
constraints that, along the moduli space, allow one to diagonalize them (in $a,b$) for all $\alpha$, $\dot\alpha$. 

Thus, the moduli space is labeled by the eigenvalues $M_{\alpha\dot\alpha a}^a$, $a=1,\cdots, p$, 
which satisfy the constraints (that follow from \eqref{wweff})
\bg\label{constmes}h\det_{\alpha\dot\alpha} M_{\alpha\dot\alpha a}^a=
\epsilon_{M,p}(r, l=0)\sim\left(h^p\Lambda_1^{3M+p}\right)^{1\over M}\nd
i.e. they lie on the deformed conifold, with deformation parameter $\epsilon_{M,p}(r, l=0)$. 
$r$ is an index that labels the $M$ vacua related by a broken $Z_M$ symmetry, as above. 
The role of the parameter $l$ will become clear shortly.
%%%

\begin{figure}[htb]
        \begin{center}
\includegraphics[height=6cm]{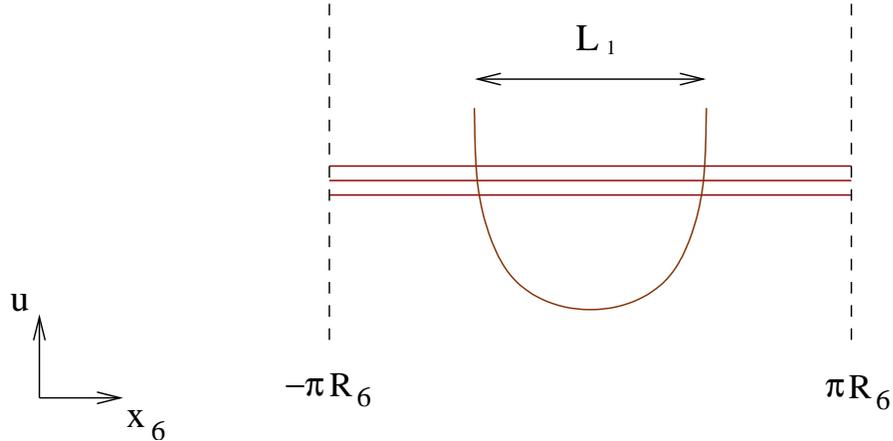}
        \end{center}
        \caption{The quantum moduli space of the brane system with $M>p>0$ is described by $p$ $D4$-branes wrapping the $x^6$ circle in the vicinity of the curved fivebrane of figure \ref{figure 7}.} \label{figure 8}
        \end{figure}

To describe the moduli space of vacua in the brane language we need to turn on $g_s$ effects
in the system of figure \ref{figure 2}. This involves replacing the $NS5$-branes connected by $M$ $D4$-branes
by the curved fivebrane \eqref{quantumfive}. The $p$ $D4$-branes in \ref{figure 2} then propagate in the vicinity
of this fivebrane (see figure \ref{figure 8}). Hence, their moduli space is the deformed conifold (as implied by T-duality).
We conclude that the quantum generalization of the classical moduli space \eqref{clasmod} is
\bg\label{quantmod}\mathcal M=\oplus_{r=1}^M {\rm Sym}_p(\mathcal C_{r, l=0})\nd
where $\mathcal C_{r, l=0}$ is the deformed conifold
\bg\label{defcon}\det z_{\alpha\dot\alpha}=\epsilon\nd
with deformation parameter $\epsilon=\zeta^2$. We see that the structure of the moduli space agrees with 
that found in gauge theory. 

\subsection{$p=M$}

The gauge theory analysis of \cite{DymarskyXT}leads in this case to a moduli space of the form 
\bg\label{quantmodmm}\mathcal M=\oplus_{l=0}^1\oplus_{r=1}^M  {\rm Sym}_{M(1-l)}(\mathcal C_{r, l})\nd
It is obtained by noting that the $SU(M+p)$ factor in \eqref{ggroup} has 
equal numbers of colors and flavors. Thus, the $SU(N_1)$ dynamics leads at low energies to a $\sigma$-model for the mesons $M$ \eqref{mesmat}, and baryons $\mathcal A=A^{N_1}$, $\mathcal B=B^{N_1}$. The classical moduli space, which is labeled by $M$, $\mathcal A$, $\mathcal B$, subject to the relation $\det M=\mathcal A\mathcal B$, is deformed in the quantum theory to 
\bg\label{defmodspace}\det M-\mathcal A\mathcal B=\Lambda_1^{2N_1}\nd
Adding the effect of the superpotential $W_0$ \eqref{www}, which is quadratic in the mesons, leads to 
two types of vacua. The mesonic (or $l=0$ in \eqref{quantmodmm}) vacua have $\mathcal A=\mathcal B=0$ and 
$\det M=\Lambda_1^{2N_1}$. The $SU(N_2)$ D-terms lead then to a moduli space described 
by the eigenvalues of $M$, as in \eqref{constmes}, \eqref{quantmod}. The baryonic ($l=1$) vacua are obtained 
by setting the mesons $M=0$; the baryons then satisfy the constraint $\mathcal A\mathcal B=-\Lambda_1^{2N_1}$. 
The low energy theory is pure $\mathcal N=1$ $SU(M)$ gauge theory, which gives rise to the $M$ isolated 
vacua labeled by $r$ in \eqref{quantmodmm}. Note that while in the classical theory the baryonic vacuum 
is identical to the origin of the mesonic branch, in the quantum theory the two are distinct, due to the
deformation \eqref{defmodspace}. The classical result is recovered in the limit $\Lambda_1\to 0$. 

We now turn to the brane description of the vacua \eqref{quantmodmm}. The mesonic $(l=0)$ branch is
described in the same way as for $p<M$, by the configuration of figure \ref{figure 8} (the quantum version of 
figure \ref{figure 2}), with $p=M$. The baryonic vacua are also easy to describe, following the discussion of 
section \ref{section 2}. We saw there that the classical baryonic vacuum of the gauge theory, \eqref{susyaone}, 
\eqref{susyatwo} is described  by $D4$-branes with non-zero winding (see figure \ref{figure 4}). It is natural to expect 
that something similar happens here. 

\begin{figure}[htb]
        \begin{center}
\includegraphics[height=3cm]{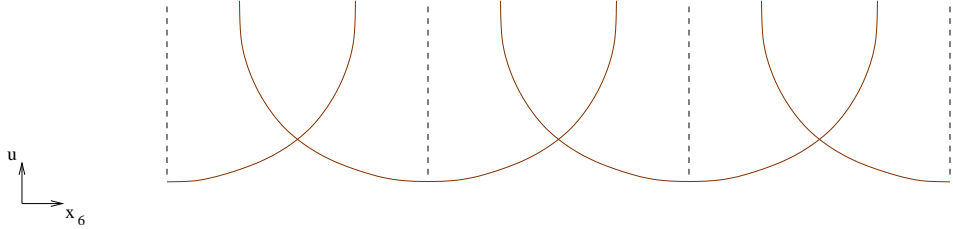}
        \end{center}
        \caption{The baryonic vacua of the brane system with $p=M$, viewed in the covering space of the $x^6$ circle. Vertical dashed lines are separated by $2\pi R_6$ and are identified on the circle.} \label{figure 9}
        \end{figure}

In more detail, the baryonic vacua are described by the quantum version of a brane configuration in which
$M$ branes connect the $NS$ and $NS'$-branes while winding once around the circle. The classical 
configuration is indistinguishable from that of figure \ref{figure 1} (with $p=M$), which can also be thought of as the
origin of the mesonic branch of figure \ref{figure 2}, but quantum mechanically the two are different. While the mesonic
branch is replaced by the configuration of figure \ref{figure 8}, a baryonic vacuum gives rise to that of figure \ref{figure 9}. 
%%%
In the covering space, it is again described by the profile \eqref{quantumfive}, but with the boundary conditions 
\eqref{boundcond} replaced by 
\bg\label{bcondbar}\Delta x^6(u_\infty)=L_1+2\pi R_6\nd
As in the discussion of section \ref{section 2}, the fact that the curved fivebrane \eqref{quantumfive} winds once
around the circle implies that unlike the mesonic branch of figure \ref{figure 8}, here there are no mobile 
$D4$-branes and the vacuum is isolated.  The dynamically generated scale in the baryonic 
vacuum of figure \ref{figure 9} differs from that of the mesonic one (figure \ref{figure 8}) as well.  In general, the scale 
is given by (see eq. \eqref{xiuv}),
\bg\label{dynscale}\zeta=u_\infty\exp\left(-\Delta x_6(u_\infty)/2\lambda_M\right)\nd
In the vacua of figures 8,  9 one has 
\bg\label{delxsix}\Delta x_6(u_\infty)=L_1+2\pi R_6l\nd
with the winding number $l=0(1)$ in the mesonic (baryonic) branch. 
Plugging \eqref{delxsix} into \eqref{dynscale} we find that 
\bg\label{xill}\zeta_l=u_\infty\exp\left(-{\Delta x^6(u_\infty)\over2\lambda_M}\right)=
\zeta_0\exp\left(-{2\pi R_6 l\over 2\lambda_M}\right)=\zeta_0 I^{l\over2M}\nd
with
\bg\label{imp}I=\exp\left(-{2\pi R_6\over l_s g_s}\right)\nd
This expression for the scale is the same as that obtained in gauge theory \cite{DymarskyXT}.  
We will discuss the general relation in the next subsection. 

An interesting feature of the brane configuration of figure \ref{figure 9} is that there are actually {\it two} different values of the UV cutoff $u_\infty$ for which the two ``arms'' of the curved fivebrane are separated on the $x^6$ circle by the distance $L_1$. One is the value drawn in figure \ref{figure 9}, which corresponds to \eqref{bcondbar} and describes a fivebrane that winds once around the circle. The second is obtained by lowering the value of $u_\infty$ until the distance becomes $L_1$ again, this time with no winding. In terms of the dynamically generated scale \eqref{xill} the two values are given by $\zeta_1\exp(L_1+2\pi R_6)/2\lambda_M$ and  
$\zeta_1\exp(L_1/2\lambda_M)$, respectively. For the second (lower) value of the cutoff, for $u<u_\infty$ the brane configuration is identical to the one depicted in figure \ref{figure 7}, which describes the vacuum of the theory with $p=0$. Thus, we see that the two are equivalent at long distances; the low energy theory is in both cases $\mathcal N=1$ pure  $U(M)$ SYM theory.

This infrared equivalence between the $U(2M)\times U(M)$ and $U(M)$ theories can be thought of as a consequence of Seiberg duality. Seiberg duality is usually realized in IIA string theory via motions of fivebranes \cite{ElitzurFH}. Here, this motion occurs dynamically, as a function of the RG scale $u$. The situation is under better control than in \cite{ElitzurFH}, since the fivebrane configuration of figure \ref{figure 9} remains smooth as $u_\infty$ is decreased. Thus, in this case one does need to rely on unproven conjectures to establish the  equivalence between the baryonic vacua of the theory with $p=M$ and the vacua of the one with $p=0$. 

A few other features of the brane construction are useful to note: 
\begin{enumerate}
\item While the baryonic $(l=1)$ vacua of the theory with $p=M$ can be identified with those of the $p=0$ one, this equivalence is not true for the mesonic vacua. Indeed, in the configuration of figure \ref{figure 8}, the distance on the circle between the two arms of the curved fivebrane is strictly smaller than $L_1$ for all $u$ below the UV cutoff $u_\infty$. There is clearly no corresponding vacuum of the theory with $p=0$.

\item In section \ref{section 2} we discussed what happens when we turn on an FI term for 
$U(1)_b$ in the classical gauge theory. In the quantum theory the situation
is essentially the same. The mesonic branch of moduli space is lifted by the perturbation,
since the mobile fourbranes in figure \ref{figure 8} are no longer mutually BPS with the curved fivebrane,
a rotated version of \eqref{quantumfive}. The baryonic vacua, which contain no mobile branes, are
still supersymmetric. The curved fivebrane that describes them is the quantum version 
of the classical configuration of figure \ref{figure 4}. 

\item One could consider {\it increasing} the UV cutoff $u_\infty$ in figure \ref{figure 9}, rather than decreasing it, i.e. flowing up the RG. This relates the vacua of the theory with $p=M$ to those of theories with $p=kM$,  $k>1$. We will discuss such theories next. 

\end{enumerate}
%%%

\subsection{$p > M$}

For general $M$ and $p$, the gauge theory analysis of \cite{DymarskyXT} leads to the moduli space
\bg\label{genquantmod}\mathcal M=\oplus_{l=0}^k\oplus_{r=1}^M  {\rm Sym}_{p-lM}(\mathcal C_{r, l})\nd
where $k$ is defined in \eqref{ptilde}, and the deformation parameter of the conifold $\mathcal C_{r,l}$
is given by 
\bg\label{epsrl}\epsilon_{M,p}(r, l)=\epsilon_{M,p}(r, l=0)I(M,p)^{l\over M}\nd
with
\bg\label{instfactor}I(M,p)=h^{M+2p}\Lambda_1^{3M+p}\Lambda_2^{p-2M}=e^{2\pi i\tau}\nd
The last equality expresses the factor $I(M,p)$ in terms of the D-instanton amplitude
in type IIB string theory. In particular, in string theory this quantity is independent of $M$, $p$.

As before, the index $r$ labels vacua related by the broken $Z_M$ symmetry; the $r$ dependence 
corresponds to picking different $M$'th roots of the identity in \eqref{epsrl}. 
A natural field theory interpretation of the quantum number $l$ in \eqref{genquantmod} involves 
a series of Seiberg dualities that take $SU(M+p)\times SU(p)$ to $SU(p-(l-1)M)\times SU(p-lM)$.
From the IIB perspective, vacua with given $l$ involve $p-lM$ mobile $D3$-branes propagating 
on the deformed conifold with deformation parameter $\epsilon_{M,p}(r, l)$ \eqref{epsrl}. 

To describe the vacuum structure \eqref{genquantmod} using the IIA brane construction of figure \ref{figure 1}, we 
need to  generalize the discussion of the previous subsections to all $p$.  The parameter $l$ 
labeling different branches of moduli space \eqref{genquantmod} has a clear IIA  interpretation  -- it is the 
winding number of the $D4$-branes connecting the $NS$ and $NS'$-branes.  In a vacuum 
with given $l$, $M$ $D4$-branes stretch from the $NS$-brane to the $NS'$-brane, in the process
winding $l$ times around the circle. This leaves $p-lM$ mobile $D4$-branes wrapping the circle,
which live as before on a deformed conifold. 

The fivebranes with $D4$-branes ending on them are described quantum mechanically 
in terms of a connected curved fivebrane \eqref{quantumfive}, with the scale parameter $\zeta=\zeta_l$
\eqref{xill}, \eqref{imp}. The mobile $D4$-branes live on a deformed conifold \eqref{defcon} with deformation parameter
$\epsilon_{M,p}(r,l)=\zeta_l^2$, which can be written in the form \eqref{epsrl}, with $I(M,p)$ given by \eqref{imp}. 
This agrees with the IIB result (the last expression in \eqref{instfactor}), since one can think of \eqref{imp} as the 
amplitude of a D-instanton obtained by wrapping a Euclidean $D0$-brane around the $x^6$ circle. 
This brane is related by T-duality to the IIB D-instanton whose amplitude is given by \eqref{instfactor}. 

Note that the deformation parameter goes like $X^l$, with 
\bg\label{xxxx}X=I^{1\over M}=\exp\left(-{2\pi R_6\over\lambda_M}\right)\nd
If we choose the 't Hooft coupling at the cutoff scale $\lambda_M^{(4)}$ to be very small, as we have done 
in the discussion around \eqref{xiuv}, the parameter $X$ is very small as well. Thus, the scales of vacua with 
larger $l$ are strongly suppresed relative to those with smaller $l$. This should be contrasted with the 
situation in the IIB theory where at large 't Hooft coupling (the supergravity regime), the analog of $X$ 
\eqref{xxxx} is very close to one, and one has to consider large values of $l$ to get large suppression. 

We see that the IIA brane description reproduces the structure of the supersymmetric moduli space 
\eqref{genquantmod}, and the dependence of the deformation parameter \eqref{epsrl} on the branch (i.e on $r$
and $l$). One can also
compare the value of the superpotential in the different vacua \footnote{The value of the superpotential is
important for calculating the tension of BPS domain walls between vacua with different values of $r$
in \eqref{genquantmod}.}.  In the field theory, gluino condensation in a low energy $SU(M)$ subgroup of $G$ 
\eqref{ggroup} leads to the superpotential
\bg\label{nonzerow}W=ML_1(M,p)^{1\over M}I(M,p)^{l\over M}\nd
where 
\bg\label{lone}L_1(M,p)=h^p\Lambda_1^{3M+p}\nd
In the brane language, the superpotential was computed in \cite{WittenEP} and is given (up to a universal
overall constant) by 
\bg\label{iiasup}W\simeq M\zeta_l^2\nd
Substituting the form of $\zeta_l$ \eqref{xill} into \eqref{iiasup}, we conclude that the two expressions agree
if we take 
\bg\label{formlone}\zeta_0^2\simeq L_1(M,p)^{1\over M}\nd
This identification is natural since the right hand side is nothing but $\Lambda^3$, the 
non-perturbative superpotential of the low energy $SU(M)$ gauge theory in the vacuum with $l=0$. 
%%%
Comments:
\begin{enumerate}
\item In section \ref{section 2} we discussed the classical vacuum structure in the presence of an FI 
D-term.  From the IIA brane perspective, it is  clear that the situation in the quantum theory
is similar. If $p$ is not divisible by $M$ (i.e if $\tilde p\neq 0$ in \eqref{ptilde}), the vacuum spontaneously
breaks supersymmetry. We will discuss this case further in the next section. For $\tilde p=0$, 
the vacua with $0\le l<k$  again break supersymmetry, while the vacuum with $l=k$, which 
corresponds to the quantum generalization of the configuration of figure \ref{figure 4}, does not 
(it is $M$-fold degenerate, as in \eqref{genquantmod}). \\

\item As mentioned above, in field theory the vacua \eqref{genquantmod} with $l>0$ can be understood 
in terms of  Seiberg duality. This too has a natural interpretation in the brane construction, as we saw 
in the previous subsection for $p=M$. A vacuum with given $l$ involves $M$ $D4$-branes connecting 
the fivebranes while winding $l$ times around the $x^6$ circle, making a single curved fivebrane of the 
form \eqref{quantumfive}, with $\zeta=\zeta_l$ \eqref{xill}.   By decreasing the UV cutoff while keeping the two arms
of the fivebrane at the same distance on the $x^6$ circle  one obtains a vacuum of the theory with 
$p\to p-M$ and $l\to l-1$ (such that the number of mobile $D4$-branes, $p-lM$, remains fixed). 
Looking back at \eqref{xill} we see that 
\bg\label{cascade}u_\infty(p-M)=I^{1\over 2M} u_\infty(p)\nd
This is the IIA manifestation of the duality cascade. \\

\item There are many other aspects of the gauge theory that can be studied in the brane description, 
such as domain walls connecting different vacua, QCD strings etc. This description is also useful for
discussing generalizations of the Klebanov-Strassler construction to other cascading gauge theories. 
For example, one can replace the $NS-D4-NS'$ system in figure \ref{figure 2} by a more general one, with or without supersymmetry,
and repeat the discussion of the last two sections. 
\end{enumerate}
%%%

\section{Non-supersymmetric brane configurations}\label{section 4}

In the previous section we focused on supersymmetric vacua of the quantum theory. In this section we would 
like to comment on some aspects of the non-supersymmetric dynamics. 

\subsection{Non-supersymmetric vacua with $\xi\not=0$}

In section \ref{section 2} we discussed the classical theory with non-zero FI parameter for $U(1)_b$.
We saw that the vacuum structure depends on whether $p$ is a multiple of $M$ \eqref{pkm}. If it is, the lowest
energy state is supersymmetric; it is described by the field configuration \eqref{susyaone}, \eqref{susyatwo}  in the
gauge theory, and by the brane configuration of figure \ref{figure 4} in the IIA language. On the other hand, if 
$\tilde p$ in \eqref{ptilde}\ does not vanish, the ground state is non-supersymmetric; it is described by  the 
brane configuration of figure \ref{figure 6} and  corresponding field configuration (discussed around \eqref{dtermpot}).

It is interesting to study the quantum generalization of this brane configuration. An important effect 
that needs to be taken into account in this case is the interaction between the $M-\tilde p$ 
$D4$-branes that wind k times around the circle, and the $\tilde p$ $D4$-branes that wind $k+1$ times.
Since the two stacks of fourbranes are no longer parallel, there is a force between their endpoints on 
the $NS5$-branes. This force is due to an incomplete cancellation between the electrostatic repulsion 
between the endpoints, which can be thought of as (like) charges on the fivebrane, and the attraction 
due to scalar exchange. The former is independent of the angle between the two stacks of $D4$-branes, 
while the latter goes like $\cos\theta$, the angle between the two stacks. 

Thus, the total force is repulsive, and goes like $1-\cos\theta$. This force was discussed in a different
context in \cite{GiveonWP}, where this repulsion played an important role in comparing the dynamics of the branes
to that of the corresponding low energy field theory. There, it gave rise to a runaway of certain pseudomoduli;
in our case, the $D4$-branes cannot escape to infinity, since the two fivebranes they are connecting are
stretched in different directions. Thus, the effect of the repulsion is to push them away from each other
by a finite distance. 

This has a natural interpretation in the low energy field theory of the brane system of figure \ref{figure 6}. As 
mentioned in section \ref{section 2}, this theory is an $SU(M-\tilde p)\times SU(\tilde p)\times U(1)_b$ gauge
theory coupled to fermions in the bifundamental representation. These fermions are classically massless,
but quantum mechanically are expected to acquire a mass due to chiral symmetry breaking.
The separation of the two stacks of $D4$-branes leads to precisely this effect. The chiral symmetry broken
by the vacuum is part of the $9+1$ dimensional Lorentz group corresponding to rotations in $(45)$ and $(89)$.

One can in principle study the quantum deformations of the configuration of figure \ref{figure 6} in more detail when the
parameters $M$ and $\tilde p$ are in particular regimes. For example, if $g_sM$ is large while $\tilde p$ is of
order one, one can replace the $NS5$-branes connected by $M-\tilde p$ $D4$-branes in figure \ref{figure 6} by a 
curved fivebrane, which looks like a rotated version of \eqref{quantumfive}, and study the shape of the $\tilde p$ 
probe $D4$-branes which end on this fivebrane and wind $k+1$ times around the circle. If both $g_sM$ and
$g_s\tilde p$ are large, we can replace them by a two center solution and look for the lowest energy 
configuration with the given boundary conditions.  We will leave these calculations to future work. 

The authors of \cite{DymarskyXT}\ proposed to use the system with non-zero FI parameter as a possible model of 
early universe cosmology. It is interesting to reexamine this proposal in the regime of validity of the IIA brane construction. 
Consider, for example, the model with $\tilde p=1$, i.e $p=kM+1$ (see \eqref{ptilde}), $k\gg 1$ and $\xi\not=0$. 
For $\xi=0$, the quantum moduli space has multiple branches \eqref{genquantmod}, most of which are 
unstable for non-zero $\xi$.  In the IIA brane picture, the mobile branes are attracted to the curved fivebrane, 
and are absorbed by it as described in section \ref{section 2} (figure \ref{figure 5}). Even if the FI parameter is not small, i.e the  
relative displacement of the $NS5$-branes in figure \ref{figure 3}, $\Delta x^7$, is comparable to the distance between the
fivebranes $L_1$, as the process of figure \ref{figure 5} takes place, the angle the curved fivebrane makes with the $x^6$
axis decreases, and thus the attractive potential felt by the mobile $D4$-branes becomes more flat.  

Consider the final step in this process, where all fourbranes but one have been absorbed by the winding
curved fivebrane, which takes the (quantum generalization of the) shape in figure \ref{figure 4}, with winding $k$. 
The remaining single mobile fourbrane is subject to a long range attractive potential proportional to 
$1-\cos\theta_k$, where $\theta_k$ is the relative angle between the mobile and bound $D4$-branes,
\bigskip
%%%
\bg\label{thetak}\tan\theta_k={\Delta x^7\over L_1+2\pi R_6 k}\nd
For large $k$ this angle goes like $1/k$,
\bg\label{smallangle}\theta_k\simeq {\Delta x^7\over 2\pi R_6 k}\nd
Since the $M$ bound fourbranes wind $k$ times around the circle, the attractive potential felt by the mobile
fourbrane goes like $V\sim kM(1-\cos\theta_k)\sim M/k$.  Thus, as mentioned above, it becomes more and
more flat as $k$ increases. It would be interesting to see whether it can be made sufficiently flat for inflation to take place. 

The inflationary potential $V$ is due to gravitational attraction between the branes. Thus, it corresponds to a 
D-term potential in the low energy effective description. Therefore, the dynamics studied  here is similar to that
discussed in \cite{BinetruyXJ, HalyoPP}, where it was noted that such models have favorable properties in
supergravity (i.e at finite $G_N$). 

In this picture, the exit from inflation occurs when the mobile $D4$-brane reaches the vicinity of the curved 
fivebrane. There, processes of the sort depicted in figure \ref{figure 5} transfer the energy of the fourbrane to the fivebrane 
and reheat the universe. The endpoint of the dynamical process is a non-supersymmetric gauge theory, with 
gauge group $SU(M-1)\times U(1)$ and fermions in the adjoint $+$ bifundamental representation. 

It is natural to ask whether the early universe cosmology of the model is likely to lead to the type of initial 
conditions assumed in the above discussion. We will only comment on this issue here, leaving a more detailed
study for future work (see \cite{AbelCR, CraigKX, FischlerXH, KutasovKB} for recent discussions of some relevant
issues). At high temperature the system is expected to be in the state with the largest number of massless 
degrees of freedom, which has the lowest free energy. For the moduli space \eqref{genquantmod} this is the branch 
with the largest number of mobile $D4$-branes, i.e the one with $l=0$ (figure \ref{figure 8}). At zero temperature and 
$\xi\not=0$ 
this is not a true minimum of the energy function, but at high temperature this instability is washed out by thermal 
effects. As the temperature decreases, it becomes less stable, and eventually more and more of the mobile 
$D4$-branes undergo the process of figure \ref{figure 5} and collapse onto the curved fivebrane. Thus, an initial state of
the sort assumed in the discussion of inflation above is not particularly unnatural in the early universe evolution
of this system. 
%%%
\subsection{Adding ${\bar D}$-branes to KS}

The authors of \cite{KachruGS}\ proposed that adding anti $D3$-branes to the type IIB brane system of 
\cite{KlebanovHB}\ leads to the appearance of metastable states in which the antibranes expand into 
an $NS5$-brane which can only annihilate via quantum tunneling. Much about these states remains 
mysterious. In the IIB gravity regime, the approximations employed in \cite{KachruGS}\ to establish their 
existence are not obviously reliable. If  these states do exist, 
there is the question whether they should be thought of as metastable states in the Klebanov-Strassler 
gauge theory, or as states in a bigger theory that also contains the supersymmetric KS states.  

In this subsection we will study these issues in the IIA description. Our conclusions will not 
be directly applicable to the IIB regime, or to the gauge theory, since the different regimes  are related 
by large continuous deformations, which may well change the energy landscape. Nevertheless, it seems 
useful to address these questions in any regime where they can be analyzed reliably. 

We start with the brane system studied in the previous sections, with 
\bg\label{pbar}p=kM-\bar p;\hskip 1cm 0<\bar p<M\nd
We saw that this system has a rich moduli space of vacua \eqref{genquantmod}, labeled among other things
by the number of mobile $D4$-branes $p-lM$, $l=0,\cdots, k-1$. Since this number never vanishes, all 
the vacua \eqref{genquantmod}\ belong in this case to mesonic branches. 

Following \cite{KachruGS}, we  start with the vacuum with $l=k-1$, which has $M-\bar p$ mobile $D4$-branes,
and add $\bar p$ pairs of $D4$ and $\bar{D4}$-branes wrapping the circle. The brane configuration now 
contains $M$ $D4$-branes and $\bar p$ $\bar{D4}$-branes, and there are two possible things that can 
happen to it:
\begin{enumerate}
\item The antibranes can annihilate with some of the branes. This takes us back to the mesonic
supersymmetric vacuum with $M-\bar p$ mobile $D4$-branes. \\

\item The $M$ $D4$-branes can combine with the curved fivebrane \eqref{quantumfive}, and increase its
winding from $k-1$ to $k$. This describes the baryonic vacuum of the theory with $p=kM$, but now
we also have $\bar p$ $\bar{D4}$-branes propagating in the vicinity of the curved fivebrane. 
\end{enumerate}

\begin{figure}[htb]
        \begin{center}
\includegraphics[height=4cm]{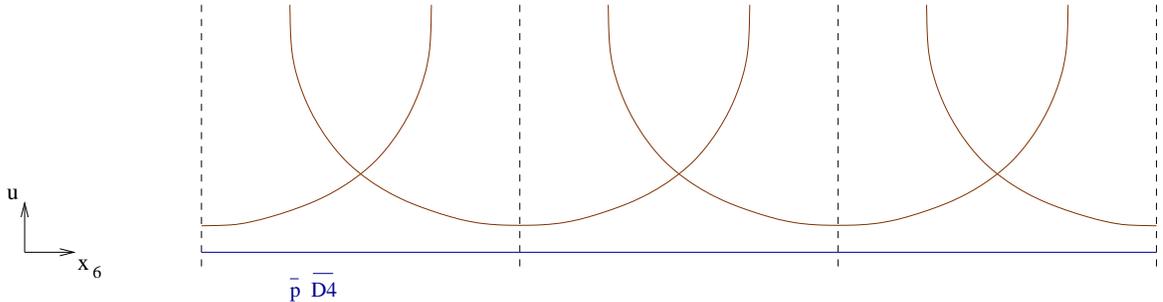}
        \end{center}
        \caption{The baryonic branch of the brane system with $p=kM$,  with $\bar p$ $\bar{D4}$-branes wrapping the circle ($k=1$ in the figure).} \label{figure 10}
        \end{figure}

\noindent
The second possibility gives rise to the metastable state of \cite{KachruGS}.
The $\bar{D4}$-branes, which wrap the $x^6$ circle,  are T-dual to the $\bar{D3}$-branes discussed in 
\cite{KachruGS}. Placing the $\bar{D3}$-branes at the tip of the conifold corresponds in the IIA language to 
placing the $\bar{D4}$-branes at $u=0$ (see figure \ref{figure 10}). In the IIB description it was argued in 
\cite{KachruGS}\ that the antibranes expand into an $NS5$-brane carrying $\bar{D3}$-brane charge. The 
IIA analog of this phenomenon is the following. 

\bigskip

\begin{figure}[htb]
        \begin{center}
\includegraphics[height=3.5cm]{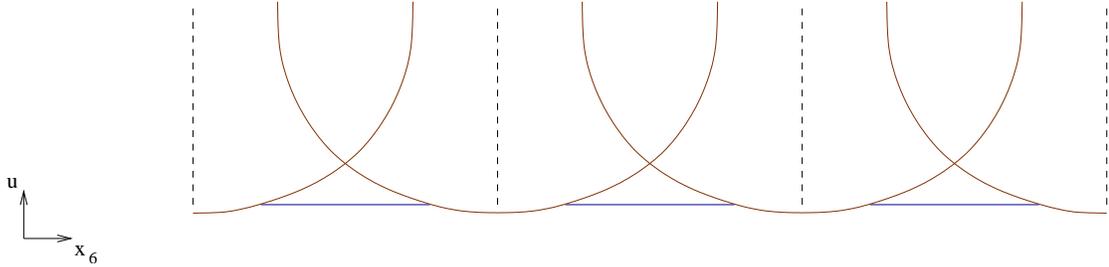}
        \end{center}
        \caption{The configuration of figure \ref{figure 10} is unstable to decay to that depicted here. }\label{figure 11}
        \end{figure}

While the configuration of figure \ref{figure 10} is stationary, it is not stable. The $\bar{D4}$-branes are attracted to the
curved fivebrane (which carries fourbrane charge), and if one displaces them infinitesimally from $u=0$,
will start moving towards the fivebrane. Consider, for example, the case $\bar p=1$. The lowest energy
configuration of the single $\bar{D4}$-brane is qualitatively described by the configuration of figure \ref{figure 11}.   
It can be determined in the probe approximation; we will not describe the details here. 
The $D4$-brane flux carried by the bottom of the fivebrane in figure \ref{figure 11} is $M-\bar p$; the location of the 
$\bar{D4}$-brane is determined by balancing the geometric and electrostatic forces acting on it. 

Since the $\bar{D4}$-brane is displaced from the origin of the $\mathbb R^4$ labeled by $(v,w)$ \eqref{vvww}, the configuration
of figure \ref{figure 11} breaks the $U(1)$ symmetry of the curved fivebrane \eqref{quantumfive}, which acts as (opposite) 
rotations in $v$, $w$. The interpretation of this symmetry in the gauge theory was discussed in \cite{AharonyMI}. 
Its breaking gives rise to a Nambu-Goldstone boson, which corresponds to slow motions of the $\bar{D4}$-brane
on the circle of fixed $u(x^6)$ corresponding to its shape. 

If the number of $\bar D4$-branes, $\bar p$, is larger than one,\footnote{But much lower than $M$, so that we can
neglect their backreaction on the shape of the fivebrane.} each of the $\bar{D4}$-branes can be analyzed as 
above. Since the different $\bar{D4}$-branes repel each other \cite{GiveonSR}, they arrange themselves into a discretized
tube connecting the two sides of the curved fivebrane. This is the IIA manifestation of the $NS5$-brane carrying
$\bar p$ units of $\bar{D}$-brane charge of \cite{KachruGS}. The configuration of figure \ref{figure 11} is locally stable, but 
can decay via tunneling to the supersymmetric mesonic branch with $M-\bar p$ mobile $D4$-brane described
above. 

\begin{figure}[htb]
        \begin{center}
\includegraphics[height=2cm]{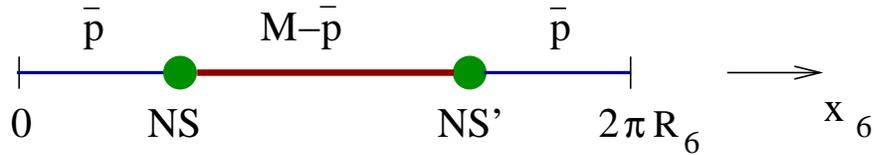}
        \end{center}
        \caption{The low energy description of the metastable vacuum of figure \ref{figure 11} consists (classically) 
of $M-\bar p$ $D4$-branes (red) and $\bar p$ $\bar{D4}$-branes (blue) stretched between 
the $NS5$-branes.}  \label{figure 12}
        \end{figure}

The dynamics described by the brane configuration of figure \ref{figure 11} in various energy regimes can be
understood by starting at small $u$ (low energy) and studying the configuration as we increase $u$. 
For $u$ below the position of the antibranes, the brane configuration is identical to that of figure \ref{figure 7}, 
ie it corresponds to pure $\mathcal N=1$ SYM with gauge group $SU(M-\bar p)$. As we increase $u$, we 
get to the position of the $\bar{D4}$-branes (blue line in figure \ref{figure 11}). Above the corresponding energy, 
we can think of the brane system as describing the quantum vacuum of the brane system of figure \ref{figure 12}. 

The effective gauge theory in this regime is an $SU(M-\bar p)\times SU(\bar p)\times U(1)$ gauge theory
with fermions in the adjoint and bifundamental representation of the gauge group. The bifundamental
fermions are classically massless (figure \ref{figure 12}), but quantum mechanically they acquire a mass via chiral 
symmetry  breaking. This is the field theory analog of the fact that the antibranes are located at a finite 
value of $u$ in figure \ref{figure 11}. Continuing to larger $u$, the brane configuration approaches the baryonic 
vacuum of the theory with $p=lM$, with $l$ increasing up to $k$ at the UV cutoff scale $(u=u_\infty)$. 
The $\bar{D4}$-brane gives rise to a localized perturbation of the curved fivebrane \eqref{quantumfive}.

Interestingly, the effective field theory that describes the metastable SUSY breaking state of \cite{KachruGS},
which corresponds to the brane configuration of figure \ref{figure 12}, is the same as the low energy theory of the
supersymmetric system with non-zero FI parameter $\xi$ (discussed after eq. \eqref{ptilde}), with $\bar p$ here 
playing the role of $\tilde p$ there. From the brane perspective, this is very natural -- the two are related 
by a continuous deformation. 
Indeed, starting with the configuration of figure \ref{figure 6}, one can move the fivebranes towards each other,
such that the winding number $k$ decreases. It is clear from the figure that no states go to zero 
mass in the process; thus, the low energy theory is unchanged by this deformation. Eventually, the 
winding number of the $M-\tilde p$ $D4$-branes vanishes. If we go once more around
the circle, these branes reverse their orientation, and we end up with a configuration similar to that
of figure \ref{figure 12}, with the two $NS5$-branes displaced relative to each other in $x^7$. However, it is clear
from figure \ref{figure 12} and the analysis of \cite{gkunpub} that this displacement also does not change the low
energy spectrum and dynamics. 

Thus, we see that the brane systems of figure \ref{figure 6}, and figure \ref{figure 11} correspond  to different UV completions of the 
$SU(M-\bar p)\times SU(\bar p)\times U(1)$ gauge theory described above. In particular, in figure \ref{figure 6} 
(and 12) supersymmetry is broken in the ground state, while in figure \ref{figure 11} the same low energy theory 
arises as an effective infrared theory in a metastable ground state. 

An interesting and widely discussed question is whether the metastable state of \cite{KachruGS}\ is a state in the cascading gauge theory (see e.g.
\cite{BenaXK, DymarskyPM, BlabackNF, MassaiJN}). In the IIA regime the answer appears to be negative for the following reason. The gauge theory provides a low energy description of the brane system of figure \ref{figure 1}, or its quantum version discussed in section \ref{section 3}. While one can arrange the parameters of the model such that the metastable state of figure \ref{figure 1} has a small energy density, the height of the barrier for the tunneling to the supersymmetric state is determined by the energy (density) of $\bar p$ $D/\bar D$ pairs wrapping the circle. For $\bar p=1$ this energy is (in string units) $E\sim R_6/g_s$. Using \eqref{couplbar}\ one can write it as $E\sim 1/g^2 k$, where $g$ is the four dimensional gauge coupling of the low energy theory. Thus, for finite $g, k,$ the height of the barrier between the supersymmetric and non-supersymmetric vacua is finite in string units, and hence the tunneling between the two goes to zero in the gauge theory limit. This should be contrasted with the situation in brane constructions of metastable vacua that are visible in the gauge theory, such as that of \cite{GiveonEW}, where all energy scales, including the height of the barrier, can be taken to be small. Thus, we conclude that while the configuration of figure \ref{figure 11} is metastable in the full string theory, it is stable in the low energy theory. It corresponds to a different superselection sector of the theory on the branes from the supersymmetric vacua. 
%%%
\newpage
\section{$\mathcal N=2$ cascade}\label{section 5}

$\mathcal N=2$ supersymmetric gauge theories are known to exhibit cascading behavior similar to that found for $\mathcal N=1$ in \cite{KlebanovHB}\ (see e.g \cite{PolchinskiMX, BeniniIR, DasguptaSW}). At first sight this is puzzling, since $\mathcal N=2$ supersymmetric QCD does not exhibit Seiberg duality. As we saw above, the type IIA description provides a useful guide for studying the classical and quantum vacuum structure of cascading gauge theories. In this section we will use it to shed light on the $\mathcal N=2$ duality cascade. 

\begin{figure}[htb]
        \begin{center}
\includegraphics[height=2cm]{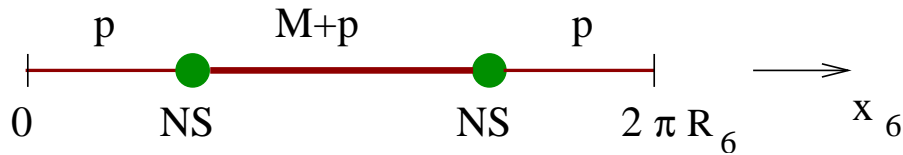}
        \end{center}
        \caption{The IIA brane configuration that realizes the $\mathcal N=2$ supersymmetric cascading gauge theory.}  \label{figure 13}
        \end{figure}

The brane configuration corresponding to the gauge theory we are interested in is a close analog of that of figure \ref{figure 1}, and is depicted in figure \ref{figure 13}. The different branes are oriented as in section \ref{section 2} (see \eqref{braneor}); the fact that the $NS5$-branes are parallel implies that this configuration preserves eight supercharges, or $\mathcal N=2$ supersymmetry in the $3+1$ dimensions $(0123)$. The low energy theory is in this case an $\mathcal N=2$ SYM theory with the gauge group\footnote{We will include in the gauge group the $U(1)$ factors, which were omitted in \eqref{ggroup}.} and matter content \eqref{ggroup}\ -- \eqref{transfab}. The superpotential \eqref{www}\ is now absent and is replaced by the standard $\mathcal N=2$ superpotential that couples the adjoints in the vector multiplet of $G$ \eqref{ggroup}\ to the bifundamentals \eqref{transfab}. If one breaks $\mathcal N=2$ SUSY by giving a mass to the adjoints (which corresponds in the brane picture to a relative rotation of the two fivebranes in $(v,w)$) one can recover \eqref{www}\ by integrating them out.   

In the rest of this section we will repeat the discussion of sections \ref{section 2}, \ref{section 3} for the $\mathcal N=2$ supersymmetric case, and describe the classical and quantum supersymmetric vacuum structure of the brane system of figure \ref{figure 13}. It should be clear from the $\mathcal N=1$ analysis above, and from the study of many other systems reviewed in \cite{GiveonSR}, that the results apply to (and can be stated in terms of) the low energy $\mathcal N=2$ SQCD. The brane picture merely provides a useful language for describing the vacuum structure. 

\subsection{Classical moduli space} 

The basic fact that governs the classical moduli space of the brane configuration of figure \ref{figure 13} is that fourbranes stretched between the two fivebranes (``fractional branes'') are free to move along the fivebranes, in the $v$ plane \eqref{vvww}, while fourbranes that wrap the whole circle (``regular branes'') are free to move in the whole transverse $\mathbb R^5$ labeled by $(45789)$. An example of a branch of the classical moduli space is  the Coulomb branch for the two gauge groups, which corresponds in figure \ref{figure 13} to displacing the $M+p$ coincident $D4$-branes to arbitrary positions $v_i$, $i=1,\cdots, M+p$, and the $p$ $D4$-branes connecting the fivebranes on the other side of the circle to $\tilde v_a$, $a=1,\cdots, p$. At  a generic point in this moduli space (with all $v$, $\tilde v$ distinct) the gauge group is broken to $U(1)^{M+2p}$. When one of the $v$'s and one of the $\tilde v$'s coincide, a bifundamental hypermultiplet goes to zero mass and a new branch of moduli space opens up. In the brane language it corresponds to the two fractional branes connecting into a regular brane, which can move off the fivebranes into the aforementioned $\mathbb R^5$. 

\begin{figure}[htb]
        \begin{center}
\includegraphics[height=6cm]{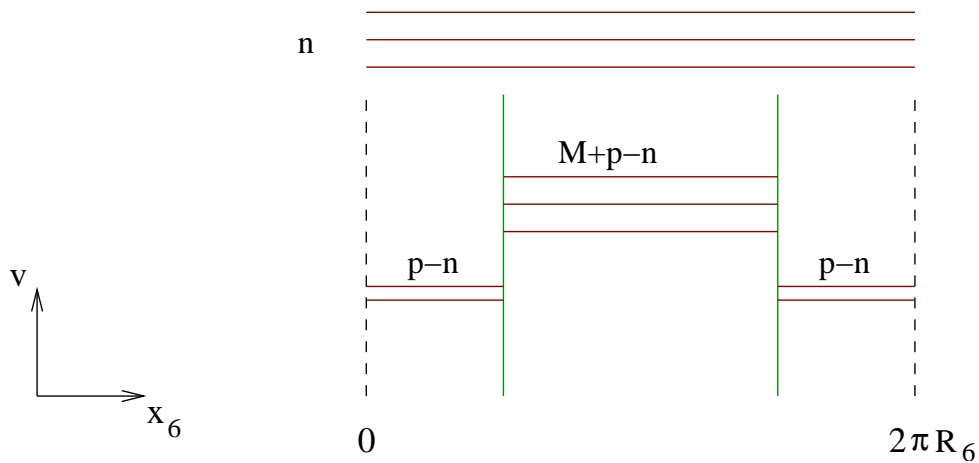}
        \end{center}
        \caption{The brane description of $\mathcal M_n$, a component of the classical moduli space, has $n$ fourbranes wrapping the circle moving in the transverse $\mathbb R^5$ (in general away from the fivebranes), and $M+p-n$ resp. $p-n$ fourbranes connecting the fivebranes and distributed in the $v$ plane.}  \label{figure 14}
        \end{figure}

The full classical moduli space is a direct sum of spaces $\mathcal M_n$, $n=0,1,2,\cdots, p$, which are described in the brane language by the configuration of figure \ref{figure 14}. As is clear from the picture, at a generic point in the moduli space the low energy theory includes $\mathcal N=4$ SYM with gauge group $U(1)^n$, and pure $\mathcal N=2$ SYM with gauge group $U(1)^{M+2p-2n}$. The different $\mathcal M_n$ intersect on subspaces where some charged hypermultiplets go to zero mass, and have other singular points at which charged vector multiplets become massless and enhance the gauge group. 

As in the $\mathcal N=1$ case, the FI coupling $\xi$ for $U(1)_b$ corresponds in the brane picture of figure \ref{figure 13} to a relative displacement of the two $NS5$-branes in $x^7$. In general, this leads to non-supersymmetric vacua of the sort discussed in section \ref{section 2} (around figure \ref{figure 3}), while for $p=kM$ with integer $k$ one finds supersymmetric vacua of the sort discussed around figure \ref{figure 4}. These vacua involve fourbranes connecting fivebranes while winding ($k$ times) around the $x^6$ circle. 

In the $\mathcal N=2$ case one can also displace the fivebranes in the $(89)$ plane. This corresponds in the gauge theory to turning on a linear superpotential $W=\lambda{\rm Tr}\Phi$ for the chiral superfield in the $U(1)_b$ vector multiplet. It has a similar effect on the vacuum structure to that of the FI term. In fact, $(\xi,\lambda)$ transform as a triplet under the $SU(2)_R$ symmetry of the $\mathcal N=2$ SYM theory, which corresponds in the brane language to the rotation symmetry  $SO(3)_{789}$.

\subsection{Quantum moduli space}

Going from classical to quantum gauge theory corresponds in the IIA brane system to turning on a finite string coupling $g_s$. When one does that, the system of two $NS$-branes connected by $N$ $D4$-branes becomes a single connected $NS5$-brane carrying $D4$-brane charge \cite{WittenMT}\footnote{In \cite{WittenMT}, $g_s$ was taken to be large. In this limit the bulk spacetime becomes eleven dimensional, and the fivebrane in question becomes an $M5$-brane. As discussed in \cite{AharonyMI}, one can alternatively consider the limit $g_s\ll1$, $g_sN\gg1$, in which the right description is in terms of an $NS5$-brane in weakly coupled type IIA string theory.} For example, the Coulomb branch discussed above, which corresponds to figure \ref{figure 14} with $n=0$, is described by a curved fivebrane which looks asymptotically (at large $v$) like a pair of curved $NS5$-branes with the profile $z\sim \pm\lambda_M\ln v$ (see the discussion around eq. \eqref{defzz}\ for the notation), connected by $M+p$ respectivelly $p$ tubes.  The precise form of the curved fivebrane is described in \cite{WittenMT}. 

In the $\mathcal N=1$ supersymmetric case we saw (in section \ref{section 3}) that the quantum vacuum structure is richer than the classical one. The basic reason for that is that configurations which are identical in the classical limit become distinct at finite $g_s$. In particular, the classical configuration of $M$ $D4$-branes connecting the fivebranes with $M$ additional fourbranes wrapping the circle and intersecting the fivebranes, can be viewed as the classical limit  of either the Higgs  branch (figure \ref{figure 8}) or the baryonic branch (figure \ref{figure 9}). We expect the same to happen in the $\mathcal N=2$ case. 

Consider, for example,\footnote{It is easy to generalize the discussion to other branches of the moduli space.} the branch of moduli space with $n=p$ in figure \ref{figure 14}. In this branch, the theory generically reduces at low energies to a direct product of $\mathcal N=2$ SYM with gauge group $SU(M)$ along its Coulomb branch, and $p$ copies of $U(1)$ $\mathcal N=4$ SYM. The configuration of figure \ref{figure 14} describes the classical moduli space; quantum mechanically, the fourbranes connecting the two fivebranes become finite tubes. Together with the $NS5$-branes they make the curved fivebrane \cite{WittenMT}
\bg\label{curvefive}t^2+B(v)t+1=0,\nd
where $t=\exp(-z/R)$ and $B(v)=v^M+u_2v^{M-2}+\cdots +u_M$. As in the $\mathcal N=1$ case, we can introduce a UV cutoff by taking $|v|$ to be bounded, $|v|\le v_\infty$, and demand that the distance between the two arms of the curved fivebrane at the cutoff scale is equal to some fixed length $L_1<2\pi R_6$, ``the distance between the fivebranes''.  If the moduli $u_2,\cdots, u_M$ are small relative to the cutoff scale $v_\infty$, one has 
\bg\label{distarms}L_1\simeq 2\lambda_M\ln (v_\infty/\zeta), \nd
with $\zeta$ a scale that was set to one before. 

Following the discussion of the $\mathcal N=1$ case, one can obtain additional branches of the quantum moduli space by taking $lM$ of the $p$ mobile $D4$-branes to coincide with the $M$ $D4$-branes stretched between the fivebranes, and consider the quantum configuration corresponding to $M$ fourbranes connecting the two $NS$-branes while winding $l$ times around the circle, together with $p-lM$ mobile $D4$-branes in the bulk of the $\mathbb R^5$.  For $p$ of the form \eqref{ptilde}, the maximal value of $l$ is $l_{\rm max}=k$, and if $\tilde p=0$, one has in that case a close analog of the baryonic branch of the $\mathcal N=1$ supersymmetric theory of section \ref{section 3}. The low energy theory in this branch is pure $\mathcal N=2$ SYM with gauge group $SU(M)$, and the moduli space is its Coulomb branch. The curved fivebrane is again described by \eqref{curvefive}, but now the distance between the two arms at the UV cutoff scale, which enters \eqref{distarms},  is $L_1+2\pi k R_6$, as in the $\mathcal N=1$ discussion. Hence the fivebrane winds $k$ times around the $x^6$ circle.

An important difference with respect to the $\mathcal N=1$ discussion is that for $\mathcal N=2$, every time the curved fivebrane winds around the circle it intersects itself at $2M$ points \footnote{To find these points one needs to calculate the intersections of the curve \eqref{curvefive}\ with another copy of this curve, in which $t\to It$ (see \eqref{imp}\ for the definition of $I$).}. This self intersection is very similar to the one discussed in appendix A. As there, each intersection point supports a $U(1)$ vector multiplet and a massless charged hypermultiplet. 

The rest of the discussion is similar to the $\mathcal N=1$ case. The fivebrane \eqref{curvefive}\ that winds $l$ times around the circle describes a particular branch of the moduli space of the theory corresponding to figure \ref{figure 13} with $p=kM$. By decreasing the value of the UV cutoff $v_\infty$ one can also view it as a vacuum of the theory with $p=(k-1)M, (k-2)M$, etc, together with $2M, 4M,\cdots$ decoupled sectors consisting of a vector multiplet and a charged hypermultiplet . If we neglect these decoupled sectors, we conclude that the theories with $p=lM$ with different values of $l$ share part of their moduli space of  vacua. Of course, there are some branches of the moduli space that are different as well. That was already the case in the $\mathcal N=1$ case \cite{DymarskyXT}, but for $\mathcal N=2$ there are more branches of moduli space, and naturally more of them are different in theories with different values of $l$. 

To understand the origin of the $\mathcal N=2$ duality cascade from the point of view of the low energy gauge theory, consider the simplest case $p=M$. The vacuum structure of the resulting $U(2M)\times U(M)$ gauge theory can be analyzed by studying first the limit where the $U(M)$ gauge coupling is very small. Then we have a $U(2M)$ $\mathcal N=2$ SQCD with $N_f=2M$ flavors. As we review in appendix A, this theory has a baryonic branch, whose root is described at low energies by a $U(1)^{2M}$ gauge theory with $2M$ hypermultiplets charged under the different $U(1)$ factors \cite{ArgyresEH}, see eq. (A.1). These fields are all singlets under the $SU(N_f)$ global symmetry. Thus, gauging $U(M)$ does not influence them, and the full low energy theory at the root of the baryonic branch is a direct product of the above abelian sector and the Coulomb branch of pure $U(M)$ $\mathcal N=2$ SYM. This picture is in complete agreement with the brane description above. The baryonic branch of the moduli space is described by a curved fivebrane \eqref{curvefive}\ that winds once around the circle. The abelian factors live at the $2M$ self intersections of this curve, while the small $v$ shape of the fivebrane describes the Coulomb branch of the low energy $U(M)$ pure SYM. Clearly, one can iterate this procedure to describe the vacuum structure of theories with larger $p$, as was done in \cite{DymarskyXT}\ for the $\mathcal N=1$ case.

To summarize, if one neglects the abelian sectors, one finds that the $U(lM)\times U((l-1)M)$ gauge theories at the root of their baryonic branches are all equivalent, and flow in the IR to pure $U(M)$ $\mathcal N=2$ SYM; this equivalence is manifest in the brane description.  This is the origin of the cascading behavior seen in the IIB description in  \cite{PolchinskiMX,BeniniIR,DasguptaSW}. The cascading geometries in these papers appear to describe the dual of the curved fivebrane, whereas the abelian factors that distinguish theories with different values of $l$ presumably correspond to singletons, that live at the boundary of the space.

As mentioned above, the full quantum moduli space of the $\mathcal N=2$ gauge theory with general $p$ is quite intricate. For example, starting with the classical moduli space $\mathcal M_n$ of figure \ref{figure 14}, we can take $l_1(M+p-n)$ of the mobile $D4$-branes and attach them to the $M+p-n$ $D4$-branes stretched between the fivebranes, making them wind $l_1$ times around the circle; similarly we can attach $l_2(p-n)$ of the remaining mobile $D4$-branes to the $p-n$ stretched $D4$-branes in figure \ref{figure 14}, and make them wind $l_2$ times around the circle. This gives new branches of moduli space labeled by $(l_1, l_2, n)$, which satisfy
\bg\label{newbranches}l_1(M+p-n)+l_2(p-n)\le n\nd
The discussion of this section can be generalized to these vacua as well.

\chapter{Conclusion}

Aiming at making this thesis self contained, we reviewed in Chapter 2 the building blocks of string theory required to understand the results presented in this text; emphasizing on the interplay between brane dynamics and supersymmetric gauge theories.  We then revisited the original work of Sen \cite{senF} and  Banks-Douglas and Seiberg \cite{bds} on embedding four-dimensional $\mathcal N=2$ superconformal field theory \cite{sw1,sw2} in F-theory \cite{vafaF} with geometry $\mathbb T^2/\mathbb Z_2\times \mathbb R^4\times \mathbb R^{0123}$.  We presented three possible background deformations of the original Sen's F-theory construction, each of which leading to a plethora of new dualities and physical phenomena that were analyzed in great detail in \cite{DasguptaSW}.

In Chapter 3, we focused on one particular construction consisting in interchanging the Higgs branch geometry $\mathbb R^4$ by $\mathbb R^4/\mathbb Z_k$,  letting multiple D3-branes probe parallel seven-branes wrapped on $k$-center Taub-NUT spaces. The resulting string theory is type IIB on $\mathbb T^2/\mathbb Z_2\times \mathbb R^4/\mathbb Z_2\times \mathbb R^{0123}$ or equivalently F-theory on $K3\times \mathbb R^4/\mathbb Z_2\times \mathbb R^{0123}$.  This construction enabled us to embed a class of Gaiotto linear model \cite{gaiotto} in F-theory as well as the various T-dual brane networks of \cite{bbt}.  However, much work remains to be done.  In particular, we have not shown explicitly how to see the weakly coupled regime emerge from the strong coupling limit of our construction, as one would expect it from generalizations of S-duality to $\mathcal N=2$ SYM theories.  This is partly due to the fact that only aspects of Argyres-Seiberg duality were captured by our model in Chapter 3 as we have not yet included color branes in the story.  How to capture Gaiotto $\mathcal N=2$ SCFT with more then 24 flavors still remains an open question given the limitation of F-theory to contain at most 24 flavors branes.  In the non-conformal limit, we proposed a geometry which seemed to lead to a cascade mechanism in $\mathcal N=2$ supersymmetric gauge theory.  This statement was supported by mapping the corresponding non-conformal $\mathcal N=2$ construction to the $\mathcal N=1$ Klebanov-Strassler \cite{KlebanovHB} geometry by nontrivially fibering the Taub-NUT space over the compactified $u$-plane.  Limited by the complexity of the type IIB/F-theory language, we could not provide in this language further description of the cascade dynamic of non-conformal supersymmetric $\mathcal N=2$ gauge theories in four-dimensions.   

Still intrigued by this phenomenon, we decide to provide further evidence in Chapter 4 for an $\mathcal N=2$ cascade behavior by turning to type IIA / M-theory language.  
The main conclusion of Chapter 4 is that the IIA brane description provides a useful qualitative and quantitative guide to the dynamics of cascading gauge theories with various amounts of supersymmetry. In particular, we saw that for the $\mathcal N=1$ cascading theory of \cite{KlebanovHB}, the classical and quantum moduli spaces of supersymmetric vacua agree.
%including the dependence of the dynamically generated scale \eqref{epsrl}\ on the parameters labeling different  branches of moduli space (which is given in the brane picture by \eqref{xill}, \eqref{imp}). 
The brane picture makes it clear that the cascade utilizes a weak form of Seiberg duality, which involves deformed SQCD, and can be proven regardless of whether the stronger version of the duality holds. 
We also saw that the brane picture provides a useful guide to the non-supersymmetric dynamics of the theory.  In particular, we discussed the stable non-supersymmetric vacuum obtained for non-zero FI parameter and generic number of fractional and regular D4-branes and the dynamics as one approaches it from vacua on the classical pseudo moduli space. It would be interesting to find the IIB geometry corresponding to the stable non-supersymmetric vacuum of figure (\ref{figure 6}).

Futhermore, we saw in Chapter 4 that the metastable state described in IIB language in \cite{KachruGS}\ 
has a IIA analog. The fact that this state exists in the regime of parameter states where the IIA description
is reliable supports the construction of \cite{KachruGS}. In the IIA regime this state is clearly metastable,
and decays to the same supersymmetric state as in the proposal of \cite{KachruGS}. An interesting open
question is whether this state exists also in the gauge theory. From the IIA point of view this appears
to be unlikely. To get it we added a $D4/\bar{D4}$ pair to the theory with $p=kM-\bar p$. This seems to 
lead to a system with more degrees of freedom than the original $SU(p)\times SU(M+p)\times U(1)$
gauge theory. This is reflected in the fact that the height of the barrier between the non-supersymmetric and supersymmetric vacua goes to infinity in the gauge theory limit.  We also noted that the low energy dynamics of the metastable  state is closely related to that of the non-supersymmetric state at non-zero FI term. As we saw, this is very natural from the brane 
description.  

We concluded Chapter 4 by generalized the discussion to systems with $\mathcal N=2$ supersymmetry. The type IIA description clarifies why they exhibit cascading behavior despite the fact that Seiberg duality is not a symmetry of such theories. This is due to the fact that while the full theory does not exhibit Seiberg duality, certain vacua do. Thus, some of the vacua of the $\mathcal N=2$ theory with gauge group $U(M+p)\times U(p)$ are 
shared by theories with $p\to p-M, p-2M,\cdots$. Even in these vacua the equivalence is not complete -- theories with higher $p$ differ from those with lower one by a decoupled sector with an abelian gauge group coupled to charged hypermultiplets.  It would be interesting to complete this work by providing an analyse of the vacuum structure of the $SU(p)\times SU(M+p)$ $\mathcal N=2$ SYM theory \`a la \cite{DymarskyXT}.  Furthermore, understanding the dynamics of the $\mathcal N=2$ cascade in the language of type IIB still remains an unsolved problem: the interesting part would be to show how the charged hypermultiplets occur in type IIB/F-theory.  
\newpage
\appendix
\chapter{Aspects of the IIA description of $\mathcal N=2$ SQCD}
\begin{figure}[htb]
        \begin{center}
\includegraphics[height=4.5cm]{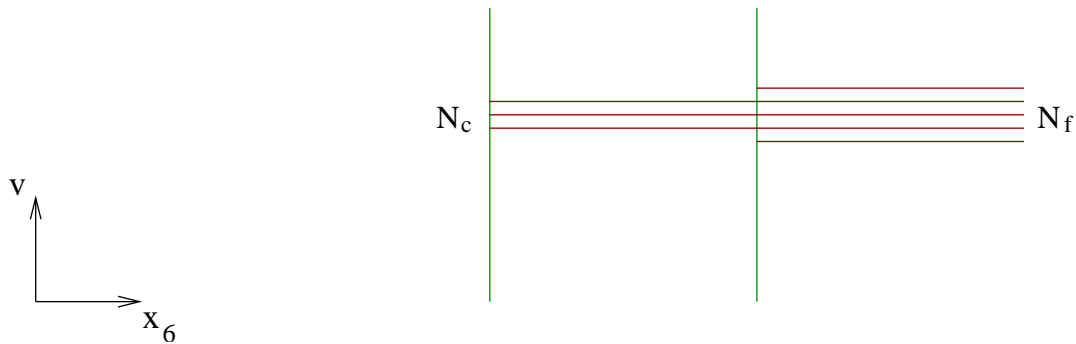}
        \end{center}
        \caption{The brane description of $\mathcal N=2$ SQCD with $N_c$ colors and $N_f$ flavors.}  \label{figure 15}
        \end{figure}

%%%
$\mathcal N=2$ SQCD with gauge group $U(N_c)$ and $N_f$ hypermultiplets in the fundamental representation of the gauge group can be described by the brane configuration of figure \ref{figure 15}.

It consists of two $NS$-branes (see \eqref{braneor}\ for the orientations of the  branes) connected by $N_c$ (``color'') $D4$-branes, which give rise to $\mathcal N=2$ SYM with gauge group $U(N_c)$. $N_f$ (``flavor'') $D4$-branes attached to one of the fivebranes give hypermultiplets in the fundamental representation of the gauge group. In order to study the full moduli space of vacua of the theory one needs to terminate the $N_f$ flavor branes on $D6$-branes, but since this is not going to be important for our purposes, we will keep them semi-infinite. 

The classical and quantum vacuum structure of $\mathcal N=2$ SQCD was analyzed in \cite{ArgyresEH}\footnote{These authors studied the case of $SU(N_c)$ gauge group, but the theory with gauged baryon number is closely related.}. Our main interest is going to be in the parameter range $N_c<N_f<2N_c$, and in the baryonic branch, in which the gauge symmetry is in general completely broken. At the origin of this branch the classical theory has an unbroken $U(N_c)$ gauge symmetry, but quantum effects are large. The authors of \cite{ArgyresEH}\ showed that in the quantum theory, the origin of the baryonic branch has an alternative weakly coupled description with gauge group 
\bg\label{ttlldd}U(\tilde N_c)\times U(1)^{N_c-\tilde N_c};\qquad \tilde N_c=N_f-N_c\nd
The matter consists of $N_f$ hypermultiplets in the fundamental of $U(\tilde N_c)$ which are not charged under the $U(1)$'s, and $N_c-\tilde N_c$ hypermultiplets $e_i$ which are singlets of $U(\tilde N_c)$ and charged under the $U(1)$'s (the latter can be normalized such that $e_i$ has charge $-\delta_{ij}$ under the $j$'th $U(1)$). 

\begin{figure}[htb]
        \begin{center}
\includegraphics[height=5cm]{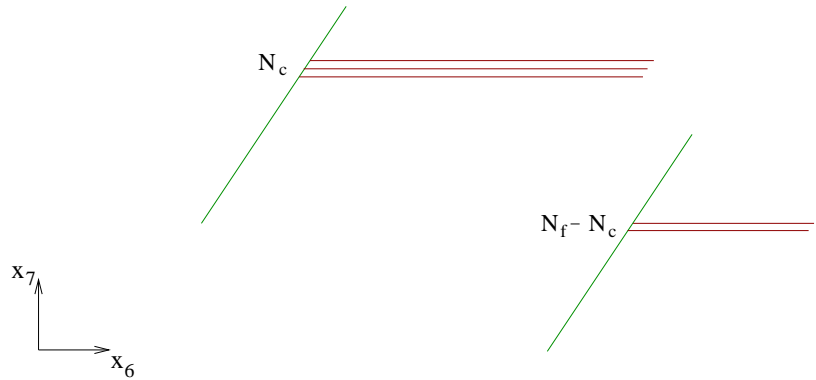}
        \end{center}
        \caption{The brane description of $\mathcal N=2$ SQCD at finite $\xi$.}  \label{figure 16}
        \end{figure}

\begin{figure}[htb]
        \begin{center}
\includegraphics[height=6cm]{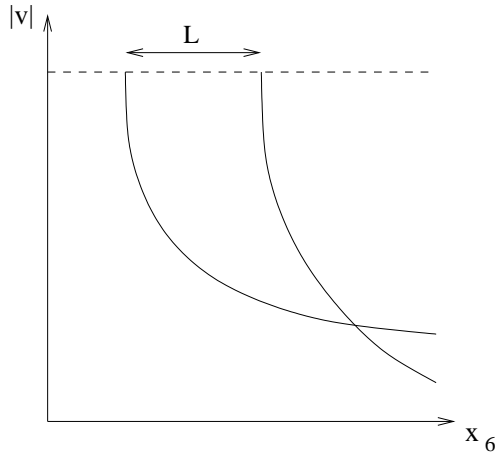}
        \end{center}
        \caption{The origin of the baryonic branch of $\mathcal N=2$ SQCD in the quantum theory. The dashed line corresponds to the UV cutoff $|v|=|v_\infty|$.}  \label{figure 17}
        \end{figure}

From the brane perspective, this can be understood as follows. As discussed in the text, one can take the theory into the baryonic branch by turning on the FI parameter $\xi$, which corresponds in the brane description to a relative displacement of the $NS$-branes in the $x^7$ direction. For finite $\xi$ the $U(N_c)$ gauge symmetry is broken and the brane system splits into two disconnected components (see figure \ref{figure 16}). As $\xi\to 0$ the $U(N_c)$ gauge symmetry is restored, and quantum effects become important. Thus, we have to replace the brane system of figure \ref{figure 16} by its finite $g_s$ analog \cite{WittenEP,WittenMT}. The two fivebranes in figure \ref{figure 16} take the forms
\bg\label{formnc}v^{N_c}=t~,\qquad v^{\tilde N_c}=\zeta^{\tilde N_c} t\nd
respectively. Here we used the freedom of choosing the origin in $x^6$, $x^{11}$ to set the coefficient of $t$ to one for one of the two fivebranes. The constant $\zeta$ can be determined by imposing the boundary conditions that at $|v|=|v_\infty|$ the two fivebranes are separated by the distance $L$, 
\bg\label{formzeta}|\zeta|^{\tilde N_c}={e^{L/ R}\over |v_\infty|^{N_c-\tilde N_c}}\nd
Viewed in the $(x_6, |v|)$ plane, the fivebranes take the form depicted in figure \ref{figure 17}.

\begin{figure}[htb]
        \begin{center}
\includegraphics[height=4.5cm]{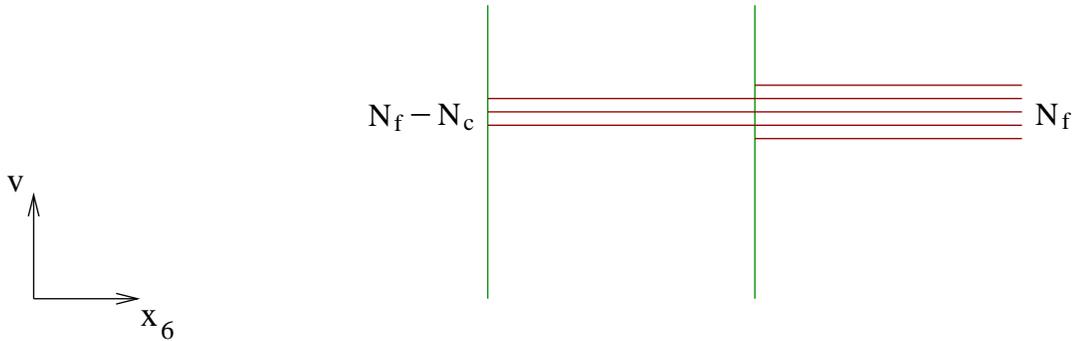}
        \end{center}
        \caption{The classical limit of the small $v$ limit of the brane configuration of figure \ref{figure 17}.}  \label{figure 18}
        \end{figure}

\noindent
The fact that the two fivebranes approach each other as $|v|$ decreases reflects the growth of the gauge coupling in the infrared. At some point the fivebranes intersect and cross, and for smaller $|v|$ (i.e. low energy), their ordering in $x^6$ is reversed. As we further lower $|v|$, the distance between the fivebranes increases, reflecting the infrared freedom of the low energy effective theory. To see what that theory is we need to take the classical limit of the resulting brane configuration, which is depicted in figure \ref{figure 18}.

\noindent
It is a $U(\tilde N_C)$ $\mathcal N=2$ SQCD with $N_f$ flavors, which is indeed not asymptotically free (and is thus weakly coupled in the IR). This theory is very similar to that found in \cite{ArgyresEH}, \eqref{ttlldd}, but it is missing the $U(1)$ factors in the gauge group and the charged hypermultiplets $e_i$. 

It is clear from figure \ref{figure 17} that these must come from the fivebrane intersection. While it seems from the figure that the two component fivebranes intersect at a single point, in fact there are $N_c-\tilde N_c$ intersection points (at finite $t$), which can be obtained by imposing both equations in \eqref{formnc}. This gives 
\bg\label{intersect}v^{N_c-\tilde N_c}=\zeta^{-\tilde N_c}\nd
which has $N_c-\tilde N_c$ solutions lying on a circle of fixed $|v|$. Comparing to \eqref{ttlldd}, it is natural to conjecture that each intersection supports a $U(1)$ vector multiplet and a charged hypermultilet. It should be possible to show this directly in string theory, but we will not attempt to do this here. 

Note that here we are interpreting the radial direction transverse to the $D4$-branes, $|v|$, as parametrizing energy, with small (large) $v$ corresponding to low (high) energies. It may seem peculiar from this point of view that some of the massless degrees of freedom, namely the $U(1)$ factors in \eqref{ttlldd}\ live at finite $v$, \eqref{intersect}.  This phenomenon is actually familiar from the study of brane systems in string theory. The non-abelian degrees of freedom associated with such systems (say, the $SU(N_c)$ part of the gauge group) typically live in the near-horizon region of the branes, while the $U(1)$ factors are localized in the interface between the near and far regions.

To summarize, the brane system of figure \ref{figure 15} provides a simple way to understand the dual description of the root of the baryonic branch of $\mathcal N=2$ SQCD \eqref{ttlldd}. This description is in the spirit of \cite{ElitzurFH}; the non-abelian factor in the dual gauge group arises from brane exchange (which happens here as a function of RG scale), and the $U(1)$ factors and charged hypermultiplets live at self-intersections of the quantum fivebrane \footnote{The brane description also makes it clear that the dual description of the root of the baryonic branch \eqref{ttlldd}\ is related to the microscopic $\mathcal N=2$ SQCD in a simpler way than the magnetic Seiberg dual theory \cite{SeibergPQ}\ is related to the microscopic electric theory in the $\mathcal N=1$ supersymmetric case. In particular, while in the former case one can derive the dual (or effective low energy) description from the microscopic one, in the latter no such derivation is known.}. Turning on a FI term in the microscopic theory corresponds in the low energy description to a FI term for the overall $U(1)$ in $U(\tilde N_c)$ and all the $U(1)$ factors in \eqref{ttlldd}, which Higgses the gauge group and gives masses to the $e_i$. In the brane description this corresponds to separating the two component fivebranes in figure \ref{figure 17} in $x^7$, so they no longer intersect, and all degrees of freedom associated with the intersections become massive.

The authors of \cite{ArgyresEH}\ also discussed what happens to the theory when one breaks $\mathcal N=2$ supersymmetry down to $\mathcal N=1$ by giving a mass to the adjoint chiral superfield in the $(S)U(N_c)$ vector multiplet. In the brane description this corresponds to rotating one of the $NS$-branes in figure \ref{figure 15} from the $v$ to the $w$ plane. Since the fivebranes are no longer parallel, the curves in figure \ref{figure 17} do not intersect in the extra dimensions. This is the brane reflection of the fact that in this case the charged chiral hypermultiplets $e_i$ get a non-zero vev,  Higgs the $U(1)$ gauge group, and lift to non-zero mass all states associated with the intersections.

\bibHeading{References}
\bibliography{Wissanji_thesis}

\begin{thebibliography}{10}

\bibitem{AbelCR}
Steven~A. Abel, Chong-Sun Chu, Joerg Jaeckel, and Valentin~V. Khoze.
\newblock {SUSY breaking by a metastable ground state: Why the early universe
  preferred the non-supersymmetric vacuum}.
\newblock {\em JHEP}, 0701:089, 2007.

\bibitem{aharonyn2}
Ofer Aharony.
\newblock {A Note on the holographic interpretation of string theory
  backgrounds with varying flux}.
\newblock {\em JHEP}, 0103:012, 2001.

\bibitem{AharonyMI}
Ofer Aharony, David Kutasov, Oleg Lunin, Jacob Sonnenschein, and Shimon
  Yankielowicz.
\newblock {Holographic MQCD}.
\newblock {\em Phys.Rev.D}, 82:106006, 2010.

\bibitem{benini2}
Riccardo Argurio, Francesco Benini, Matteo Bertolini, Cyril Closset, and
  Stefano Cremonesi.
\newblock {Gauge/gravity duality and the interplay of various fractional
  branes}.
\newblock {\em Phys.Rev.}, D78:046008, 2008.

\bibitem{plesser}
Philip~C. Argyres, M.~Ronen Plesser, and Alfred~D. Shapere.
\newblock {The Coulomb phase of N=2 supersymmetric QCD}.
\newblock {\em Phys. Rev. Lett.}, 75:1699--1702, 1995.

\bibitem{ArgyresEH}
Philip~C. Argyres, Ronen~M. Plesser, and Nathan Seiberg.
\newblock {The Moduli space of vacua of N=2 SUSY QCD and duality in N=1 SUSY
  QCD}.
\newblock {\em Nucl.Phys.B}, 471:159, 1996.

\bibitem{ArgyresSeiberg}
Philip~C. Argyres and Nathan Seiberg.
\newblock {S-duality in N=2 supersymmetric gauge theories}.
\newblock {\em JHEP}, 12:088, 2007.

\bibitem{aspinwall}
Paul~S. Aspinwall.
\newblock {Enhanced gauge symmetries and K3 surfaces}.
\newblock {\em Phys.Lett.}, B357:329--334, 1995.

\bibitem{BDGreen}
Constantin~P. Bachas, Michael~R. Douglas, and Michael~B. Green.
\newblock {Anomalous creation of branes}.
\newblock {\em JHEP}, 9707:002, 1997.

\bibitem{bk}
Dong-su Bak and Andreas Karch.
\newblock {Supersymmetric brane-antibrane configurations}.
\newblock {\em Nucl. Phys.}, B626:165--182, 2002.

\bibitem{bds}
Tom Banks, Michael~R. Douglas, and Nathan Seiberg.
\newblock {Probing F-theory with branes}.
\newblock {\em Phys. Lett.}, B387:278--281, 1996.

\bibitem{BBS}
Katrin Becker, Becker Melanie, and John~H. Schwarz.
\newblock {\em {String Theory and M-theory: A Modern Introduction}}.
\newblock Cambridge University Press, 2007.

\bibitem{BenaXK}
Iosif Bena, Mariana Grana, and Nick Halmagyi.
\newblock {On the Existence of Meta-stable Vacua in Klebanov-Strassler}.
\newblock {\em JHEP}, 1009:087, 2010.

\bibitem{bbt}
Francesco Benini, Sergio Benvenuti, and Yuji Tachikawa.
\newblock {Webs of five-branes and N=2 superconformal field theories}.
\newblock {\em JHEP}, 09:052, 2009.

\bibitem{BeniniIR}
Francesco Benini, Matteo Bertolini, Cyril Closset, and Stefano Cremonesi.
\newblock {The N=2 cascade revisited and the enhancon bearings}.
\newblock {\em Phys.Rev.D}, 79:066012, 2009.

\bibitem{bikmsv}
M.~Bershadsky, Kenneth~A. Intriligator, S.~Kachru, David~R. Morrison, V.~Sadov,
  et~al.
\newblock {Geometric singularities and enhanced gauge symmetries}.
\newblock {\em Nucl.Phys.}, B481:215--252, 1996.

\bibitem{BinetruyXJ}
P.~Binetruy and G.~R. Dvali.
\newblock {D-term inflation}.
\newblock {\em Phys.Lett.B}, 388:241, 1996.

\bibitem{BlabackNF}
Johan Blaback, Ulf~H. Danielsson, and Thomas Van~Riet.
\newblock {Resolving anti-brane singularities through time-dependence}.
\newblock {\em arXiv:1202.1132}.

\bibitem{ouyang2}
Heng-Yu Chen, Peter Ouyang, and Gary Shiu.
\newblock {On Supersymmetric D7-branes in the Warped Deformed Conifold}.
\newblock {\em JHEP}, 1001:028, 2010.
\newblock * Temporary entry *.

\bibitem{CraigKX}
Nathaniel~J. Craig, Patrick~J. Fox, and Jay~G. Wacker.
\newblock {Reheating metastable O'Raifeartaigh models}.
\newblock {\em Phys.Rev.D}, 75:085006, 2007.

\bibitem{dhhk}
Keshav Dasgupta, Carlos Herdeiro, Shinji Hirano, and Renata Kallosh.
\newblock {D3/D7 inflationary model and M-theory}.
\newblock {\em Phys. Rev.}, D65:126002, 2002.

\bibitem{DM1}
Keshav Dasgupta and Sunil Mukhi.
\newblock {F theory at constant coupling}.
\newblock {\em Phys.Lett.}, B385:125--131, 1996.

\bibitem{DMstring}
Keshav Dasgupta and Sunil Mukhi.
\newblock {BPS nature of three string junctions}.
\newblock {\em Phys.Lett.}, B423:261--264, 1998.

\bibitem{DasguptaSU}
Keshav Dasgupta and Sunil Mukhi.
\newblock {Brane constructions, conifolds and M theory}.
\newblock {\em Nucl.Phys.B}, 551:204, 1999.

\bibitem{mukdas}
Keshav Dasgupta and Sunil Mukhi.
\newblock {Brane constructions, fractional branes and Anti-de Sitter domain
  walls}.
\newblock {\em JHEP}, 9907:008, 1999.

\bibitem{DRS}
Keshav Dasgupta, Govindan Rajesh, and Savdeep Sethi.
\newblock {M theory, orientifolds and G-flux}.
\newblock {\em JHEP}, 08:023, 1999.

\bibitem{DasguptaSW}
Keshav Dasgupta, Jihye Seo, and Alisha Wissanji.
\newblock {F-theory, Seiberg-Witten Curves and $N=2$ Dualities}.
\newblock {\em JHEP}, 1202:146, 2012.

\bibitem{dw}
Ron Donagi and Edward Witten.
\newblock {Supersymmetric Yang-Mills theory and integrable systems}.
\newblock {\em Nucl.Phys.}, B460:299--334, 1996.

\bibitem{DymarskyPM}
Anatoly Dymarsky.
\newblock {On gravity dual of a metastable vacuum in Klebanov-Strassler
  theory}.
\newblock {\em JHEP}, 1105:053, 2011.

\bibitem{DymarskyXT}
Anatoly Dymarsky, Igor~R. Klebanov, and Nathan Seiberg.
\newblock {On the moduli space of the cascading SU(M+p) x SU(p) gauge theory}.
\newblock {\em JHEP}, 0601:155, 2006.

\bibitem{ElitzurFH}
Shmuel Elitzur, Amit Giveon, and David Kutasov.
\newblock {Branes and N = 1 duality in string theory}.
\newblock {\em Phys.Lett.B}, 400:269, 1997.

\bibitem{townsend2}
Roberto Emparan, David Mateos, and Paul~K. Townsend.
\newblock {Supergravity supertubes}.
\newblock {\em JHEP}, 0107:011, 2001.

\bibitem{FischlerXH}
Willy Fischler, Vadim Kaplunovsky, Chetan Krishnan, Lorenzo Mannelli, and
  Marcus A.~C. Torres.
\newblock {Meta-Stable Supersymmetry Breaking in a Cooling Universe}.
\newblock {\em JHEP}, 0703:107, 2007.

\bibitem{gaberdiel}
Matthias~R. Gaberdiel and Barton Zwiebach.
\newblock {Exceptional groups from open strings}.
\newblock {\em Nucl. Phys.}, B518:151--172, 1998.

\bibitem{gaiotto}
Davide Gaiotto.
\newblock {N=2 dualities}.
\newblock 2009.

\bibitem{GM}
Davide Gaiotto and Juan Maldacena.
\newblock {The gravity duals of N=2 superconformal field theories}.
\newblock 2009.

\bibitem{gimpol}
Eric~G. Gimon and Joseph Polchinski.
\newblock {Consistency conditions for orientifolds and d manifolds}.
\newblock {\em Phys.Rev.}, D54:1667--1676, 1996.

\bibitem{GiveonWP}
Amit Giveon, Andrey Katz, and Zohar Komargodski.
\newblock {On SQCD with massive and massless flavors}.
\newblock {\em JHEP}, 0806:003, 2008.

\bibitem{gkunpub}
Amit Giveon and David Kutasov.
\newblock {Unpublished}.

\bibitem{GiveonSR}
Amit Giveon and David Kutasov.
\newblock {Brane Dynamics and Gauge Theory}.
\newblock {\em Rev.Mod.Phys.}, 71:983--1084, 1999.

\bibitem{GiveonFK}
Amit Giveon and David Kutasov.
\newblock {Gauge symmetry and supersymmetry breaking from intersecting branes}.
\newblock {\em Nucl.Phys.B}, 778:129, 2007.

\bibitem{GiveonEW}
Amit Giveon and David Kutasov.
\newblock {Stable and Metastable Vacua in Brane Constructions of SQCD}.
\newblock {\em JHEP}, 0802:038, 2008.

\bibitem{GiveonEF}
Amit Giveon and David Kutasov.
\newblock {Stable and Metastable Vacua in SQCD}.
\newblock {\em Nucl.Phys.B}, 796:25, 2008.

\bibitem{GiveonUR}
Amit Giveon, David Kutasov, Jock McOrist, and Andrew~B. Royston.
\newblock {D-Terms and Supersymmetry Breaking from Branes}.
\newblock {\em Nucl.Phys.B}, 822:106, 2009.

\bibitem{poln3}
Mariana Grana and Joseph Polchinski.
\newblock {Gauge / gravity duals with holomorphic dilaton}.
\newblock {\em Phys.Rev.}, D65:126005, 2002.

\bibitem{GSW}
Michael~B. Green, John~H. Schwarz, and Edward Witten.
\newblock {\em {Superstring Theory, Volume 1}}.
\newblock Cambridge University Press, 1987.

\bibitem{gsvy}
Brian~R. Greene, Alfred~D. Shapere, Cumrun Vafa, and Shing-Tung Yau.
\newblock {Stringy Cosmic Strings and Noncompact Calabi-Yau Manifolds}.
\newblock {\em Nucl.Phys.}, B337:1, 1990.

\bibitem{ghm}
Ruth Gregory, Jeffrey~A. Harvey, and Gregory~W. Moore.
\newblock {Unwinding strings and t duality of Kaluza-Klein and h monopoles}.
\newblock {\em Adv.Theor.Math.Phys.}, 1:283--297, 1997.

\bibitem{HalyoPP}
Edi Halyo.
\newblock {Hybrid inflation from supergravity D-terms}.
\newblock {\em Phys.Lett.B}, 387:43, 1996.

\bibitem{HananywittenHW}
Amihay Hanany and Edward Witten.
\newblock {Type IIB superstrings, BPS monopoles, and three-dimensional gauge
  dynamics}.
\newblock {\em Nucl.Phys.B}, 492:152--190, 1997.

\bibitem{IntriligatorDD}
Kenneth~A. Intriligator, Nathan Seiberg, and David Shih.
\newblock {Dynamical SUSY breaking in meta-stable vacua}.
\newblock {\em JHEP}, 604:021, 2006.

\bibitem{KachruGS}
Shamit Kachru, John Pearson, and Herman~L. Verlinde.
\newblock {Brane/Flux Annihilation and the String Dual of a Non-Supersymmetric
  Field Theory}.
\newblock {\em JHEP}, 0206:021, 2002.

\bibitem{kleban}
Matthew Kleban and Michele Redi.
\newblock {Expanding F-Theory}.
\newblock {\em JHEP}, 0709:038, 2007.

\bibitem{KlebanovHB}
Igor~R. Klebanov and Matthew~J. Strassler.
\newblock {Supergravity and a confining gauge theory: Duality cascades and
  chiSB-resolution of naked singularities}.
\newblock {\em JHEP}, 0008:052, 2000.

\bibitem{KutasovKB}
David Kutasov, Oleg Lunin, Jock McOrist, and Andrew~B. Royston.
\newblock {Dynamical Vacuum Selection in String Theory}.
\newblock {\em Nucl.Phys.B}, 833:64, 2010.

\bibitem{KutasovKW}
David Kutasov and Alisha Wissanji.
\newblock {IIA Perspective On Cascading Gauge Theory}.
\newblock {\em arXiv:1206.0747}.

\bibitem{MassaiJN}
Stefano Massai.
\newblock {A Comment on anti-brane singularities in warped throats}.
\newblock {\em arXiv:1202.3789 [hep-th]}.

\bibitem{townsend1}
David Mateos and Paul~K. Townsend.
\newblock {Supertubes}.
\newblock {\em Phys.Rev.Lett.}, 87:011602, 2001.

\bibitem{McOristIN}
Jock McOrist and Andrew~B. Royston.
\newblock {Relating Conifold Geometries to NS5-branes}.
\newblock {\em arXiv:1101.3552}.

\bibitem{minahan1}
Joseph~A. Minahan and Dennis Nemeschansky.
\newblock {An N=2 superconformal fixed point with E(6) global symmetry}.
\newblock {\em Nucl.Phys.}, B482:142--152, 1996.

\bibitem{minahan2}
Joseph~A. Minahan and Dennis Nemeschansky.
\newblock {Superconformal fixed points with E(n) global symmetry}.
\newblock {\em Nucl.Phys.}, B489:24--46, 1997.

\bibitem{MontonenMO}
C.~Montonen and David~I. Olive.
\newblock {Magnetic Monopoles and Gauge Particles?}
\newblock {\em Phys.Lett.B}, 72:117, 1977.

\bibitem{ouyang1}
Peter Ouyang.
\newblock {Holomorphic D7 branes and flavored N=1 gauge theories}.
\newblock {\em Nucl.Phys.}, B699:207--225, 2004.

\bibitem{polchinski1}
Joseph Polchinski.
\newblock {\em {String Theory, Volume 1}}.
\newblock Cambridge University Press, 1999.

\bibitem{polchinski2}
Joseph Polchinski.
\newblock {\em {String Theory, Volume 2}}.
\newblock Cambridge University Press, 1999.

\bibitem{PolchinskiDB}
Joseph Polchinski.
\newblock {Dirichlet-Branes and Ramond-Ramond Charges}.
\newblock {\em Phys.Rec.Lett}, 75:4724--4727, 1995.

\bibitem{PolchinskiMX}
Joseph Polchinski.
\newblock {N = 2 gauge-gravity duals}.
\newblock {\em Int.J.Mod.Phys.A}, 16:707, 2001.

\bibitem{sw1}
N.~Seiberg and Edward Witten.
\newblock {Monopole Condensation, And Confinement In N=2 Supersymmetric
  Yang-Mills Theory}.
\newblock {\em Nucl. Phys.}, B426:19--52, 1994.

\bibitem{sw2}
N.~Seiberg and Edward Witten.
\newblock {Monopoles, duality and chiral symmetry breaking in N=2
  supersymmetric QCD}.
\newblock {\em Nucl. Phys.}, B431:484--550, 1994.

\bibitem{SeibergNS}
Nathan Seiberg.
\newblock {Supersymmetry and nonperturbative beta functions}.
\newblock {\em Phys.Lett.}, B206:75, 1988.

\bibitem{SeibergPQ}
Nathan Seiberg.
\newblock {Electric - magnetic duality in supersymmetric nonAbelian gauge
  theories}.
\newblock {\em Nucl.Phys.B}, 435:129, 1995.

\bibitem{senF}
Ashoke Sen.
\newblock {F-theory and Orientifolds}.
\newblock {\em Nucl. Phys.}, B475:562--578, 1996.

\bibitem{tate}
J.~Tate.
\newblock {Algorithm for determining the type of a singular fibre in an
  elliptic pencil}.
\newblock {\em {\it Modular functions of one variable IV}, Lecture notes in
  math. Springer-Verlag, Berlin}, 476, 1975.

\bibitem{tong}
David Tong.
\newblock {NS5-branes, T duality and world sheet instantons}.
\newblock {\em JHEP}, 0207:013, 2002.

\bibitem{UrangaVF}
Angel~M. Uranga.
\newblock {Brane configurations for branes at conifolds}.
\newblock {\em JHEP}, 9901:022, 1999.

\bibitem{vafaF}
Cumrun Vafa.
\newblock {Evidence for F-Theory}.
\newblock {\em Nucl. Phys.}, B469:403--418, 1996.

\bibitem{WessbaggerWB}
Julius Wess and Jonathan Bagger.
\newblock {\em {Supersymmetry and supergravity}}.
\newblock Princeton Series In Physics, 1992.

\bibitem{WittenEP}
Edward Witten.
\newblock {Branes and the dynamics of QCD}.
\newblock {\em Nucl.Phys.B}, 507:658, 1997.

\bibitem{WittenMT}
Edward Witten.
\newblock {Solutions of four-dimensional field theories via M theory}.
\newblock {\em Nucl.Phys.B}, 500:3, 1997.

\end{thebibliography}
\bibliographystyle{plain}
%\index[abbr]{IEEE@IEEE: Institute of Electrical and Electronics Engineers, Inc.}
%\index[abbr]{CDMA@CDMA: code-division multiple access}
%\index[abbr]{CTAN@CTAN: comprehensive \protect\TeX{} archive network}

%\printindex[keylist]{Index}{Index}{}
%\printindex[abbr]{KEY TO ABBREVIATIONS}{KEY TO ABBREVIATIONS}{}

\end{document}